\documentclass[twocolumn,aps,prl,showpacs,amsmath,amssymb,superscriptaddress]{revtex4-1}
\usepackage[english]{babel}
\usepackage{graphicx}
\usepackage{dcolumn}
\usepackage{bm}
\usepackage{hyperref}
\usepackage[mathlines]{lineno}
\usepackage{xcolor}
\usepackage{tikz}
\usetikzlibrary{patterns}
\usepackage{natbib}
\usepackage{braket}

\newcommand{\rg}{\mathbf{r}}

\newcommand{\bgr}[1]{\text{\boldmath $#1$}}

\begin{document}


\title{Bose-Einstein condensation of Efimovian triples in the unitary Bose gas}
\author{S. Musolino}
\email{sil.musolino@gmail.com}
\affiliation{Eindhoven University of Technology, PO Box 513, 5600 MB Eindhoven, The Netherlands}
\author{H. Kurkjian}
\affiliation{Laboratoire de Physique Th\'eorique, Universit\'e de Toulouse, CNRS, UPS, 31400 Toulouse, France}
\author{M. Van Regemortel}
\affiliation{Joint Quantum Institute and The Institute for Research in Electronics and Applied Physics, University of Maryland, College Park, 20742 MD, USA}
\author{M. Wouters}
\affiliation{TQC, Universiteit Antwerpen, Universiteitsplein 1, B-2610 Antwerp, Belgium}
\author{S. J. J. M. F. Kokkelmans}
\affiliation{Eindhoven University of Technology, PO Box 513, 5600 MB Eindhoven, The Netherlands}
\author{V. E. Colussi}
\email{colussiv@gmail.com}
\affiliation{INO-CNR BEC Center and Dipartimento di Fisica, Universit\`a di Trento, 38123 Povo, Italy}

\date{\today}

\begin{abstract}

In an atomic Bose-Einstein condensate quenched to the unitary regime, we predict the sequential formation of a significant fraction of condensed pairs and triples.  At short distances, we demonstrate the two-body and Efimovian character of the condensed pairs and triples, respectively.  As the system evolves, their size becomes comparable to the interparticle
distance, such that many-body effects become significant.  The structure of the condensed triples depends on the size of Efimov states compared with
density scales.  Unexpectedly, we find universal condensed triples in the limit where these scales are well-separated.  Our findings provide a new framework for understanding dynamics in the unitary regime as the Bose-Einstein condensation of few-body composites.  

\end{abstract}

\maketitle

\textit{Introduction.}---Bose-Einstein condensation (BEC) gives rise to such spectacular manifestations of quantum statistics as superfluidity, superconductivity, and supersolidity \cite{book:leggett,book:pitaevskii_stringari,tinkham1996introduction,book:exciton}.  The paradigmatic theories of Bogoliubov and Bardeen-Cooper-Schriefer (BCS) describe BECs of weakly coupled bosons and fermionic pairs, respectively and have been applied in many fields of physics \cite{STRINATI20181,book:exciton,PhysRevLett.80.3177,tajima2020cooper,MACKOWIAK200025,sen2011unified,art:nozieressaintjames}.  Here, the quantum statistics of the medium alters one-body dynamics, producing quasiparticles, and two-body dynamics, producing Cooper pairs, with the latter persisting even in the absence of a two-body bound state.  These guiding concepts however must be reconsidered when describing strongly-interacting systems such as liquid helium \cite{Kunitski551}, ultracold gases \cite{rev:chevy_salomon}, nuclear matter \cite{efimov1971weakly,efimov1979low,Nishida2012,Naidon2019}, and strongly-coupled polarons \cite{PhysRevLett.114.115302,PhysRevA.92.033612,PhysRevLett.119.013401,Nygaard_2014,PhysRevLett.115.125302,christianen2021efimov}.  The occurrence of a richer few-body physics including three-body bound Efimov states amongst these strongly-interacting bosons and multi-component fermions \cite{rev:naidon,Takahashi2021} necessitates the key shifting of many-body paradigms from two to three-body correlations.  Here, the fundamental question reemerges of whether bound states in vacuum (polymers:  dimers, trimers, etc.) can be bound by the medium and converted into condensed few-body composites ($n$-tuples:  pairs, triples, etc.) possessing long-range order. 

Recently, the versatility of ultracold atomic platforms was utilized to shed new light on these open problems.  Despite strong three-body losses, quasi-equilibrated states were achieved in single-component Bose gases quenched to the unitary regime $n|a|^3\gg1$, with $n$ the atomic density and $a$ the $s$-wave scattering length \cite{art:makotyn, art:klauss, art:eigen17, art:eigen18}.  Specifically, a macroscopic population of Efimov trimers was reported in Ref.~\cite{art:klauss}, following a second sweep of interactions to weak interactions.  Historically, this technique was used to measure the condensation of Cooper pairs in the BCS-BEC crossover via their conversion into weakly bound dimers \cite{art:regal2004,book:Zwerger}.  It is thus natural to ask whether the molecules measured in Ref.~\cite{art:klauss} reveal the existence of few-body condensates of pairs and triples in the unitary Bose gas.  It is unknown whether the hypothesized universality of the medium \cite{PhysRevLett.92.090402}, parametrized by the density (Fermi) scales $k_\mathrm{n}=(6\pi^2n)^{1/3}$, $E_\mathrm{n}=\hbar^2k_\mathrm{n}^2/2m$ and $t_\mathrm{n}=\hbar/E_\mathrm{n}$, produces universal pairs and triples, or conversely whether a (non-universal) sensitivity to the Efimov effect and finite-range physics is preserved at the many-body level.  Answering this question in such a nonequilibrium and strongly interacting quenched system requires a model both ergodic \cite{art:mathias} and nonperturbative \cite{art:sykes,art:corson_bohn,PhysRevLett.124.040403,art:munozdelasheras,art:silviack}, which recovers the vacuum three-body spectrum \cite{art:colussimk,art:dincao_efimov,art:3bc_victor,art:colussi2019}.  Although widely used in statistical physics \cite{vanWijland2008,kira2011semiconductor,Schuck2015}, the cumulant model was recently adapted to quantum gases and found to fulfill these requirements \cite{art:cumulant2020,KIRA2015185}.

In this Letter, we study a uniform BEC quenched to the unitary regime and develop a general theory of simultaneous atomic, pair, and triple condensation in strongly-interacting systems possessing the Efimov effect.  Within the cumulant model, we construct generalized condensate wave functions and predict significant pair and triple condensation and associated off-diagonal long-range ordering (ODLRO) occuring between depleted atoms.  We show that the Efimovian character of the triples is guaranteed at short distances, however at later times triples have a size comparable to the interparticle spacing.  Remarkably even when Fermi and Efimovian scales are well separated, medium effects lead to the persistent production of triple and pair BECs with universal populations and internal structures. 
\begin{figure}[h!]
\centering
\includegraphics[width=8.6cm]{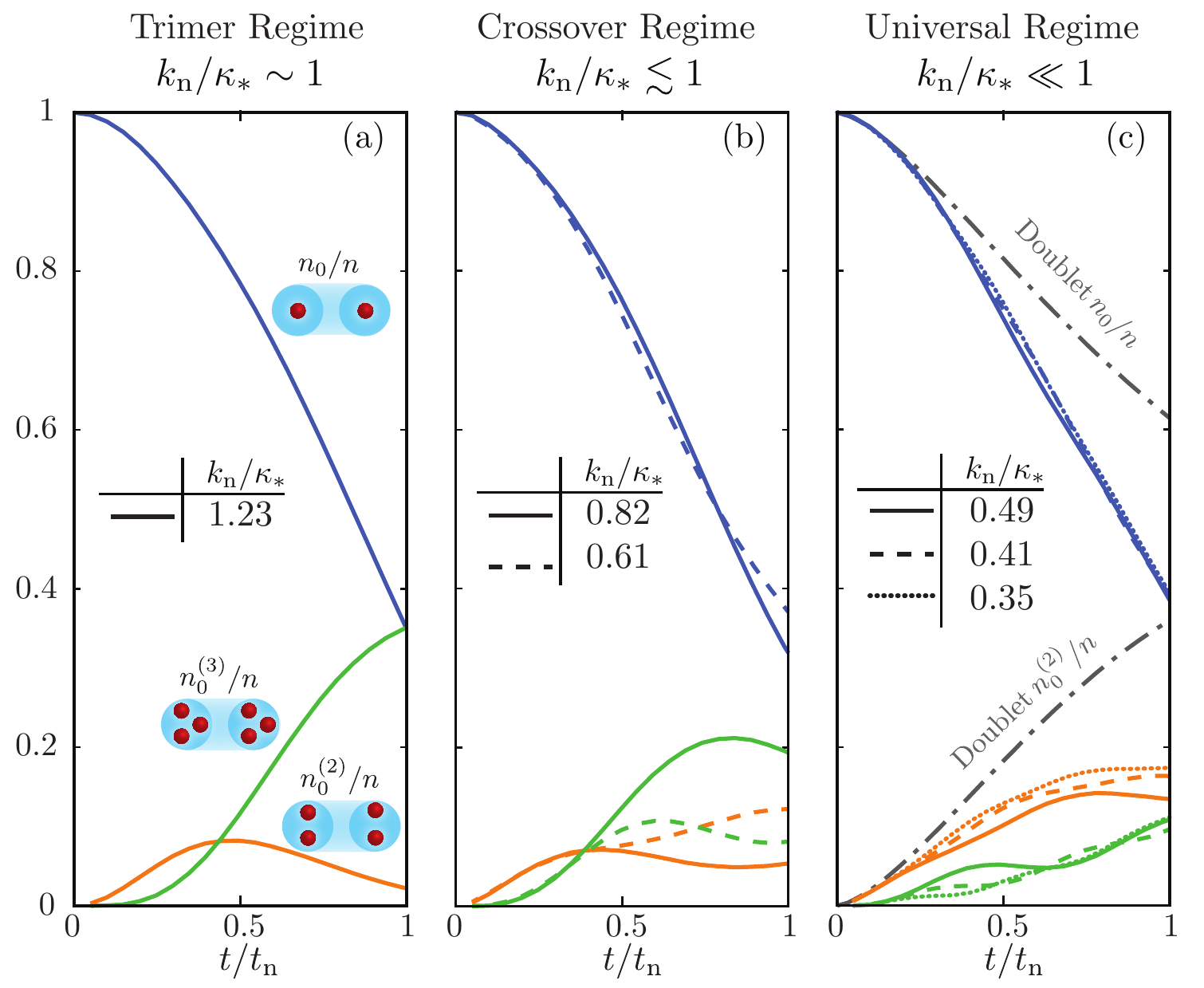}
\caption{Atomic (blue), pair (orange), and triple (green) condensate fraction dynamics in the (a) trimer, (b) crossover, and (c) universal density regimes.  The grey dot-dashed lines indicate universal results in the doublet  (Hartree-Fock Bogoliubov~\cite{book:blaizot}) model.  The comparison of Efimovian ($\kappa_*$) and Fermi ($k_\mathrm{n}$) scales strongly determines the relative populations of the condensed few-body composites. }
\label{fig:1}
\end{figure}

\textit{Model.}---We model the system of $N$ spinless bosons in a cubic volume $V$ using a single-channel Hamiltonian with pairwise $s$-wave interactions 
\begin{equation}\label{eq:mbham}
\begin{split}
\hat{H} &= \int d^3 r \hat{\psi}^\dagger(\rg) \left( - \frac{\hbar^2}{2m}  \Delta_{\rg}\right)  \hat{\psi}(\rg) \\
&+ \frac{1}{2} \int d^3 r d^3 r'   \hat{\psi}^\dagger(\rg) \hat{\psi}^\dagger(\rg') V(|\rg-\rg'|) \hat{\psi}(\rg'){\hat{\psi}}(\rg),
\end{split}
\end{equation}
where $\hat{\psi}({\bf r})=(1/\sqrt{V})\sum_{\bf k}\hat{a}_{\bf k}e^{i{\bf k}\cdot{\bf r}}$ are field operators and  $\hat{a}_{\bf k}$ annihilates a boson of momentum $\hbar{\bf k}$.
At unitarity ($|a|\to\infty$), the actual potential can be replaced by a simpler nonlocal separable potential $\hat{V}=g|\zeta\rangle\langle\zeta|$ with $s$-wave form factors $\langle {\bf k}|\zeta\rangle=\theta(\Lambda-|{\bf k}|)$, interaction strength $g=-\pi^3\hbar^2\bar{a}/m$ and cutoff $\Lambda=2/\pi\bar{a}$, where $\bar{a}=0.955r_\mathrm{vdW}$ is the mean scattering length and $r_\mathrm{vdW}$ is the van der Waals length for a particular atomic species \cite{art:chin,PhysRevA.59.1998,PhysRevA.67.013601,art:colussimk}.  This sets the three-body parameter $\kappa_*r_\mathrm{vdW}\approx 0.211$, which is the wave number of the ground-state Efimov trimer at unitarity (see \cite{SM,art:colussimk,art:silviack,art:cumulant2020}).

 We model the postquench many-body dynamics of an initially pure, noninteracting atomic condensate  using the method of cumulants whose hierarchical structure reflects the sequential growth of intrinsically higher-order correlations \cite{FRICKE1996479,PhysRevA.65.033601,kira2011semiconductor,KIRA2015185,kira2014excitation,kirancomm,art:cumulant2020}. Within the $U(1)$ symmetry-breaking picture, we study the dynamics of the singlet $\langle\hat a_{\bf k}\rangle=\delta_{{\bf k} 0}\sqrt{V}\psi_0$, which describes the atomic BEC in the $k=0$ mode.  In the frame rotating with the condensate phase $\theta_0$, we study also the doublets 
\begin{equation}
n_{\bf k} = \langle \hat{a}^\dagger_{\bf k}\hat{a}_{\bf k}\rangle,\quad c_{\bf k} =e^{-2i\theta_0} \langle \hat{a}_{\bf -k}\hat{a}_{\bf k}\rangle,
\end{equation}
describing the single-particle momentum distribution and pairing, respectively, and the triplets
\begin{equation}\label{eq:triplets}
M_{\bf k,q} =e^{i\theta_0}  \langle \hat{a}^\dagger_{\bf k-q}\hat{a}^\dagger_{\bf q}\hat{a}_{\bf k}\rangle,\quad R_{\bf k,q} =e^{-3i\theta_0} \langle \hat{a}_{\bf q-k}\hat{a}_{\bf k}\hat{a}_{\bf -q}\rangle,
\end{equation}
introducing ergodic processes and the Efimov effect \cite{art:mathias,PhysRevLett.89.210404,art:colussimk,art:cumulant2020}. Truncating the cumulant hierarchy can be justified at early times due to the sequential nature of correlation buildup~\cite{SM,art:cumulant2020}. However
the increasing importance of quadruplets, in particular for energy conservation,
limits our study to times $t\lesssim t_\mathrm{n}$.

\textit{Off-diagonal long-range ordering.}---The triplet model contains anomalous averages at the one-body level $(\psi_0)$ in the atomic condensate and at the two- ($c_{\bf k}$) and three-body ($R_{\bf k,q}$) levels {\it within} the quantum depletion.  These cumulants are intimately connected to the eigenfunctions of the reduced density matrices, signalling ODLRO and condensation \cite{art:penrose_51,art:penrose_onsager,art:yang}.  We begin from the spectral decomposition of the one-body density matrix
\begin{equation}
\rho^{(1)}(\rg_1, \rg_2; t)\! =\! \braket{\hat{\psi}^\dagger(\rg_2)\hat{\psi}(\rg_1)}
 \!=\! \sum_\nu N_\nu(t) \varphi^\ast_\nu (\rg_2, t) \varphi_\nu (\rg_1, t),
\label{eq:rho_1}
\end{equation}
where 
 $\varphi_\nu$ are orthogonal one-body eigenstates.  Only one eigenvalue $N_{\nu=0}$ is assumed to be macroscopic such that $\varphi_0$ is responsible for ODLRO at the one-body level
\begin{equation}
\lim_{|\mathbf{r}_1-\mathbf{r}_2|\to \infty}\rho^{(1)}(\rg_1, \rg_2; t) = N_0(t) \varphi_0^\ast(\rg_2, t)\varphi_0(\rg_1, t).
\label{eq:lim_ODLR_1}
\end{equation}
Within the cumulant approach, the long-range part of $\rho^{(1)}$ is simply $|\psi_0|^2$, such that $N_0=V|\psi_0|^2$ coincides with the condensate population $\langle\hat{a}_0^\dagger \hat{a}_0\rangle$ and fraction $n_0=N_0/V$.

In the presence of one-body condensation, ODLRO occurs trivially at all higher orders \cite{kira2014excitation,art:yang}. We isolate therefore the atomic condensate from the fluctuations $\hat{\psi}(\rg)= \psi_0+ \delta\hat{\psi}(\rg)$, satisfying $\langle\delta\hat{\psi}(\rg)\rangle=0$. To study intrinsically few-body ODRLO amongst fluctuations, we adapt the treatment of Yang \cite{art:yang} and spectrally decompose the corresponding $p$-body density matrices 
\begin{align}
&\braket{\delta\hat{\psi}^\dagger(\rg'_1)\dots\delta\hat{\psi}^\dagger(\rg'_p)\delta\hat{\psi}(\rg_p)\dots\delta\hat{\psi}(\rg_1)}\nonumber\\
&= \sum_\nu N_\nu^{(p)}(t) \varphi^{(p)\ast}_\nu (\rg'_1,\dots,\rg'_p,t) \varphi^{(p)}_\nu (\rg_1,\dots,\rg_p,t),\label{eq:flucrho}
\end{align}
where $ \varphi^{(p)}_\nu$ and $N^{(p)}_\nu$ are the orthogonal $p$-body eigenstates and eigenvalues, respectively.
Analogous to Eq.~\eqref{eq:lim_ODLR_1}, when the $p$-body density matrix is nonzero in the long-range limit $|\sum_{i=1}^p \textbf{r}_i-\textbf{r}_i'|/p\to\infty$, there exists intrinsic $p$-body ODLRO.  In the triplet model, this limit is dominated by the anomalous contraction $\langle\delta\hat{\psi}^\dagger\dots\delta\hat{\psi}^\dagger\rangle \langle\delta\hat{\psi}\dots\delta\hat{\psi}\rangle$, such that nonzero $c$ or $R$ cumulants generate ODLRO. The associated normalized pair and triple wave functions are
\begin{equation}\label{varphi}
\varphi^{(2)}_0({\bf r},t)=\frac{c({\bf r},t)}{\sqrt{N^{(2)}_0(t)}},\ \varphi^{(3)}_0({\bf r},\boldsymbol{\rho},t)=\frac{R({\bf r},\boldsymbol{\rho},t)}{\sqrt{N^{(3)}_0(t)}},
\end{equation}
with $N^{(2)}_0=\sum_{\bf k} |c_{\bf k}|^2$, $N^{(3)}_0=\sum_{{\bf k},{\bf q}} |R_{{\bf k},{\bf q}}|^2$, and Jacobi vectors $\rg\equiv\rg_1-\rg_2$, $\text{\boldmath$\rho$}=\rg_3-(\rg_1+\rg_2)/2$.  We note that pair or triple condensation may generate trivial ODLRO in density matrices with $p\geq4$.
If one were to study, e.g., quadruple condensation, such contributions should be removed.

\textit{Condensate fractions.}---Unlike one-body condensation, the macroscopic eigenvalues $N_0^{(p)}$ cannot be directly related to condensed fractions. To understand this, we construct composite operators annihilating condensed pairs and triples
\begin{align}
\hat{b}_0^{(p)}=\frac{1}{\sqrt{p!}}\left[\prod_{i=1}^p\int d^3 r_i \delta\hat{\psi}({\bf r}_i)\right] \varphi_0^{(p)*}({\bf r}_1,\dots,{\bf r}_p).\label{b0p}
\end{align}
Evaluating the quantum average of the commutators in the triplet model gives
\begin{align}
&\langle[\hat b_0^{(2)},\hat b_0^{(2)\dagger}]\rangle=1+\frac{2}{N_0^{(2)}}\sum_{\bf k}|c_{\bf k}|^2n_{\bf k},\label{eq:paircom}\\
&\langle[\hat b_0^{(3)},\hat b_0^{(3)\dagger}]\rangle=1+\frac{3}{N_0^{(3)}}\sum_{{\bf k},{\bf q}}|R_{\bf k,q}|^2n_{\bf k}(1+n_{\bf q}),\label{eq:tripcom}
\end{align}
which approximate the canonical relations only for weak excitations ($n_{\bf k}\ll1$) or localized pairs and triples relative to the medium.  In the opposite limit, composite bosons are created on top of densely populated Fourier modes, leading to Bose enhancement of atoms within the created composite and an overestimation of the condensed fraction \cite{PhysRevA.75.043613}.  Consequently, the rapid quantum depletion of the atomic condensate in the unitary regime yields $\hat{b}_0^{(p)}$ that are {\it approximately} bosonic only at $t\lesssim t_n$ (see \cite{SM}).  The renormalization procedure
\begin{equation}
\hat{B}_0^{(p)} = \frac{\hat{b}_0^{(p)}}{\sqrt{\langle [\hat{b}_0^{(p)},\hat{b}_0^{(p)\dagger}]\rangle}},\label{renormalization}
\end{equation}
effectively ensures that the bosonic canonical relations are preserved {\it on average} at all times, which we use to compute the pair and triple condensate fractions as $n^{(p)}_0/n=\protect\langle\hat{B}^{(p)\dagger}_0\hat{B}^{(p)}_0\protect\rangle/(N/p)$.

The postquench dynamics of the condensate fractions are shown in Fig.~\ref{fig:1}.  Compared to the doublet model, three-body processes in the triplet model lead to an accelerated depletion of the atomic condensate, reaching $n_0/n\approx 0.4$ by $t=t_\mathrm{n}$.  At early times, the formation of condensed triples follows sequentially the universal pair condensate growth, reflecting the hierarchical structure of the cumulant equations of motion (see \cite{SM,art:colussimk}).  At later times the dynamics depend strongly on the density regime, repeating log-periodically with the density typical of the Efimov effect \cite{art:dincao_efimov,art:3bc_victor,art:colussi2019}.  In the trimer regime [Fig.~\ref{fig:1}(a)], the ground-state Efimov trimer resonantly overlaps with the scale set by the density ($k_\mathrm{n}/\kappa_*\sim1$), and triple condensation dominates clearly at later times, becoming comparable to the atomic condensate fraction. Condensed atoms can be converted to pairs and triples via low energy two and three-body scattering, respectively \cite{PhysRevLett.89.210404,PhysRevLett.82.463,PhysRevLett.88.040401}.  As this overlap becomes less resonant ($k_\mathrm{n}/\kappa_*\lesssim1$), the system enters the crossover regime where particle-number oscillations between pair and triple BECs visible in Fig.~\ref{fig:1}(b) are analogous to the atom-dimer coherences of Ref.~\cite{art:donley_ram}.  In the universal regime [Fig.~\ref{fig:1}(c)], Efimovian and Fermi scales are well separated ($k_\mathrm{n}/\kappa_*\ll1$), and the oscillation becomes increasingly faster relative to $t_\mathrm{n}$.  This is the characteristic dynamical signature of the Efimov effect \cite{art:colussimk,art:3bc_victor,art:colussi2019,art:braaten_2003}.  At later times, pair condensation remains dominant while the nonuniversal oscillations fade and the condensate fractions converge universally, approaching $n_0^{(2)}/n\approx 0.2$ and $n_0^{(3)}/n\approx 0.1$ by $t=t_\mathrm{n}$.

\textit{Short-range expansions.}---We study now how the short-range behavior of the condensate wave functions $c({\bf r},t)$ and $R({\bf r},\boldsymbol{\rho},t)$ are dictated by few-body physics.  This can be understood from the corresponding cumulant equations of motion which are identical to few-body Schr\"odinger equations at momenta large compared to the many-body scales \cite{art:cumulant2020,art:colussimk,SM,book:leggett}.  For distances larger than the short range of the potential $(r_\mathrm{vdW}<r<k_\mathrm{n}^{-1},a)$, the pair and triple condensate wave functions can be expanded in terms of the zero-energy few-body scattering wave functions
\begin{align}
&c({\bf r},t)\underset{r\to0}{=} \frac{1}{4\pi}\Psi^{(2)}_0(t)\phi({\bf r}),\label{eq:cshort} \\
&R({\bf r},\boldsymbol{\rho},t)\underset{R\to0}{=}\frac{2^{3/2}}{3^{1/4}s_0}\Psi^{(3)}_0(t)\Phi(R,{\bf \Omega}),\label{eq:Rshort}
\end{align}
which define the macroscopic order parameters $\Psi^{(p)}_0$ (see the Supplemental Material~\cite{SM}).  Here $\phi({\bf r})=1/r-1/a$ is the zero-energy two-body scattering state, and 
\begin{equation}
\Phi(R, \bgr\Omega) = \frac{1}{R^2}\sin\left[s_0\log \frac{R}{R_t}\right] \frac{\phi_{is_0}(\bgr\Omega)}{\sqrt{\langle \phi_{is_0}|\phi_{is_0}\rangle}}
\end{equation}
is the zero-energy three-body scattering state for hyperraddius $R=\sqrt{r^2/2+2\rho^2/3}$ and hyperangles ${\bf \Omega}=\{ \hat{\boldsymbol{\rho}},{\bf \hat{r}},\alpha=\arctan(r/\rho)\}$ \cite{art:werner_bosons}.  Here,  $s_0\approx 1.00624$, $R_t=\sqrt{2}\exp\{\text{Im} \ln[\Gamma(1+is_0)]/s_0\}/\kappa_*$, and $\Gamma$ is the gamma function.  The hyperangular function describing $s$-wave pairwise scatterings is $\phi_{s}({\bf \Omega})=(1+\hat{P}_{13}+\hat{P}_{23})\sin(s(\pi/2-\alpha))/\sin(2\alpha)\sqrt{4\pi}$ where $\hat{P}_{ij}$ swaps particles $i$ and $j$ \footnote{The normalization constant is $\protect\langle\phi_s|\phi_s\protect\rangle\equiv \int d{\bf \Omega}|\phi_{s}({\bf \Omega})|^2=-\frac{12\pi}{s}\sin\left(\frac{s^\ast\pi}{2}\right)$ $\left[\cos\left(\frac{s\pi}{2}\right)-\frac{s\pi}{2}\sin\left(\frac{s\pi}{2}\right)-\frac{4\pi}{3\sqrt{3}}\cos\left(\frac{s\pi}{6}\right)\right]$.}.  
From Eqs.~\eqref{eq:cshort} and \eqref{eq:Rshort} we see then at unitarity that the pairs have a universal behavior $\sim 1/r$ at short distances, whereas the triples have an Efimovian character, diverging as $1/R^2$ and oscillating log periodically with the three-body parameter.

 \begin{figure}
\centering
\includegraphics[width=8.6cm]{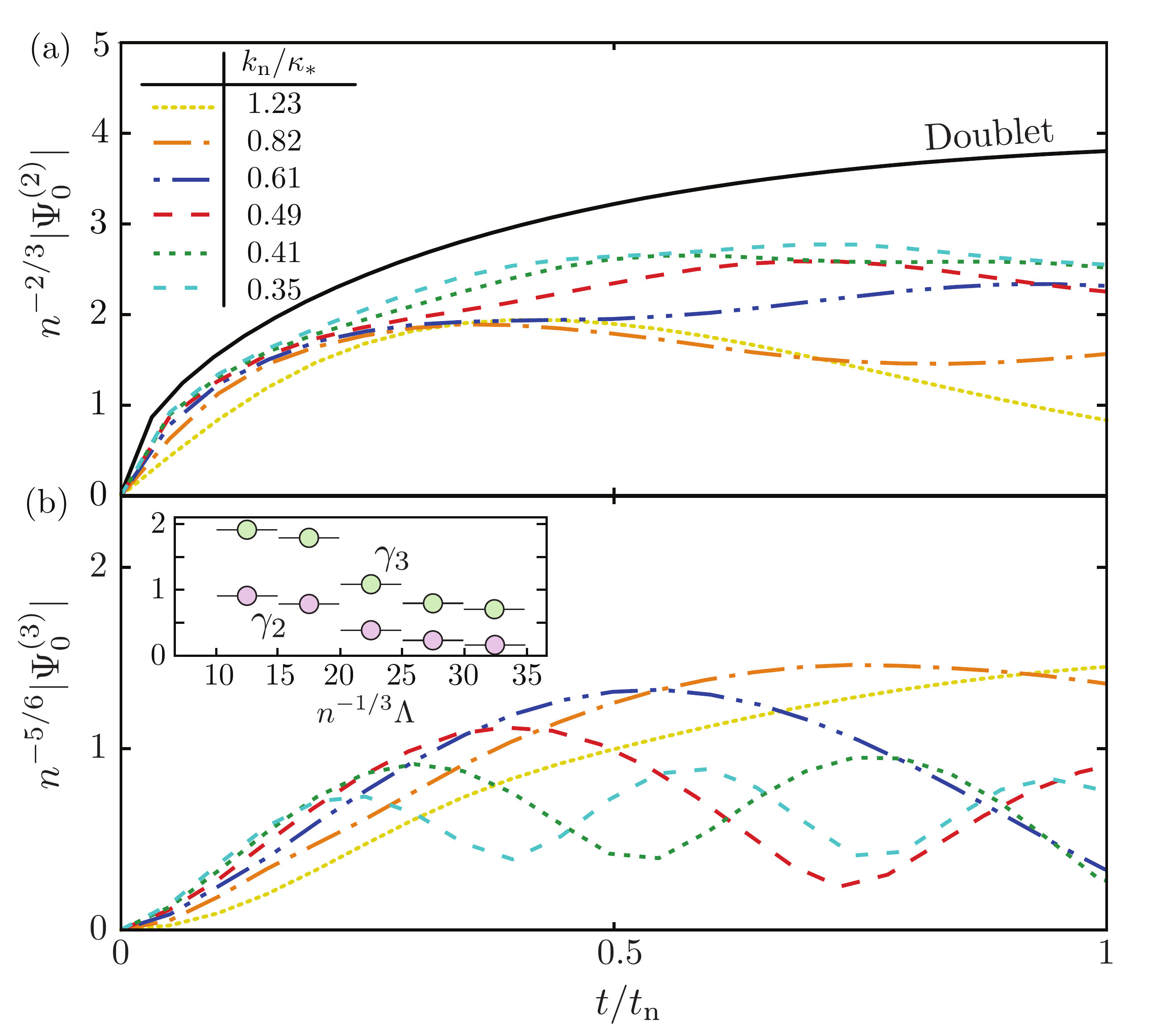}
\caption{Macroscopic (a) pair and (b) triple order parameter dynamics over a range of densities within the triplet model.  Black solid line: universal results within the doublet model.  (inset) Residual exponents $\Psi_0^{(p)}\propto \Lambda^{\gamma_p}$ evaluated at fixed $t=0.05t_\mathrm{n}$ converge as expected to 0 as the system becomes increasingly dilute with respect to the range of the interaction ($\Lambda/k_\mathrm{n}\to\infty$).}
\label{fig:2}
\end{figure}
At short distances, the total probability to measure clustered pairs and triples is encoded in the two- and three-body contact densities $\mathcal{C}_2$ and $\mathcal{C}_3$, respectively, central to a set of universal relations between system properties \cite{art:werner_bosons,art:braaten_c3_first}.  In the presence of pair and triple condensation, these clusters can be divided into contributions from the order parameters and higher-order cumulants
\begin{align}
&\mathcal{C}_2=\frac{m^2 g^2}{\hbar^4}\langle(\hat{\psi}^\dagger)^2 \hat{\psi}^2\rangle= |\Psi_0^{(2)}|^2+  \delta\mathcal{C}_2\label{eq:contactconnect2},\\
&\mathcal{C}_3=-\frac{m^2g^2}{2\hbar^4\Lambda^2}\left(H'+\frac{J'}{a\Lambda}\right)\langle(\hat{\psi}^\dagger)^3\hat{\psi}^3\rangle=|\Psi_0^{(3)}|^2+ \delta \mathcal{C}_3,\label{eq:contactconnect3}
\end{align}
where $H'$ and $J'$ are log-periodic functions of $\Lambda$, $\hat{\psi}=\hat{\psi}({\bf 0})$ are local field operators, and $\delta \mathcal{C}_p$'s are contributions absent in the triplet model (see the Supplemental
Material~\cite{SM}).  This establishes the square modulus of $\Psi_0^{(p)}$ as a probability density, analogous to $\psi_0$ at the one-body level.   The dynamics shown in Fig.~\ref{fig:2} can be understood from Refs.~\cite{art:cumulant2020,art:colussi2019,art:sykes,art:3bc_victor},  namely, early-time growths $|\Psi_0^{(p)}(t)|\propto t^{(p-1)/2}$ with primary ($p=3$) and secondary ($p=2$) visibility of nonuniversal trimer oscillations in the crossover regime ($k_\mathrm{n}/\kappa_*=0.82$ and $0.61$).

\begin{figure}
  \centering 
\includegraphics[width=8.6cm]{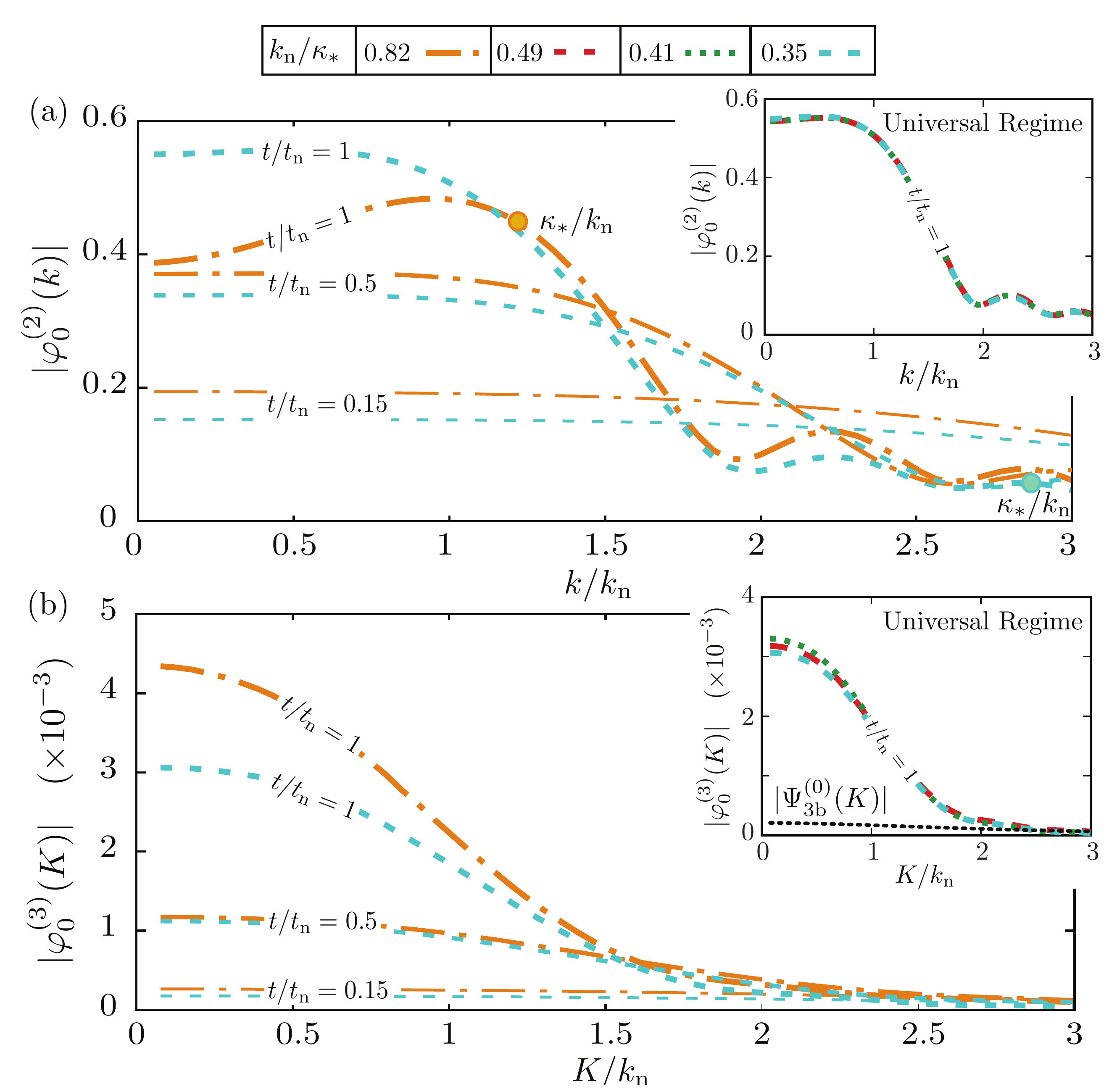}
\caption{Internal structures of the normalized (a) pair and (b) triple condensate wave functions as the system evolves for densities within the crossover (orange) and universal (light blue) regimes.  The filled circles in (a) indicate the three-body parameter.  (insets) Normalized pair and triple condensate wave functions in the universal regime at $t=t_\mathrm{n}$ compared to the vacuum ground-state Efimov trimer $|\Psi_\mathrm{3b}^{(0)}(K)|$ (black dotted) for density $k_\mathrm{n}/\kappa_*=0.35$.}
\label{fig:3}
\end{figure}

\textit{Internal structure.}---We study now the longer-range internal structure of the pair and triple condensate wave functions, focusing on the interplay between Efimovian and Fermi scales.   Fig.~\ref{fig:3} shows the triplet model results for the normalized pair and triple condensate wave functions at $t/t_\mathrm{n}=0.15, 0.5, 1$.  To visualize the triple condensate wave function, we average $R_{\bf k,q}$ over internal configurations at a fixed hypermomentum $K^2=k^2+q^2+kq\cos\theta$ where $\cos\theta={\bf \hat k}\cdot{\bf \hat q}$. The corresponding $\varphi_0^{(3)}(K)$ captures variations of the coherent tripling physics with changes in the overall three-body momentum scale \cite{SM}.

At early stages of evolution $(t=0.15t_\mathrm{n})$, the pair and triple wave functions are relatively constant over the range of momenta considered consistent with the buildup of local correlations between nearby particles \cite{art:sykes}.  Accordingly, the small amount of clustered pairs and triples are dominated by few-body physics.  This explains why their condensation dynamics shown in Fig.~\ref{fig:1} follow the corresponding contact growth laws.  We note that the local, structural origin of these laws was not recognized in Ref.~\cite{art:dincao_efimov}.

At later times, both pair and triple wave functions become increasingly nonlocal.  From the insets of Fig.~\ref{fig:3} we see that as the system approaches more deeply the universal regime, both condensate wave functions acquire a universal form by $t=t_\mathrm{n}$.  Together with the universal behavior of the triple condensed fractions in Fig.~\ref{fig:1}(c), this remarkable finding suggests the existence of a condensate composed of {\it universal} Efimovian triples at later times despite strongly non-universal short-distance behavior [see Fig.~\ref{fig:2}(b)].  In the crossover regime results in Fig.~\ref{fig:3}(a), we find the development of a peak at momenta $k_\mathrm{n}\sim\kappa_*$ reminiscent of the Cooper pair in the BEC-BCS crossover \cite{art:parish}.  The absence of this peak in simulations of the doublet model and universal regime of the triplet model [Fig.~\ref{fig:3}(a)] ties it to the Efimov effect.

\begin{figure}
  \centering
\includegraphics[width=8.6cm]{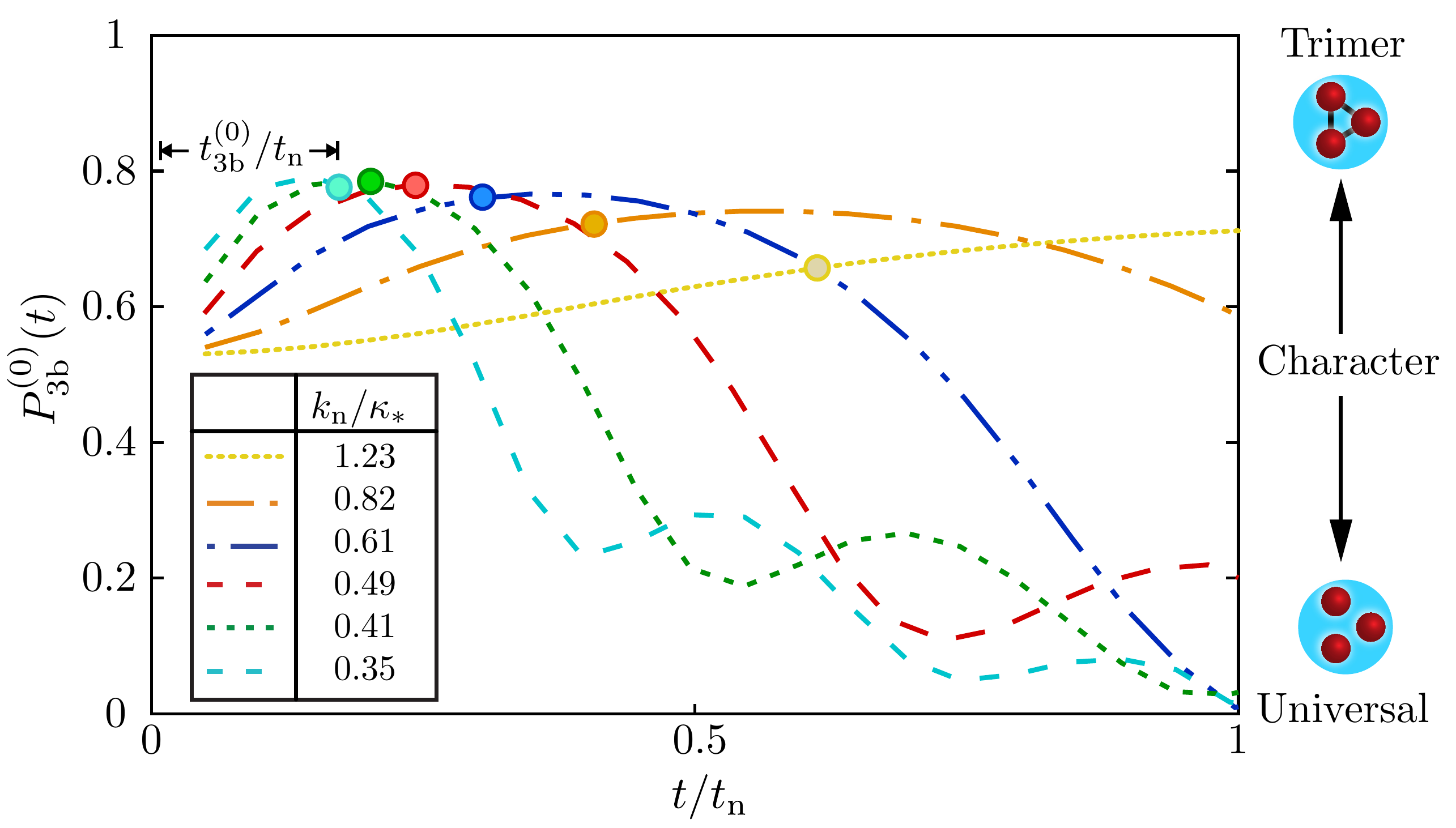}
\caption{Evolution of the overlap (Eq.~\eqref{eq:p3b}) between ground-state Efimov trimer and triple condensate wave function for varying densities, with $P^{(0)}_\mathrm{3b}(t=0)$ undefined.  The numerical upper limit is comparable to the total probability ($\sim\%82$) projected onto the Efimov channel in the quenched three-body problem with characteristic timescale $t_\mathrm{3b}^{(0)}=m/\hbar\kappa_*^2$ (filled circles) \cite{wernerthesis}.  The illustrations indicate the presence or absence of a trimer character in the triple condensate.}
\label{fig:4}
\end{figure}
To study the presence and role of the ground-state Efimov trimer $| \Psi_\mathrm{3b}^{(0)}\rangle$ in the triple condensate $|\varphi_0^{(3)}\rangle$, we evaluate the overlap
\begin{equation}
P^{(0)}_\mathrm{3b}(t) = |\braket{ \varphi_0^{(3)}| \Psi_\mathrm{3b}^{(0)}}|^2,
\label{eq:p3b}
\end{equation}
as shown in Fig.~\ref{fig:4} (see the Supplemental Material~\cite{SM}). In the universal regime, the ground-state Efimov trimer is localized relative to the Fermi scales.  At all times in this regime, $P^{(0)}_\mathrm{3b}$ reflects therefore the short-range behavior of the triple condensate wave function encapsulated by $\Psi_0^{(3)}$, contributing the characteristic trimer oscillations visible in Fig.~\ref{fig:4}.  After a small initial increase, the rapid decrease of $P_\mathrm{3b}^{(0)}$ in this regime reflects the local to nonlocal transition of the triple condensate wave function, which bears little resemblance to the trimer as shown in the inset of Fig.~\ref{fig:3}(b).  This is responsible also for the decreased visibility of the trimer oscillations in Fig.~\ref{fig:1}(c).  Consequently, this local to nonlocal structural transition, generated by the medium effect of the strong quantum-depletion, is the underlying mechanism by which macroscopic observables, such as the condensate fractions in Fig.~\ref{fig:1}(c), display a universal scale invariance.  This occurs despite continued local sensitivity to the three-body parameter [Eq.~\eqref{eq:Rshort}] which becomes less relevant as the coherent physics begins to occur predominantly on the Fermi scale.  In the trimer regime, we find no decrease of $P_\mathrm{3b}^{(0)}$ for $t\lesssim t_n$, and the early-time increase is more gradual.  From Fig.~\ref{fig:3}(b) it is clear that although the coherent physics occurs predominantly on the Fermi scale at later times as before, when one has $k_\mathrm{n}\sim\kappa_*$ the condensed triples are increasingly dominated by the ground-state Efimov trimer.  In short, the non-universal, trimer character of the triple condensate increases at later times in the trimer regime, whereas it decreases in the universal regime where one finds triples {\it without} a vacuum equivalent.

\textit{Conclusion.}--- Using the cumulant model that includes three-body correlations, we have shown that novel types of few-body condensates are generated within the quantum depletion of a quenched unitary Bose gas.  Crucially, the regime of universal pair and triple condensation demonstrates a strongly-interacting many-body system behaving universally even in the presence of non-universal few-body physics such as the Efimov effect.  We expect the molecular fractions produced following an interaction sweep back to weak interactions \cite{art:regal2004,art:klauss} to reflect the few-body composites present in the unitary regime.  However the highlighted difficulties of counting composite bosons extended in the medium requires a precise modeling of the 
projection and remains the subject of future work \cite{PhysRevLett.95.110404,art:silviack}.  Additionally, the tripling fluctuations discussed in this Letter raise interesting prospects for measuring non-Gaussian many-body states \cite{Clement2021,Schweigler2017}.

\begin{acknowledgments}

\textit{Acknowledgments.}---We thank Jinglun Li, Denise Ahmed-Braun, Paul Mestrom, Thomas Secker, and Gijs Groeneveld for fruitful discussions.  M.V.R. acknowledges support by MURI-AFOSR FA9550- 19-1-0399.  M.W. acknowledges financial support from the FWO-Vlaanderen under the project G016219N.  V.E.C. acknowledges financial support from Provincia Autonoma di Trento and the Italian MIUR under the PRIN2017 projectCEnTraL.  S.M. and S.J.J.M.F.K acknowledge financial support by the Netherlands Organisation for Scientific Research (NWO) under Grant No. 680-47-623.

\end{acknowledgments}

\bibliographystyle{apsrev4-1}
\bibliography{biblio}

\begin{thebibliography}{88}%
\makeatletter
\providecommand \@ifxundefined [1]{%
 \@ifx{#1\undefined}
}%
\providecommand \@ifnum [1]{%
 \ifnum #1\expandafter \@firstoftwo
 \else \expandafter \@secondoftwo
 \fi
}%
\providecommand \@ifx [1]{%
 \ifx #1\expandafter \@firstoftwo
 \else \expandafter \@secondoftwo
 \fi
}%
\providecommand \natexlab [1]{#1}%
\providecommand \enquote  [1]{``#1''}%
\providecommand \bibnamefont  [1]{#1}%
\providecommand \bibfnamefont [1]{#1}%
\providecommand \citenamefont [1]{#1}%
\providecommand \href@noop [0]{\@secondoftwo}%
\providecommand \href [0]{\begingroup \@sanitize@url \@href}%
\providecommand \@href[1]{\@@startlink{#1}\@@href}%
\providecommand \@@href[1]{\endgroup#1\@@endlink}%
\providecommand \@sanitize@url [0]{\catcode `\\12\catcode `\$12\catcode
  `\&12\catcode `\#12\catcode `\^12\catcode `\_12\catcode `\%12\relax}%
\providecommand \@@startlink[1]{}%
\providecommand \@@endlink[0]{}%
\providecommand \url  [0]{\begingroup\@sanitize@url \@url }%
\providecommand \@url [1]{\endgroup\@href {#1}{\urlprefix }}%
\providecommand \urlprefix  [0]{URL }%
\providecommand \Eprint [0]{\href }%
\providecommand \doibase [0]{http://dx.doi.org/}%
\providecommand \selectlanguage [0]{\@gobble}%
\providecommand \bibinfo  [0]{\@secondoftwo}%
\providecommand \bibfield  [0]{\@secondoftwo}%
\providecommand \translation [1]{[#1]}%
\providecommand \BibitemOpen [0]{}%
\providecommand \bibitemStop [0]{}%
\providecommand \bibitemNoStop [0]{.\EOS\space}%
\providecommand \EOS [0]{\spacefactor3000\relax}%
\providecommand \BibitemShut  [1]{\csname bibitem#1\endcsname}%
\let\auto@bib@innerbib\@empty
\bibitem [{\citenamefont {Leggett}(2006)}]{book:leggett}%
  \BibitemOpen
  \bibfield  {author} {\bibinfo {author} {\bibfnamefont {A.}~\bibnamefont
  {Leggett}},\ }\href {https://books.google.nl/books?id=HnlPAwAAQBAJ} {\emph
  {\bibinfo {title} {Quantum Liquids: Bose condensation and Cooper pairing in
  condensed-matter systems}}},\ Oxford Graduate Texts\ (\bibinfo  {publisher}
  {OUP Oxford},\ \bibinfo {year} {2006})\BibitemShut {NoStop}%
\bibitem [{\citenamefont {Pitaevskii}\ and\ \citenamefont
  {Stringari}(2016)}]{book:pitaevskii_stringari}%
  \BibitemOpen
  \bibfield  {author} {\bibinfo {author} {\bibfnamefont {L.}~\bibnamefont
  {Pitaevskii}}\ and\ \bibinfo {author} {\bibfnamefont {S.}~\bibnamefont
  {Stringari}},\ }\href {https://books.google.nl/books?id=yHByCwAAQBAJ} {\emph
  {\bibinfo {title} {Bose-Einstein Condensation and Superfluidity}}},\
  International Series of Monographs on Physics\ (\bibinfo  {publisher} {OUP
  Oxford},\ \bibinfo {year} {2016})\BibitemShut {NoStop}%
\bibitem [{\citenamefont {Tinkham}(1996)}]{tinkham1996introduction}%
  \BibitemOpen
  \bibfield  {author} {\bibinfo {author} {\bibfnamefont {M.}~\bibnamefont
  {Tinkham}},\ }\href@noop {} {\emph {\bibinfo {title} {Introduction to
  Superconductivity}}}\ (\bibinfo  {publisher} {McGraw-Hill, New York},\
  \bibinfo {year} {1996})\BibitemShut {NoStop}%
\bibitem [{\citenamefont {Moskalenko}\ and\ \citenamefont
  {Snoke}(2000)}]{book:exciton}%
  \BibitemOpen
  \bibfield  {author} {\bibinfo {author} {\bibfnamefont {S.~A.}\ \bibnamefont
  {Moskalenko}}\ and\ \bibinfo {author} {\bibfnamefont {D.~W.}\ \bibnamefont
  {Snoke}},\ }\href@noop {} {\emph {\bibinfo {title} {Bose-Einstein
  condensation of excitons and biexcitons: and coherent nonlinear optics with
  excitons}}}\ (\bibinfo  {publisher} {Cambridge University Press},\ \bibinfo
  {year} {2000})\BibitemShut {NoStop}%
\bibitem [{\citenamefont {Strinati}\ \emph {et~al.}(2018)\citenamefont
  {Strinati}, \citenamefont {Pieri}, \citenamefont {R\"opke}, \citenamefont
  {Schuck},\ and\ \citenamefont {Urban}}]{STRINATI20181}%
  \BibitemOpen
  \bibfield  {author} {\bibinfo {author} {\bibfnamefont {G.~C.}\ \bibnamefont
  {Strinati}}, \bibinfo {author} {\bibfnamefont {P.}~\bibnamefont {Pieri}},
  \bibinfo {author} {\bibfnamefont {G.}~\bibnamefont {R\"opke}}, \bibinfo
  {author} {\bibfnamefont {P.}~\bibnamefont {Schuck}}, \ and\ \bibinfo {author}
  {\bibfnamefont {M.}~\bibnamefont {Urban}},\ }\href {\doibase
  https://doi.org/10.1016/j.physrep.2018.02.004} {\bibfield  {journal}
  {\bibinfo  {journal} {Phys. Rep.}\ }\textbf {\bibinfo {volume} {738}},\
  \bibinfo {pages} {1} (\bibinfo {year} {2018})}\BibitemShut {NoStop}%
\bibitem [{\citenamefont {R\"opke}\ \emph {et~al.}(1998)\citenamefont
  {R\"opke}, \citenamefont {Schnell}, \citenamefont {Schuck},\ and\
  \citenamefont {Nozi\`eres}}]{PhysRevLett.80.3177}%
  \BibitemOpen
  \bibfield  {author} {\bibinfo {author} {\bibfnamefont {G.}~\bibnamefont
  {R\"opke}}, \bibinfo {author} {\bibfnamefont {A.}~\bibnamefont {Schnell}},
  \bibinfo {author} {\bibfnamefont {P.}~\bibnamefont {Schuck}}, \ and\ \bibinfo
  {author} {\bibfnamefont {P.}~\bibnamefont {Nozi\`eres}},\ }\href {\doibase
  10.1103/PhysRevLett.80.3177} {\bibfield  {journal} {\bibinfo  {journal}
  {Phys. Rev. Lett.}\ }\textbf {\bibinfo {volume} {80}},\ \bibinfo {pages}
  {3177} (\bibinfo {year} {1998})}\BibitemShut {NoStop}%
\bibitem [{\citenamefont {Tajima}\ \emph {et~al.}()\citenamefont {Tajima},
  \citenamefont {Tsutsui}, \citenamefont {Doi},\ and\ \citenamefont
  {Iida}}]{tajima2020cooper}%
  \BibitemOpen
  \bibfield  {author} {\bibinfo {author} {\bibfnamefont {H.}~\bibnamefont
  {Tajima}}, \bibinfo {author} {\bibfnamefont {S.}~\bibnamefont {Tsutsui}},
  \bibinfo {author} {\bibfnamefont {T.~M.}\ \bibnamefont {Doi}}, \ and\
  \bibinfo {author} {\bibfnamefont {K.}~\bibnamefont {Iida}},\ }\href@noop {}
  {}\Eprint {http://arxiv.org/abs/2012.03627} {arXiv:2012.03627
  [cond-mat.quant-gas]} \BibitemShut {NoStop}%
\bibitem [{\citenamefont {Maćkowiak}\ and\ \citenamefont
  {Tarasewicz}(2000)}]{MACKOWIAK200025}%
  \BibitemOpen
  \bibfield  {author} {\bibinfo {author} {\bibfnamefont {J.}~\bibnamefont
  {Maćkowiak}}\ and\ \bibinfo {author} {\bibfnamefont {P.}~\bibnamefont
  {Tarasewicz}},\ }\href {\doibase
  https://doi.org/10.1016/S0921-4534(99)00546-8} {\bibfield  {journal}
  {\bibinfo  {journal} {Physica C: Superconductivity}\ }\textbf {\bibinfo
  {volume} {331}},\ \bibinfo {pages} {25} (\bibinfo {year} {2000})}\BibitemShut
  {NoStop}%
\bibitem [{\citenamefont {Sen$'$kov}\ and\ \citenamefont
  {Zelevinsky}(2011)}]{sen2011unified}%
  \BibitemOpen
  \bibfield  {author} {\bibinfo {author} {\bibfnamefont {R.}~\bibnamefont
  {Sen$'$kov}}\ and\ \bibinfo {author} {\bibfnamefont {V.}~\bibnamefont
  {Zelevinsky}},\ }\href {https://doi.org/10.1134/S1063778811090110} {\bibfield
   {journal} {\bibinfo  {journal} {Phys. At. Nucl.}\ }\textbf {\bibinfo
  {volume} {74}},\ \bibinfo {pages} {1267} (\bibinfo {year}
  {2011})}\BibitemShut {NoStop}%
\bibitem [{\citenamefont {Nozi{\`e}res}\ and\ \citenamefont
  {Saint~James}(1982)}]{art:nozieressaintjames}%
  \BibitemOpen
  \bibfield  {author} {\bibinfo {author} {\bibfnamefont {P.}~\bibnamefont
  {Nozi{\`e}res}}\ and\ \bibinfo {author} {\bibfnamefont {D.}~\bibnamefont
  {Saint~James}},\ }\href {\doibase 10.1051/jphys:019820043070113300}
  {\bibfield  {journal} {\bibinfo  {journal} {{J. Phys.}}\ }\textbf {\bibinfo
  {volume} {43}},\ \bibinfo {pages} {1133} (\bibinfo {year}
  {1982})}\BibitemShut {NoStop}%
\bibitem [{\citenamefont {Kunitski}\ \emph {et~al.}(2015)\citenamefont
  {Kunitski}, \citenamefont {Zeller}, \citenamefont {Voigtsberger},
  \citenamefont {Kalinin}, \citenamefont {Schmidt}, \citenamefont
  {Sch{\"o}ffler}, \citenamefont {Czasch}, \citenamefont {Sch{\"o}llkopf},
  \citenamefont {Grisenti}, \citenamefont {Jahnke}, \citenamefont {Blume},\
  and\ \citenamefont {D{\"o}rner}}]{Kunitski551}%
  \BibitemOpen
  \bibfield  {author} {\bibinfo {author} {\bibfnamefont {M.}~\bibnamefont
  {Kunitski}}, \bibinfo {author} {\bibfnamefont {S.}~\bibnamefont {Zeller}},
  \bibinfo {author} {\bibfnamefont {J.}~\bibnamefont {Voigtsberger}}, \bibinfo
  {author} {\bibfnamefont {A.}~\bibnamefont {Kalinin}}, \bibinfo {author}
  {\bibfnamefont {L.~P.~H.}\ \bibnamefont {Schmidt}}, \bibinfo {author}
  {\bibfnamefont {M.}~\bibnamefont {Sch{\"o}ffler}}, \bibinfo {author}
  {\bibfnamefont {A.}~\bibnamefont {Czasch}}, \bibinfo {author} {\bibfnamefont
  {W.}~\bibnamefont {Sch{\"o}llkopf}}, \bibinfo {author} {\bibfnamefont
  {R.~E.}\ \bibnamefont {Grisenti}}, \bibinfo {author} {\bibfnamefont
  {T.}~\bibnamefont {Jahnke}}, \bibinfo {author} {\bibfnamefont
  {D.}~\bibnamefont {Blume}}, \ and\ \bibinfo {author} {\bibfnamefont
  {R.}~\bibnamefont {D{\"o}rner}},\ }\href {\doibase 10.1126/science.aaa5601}
  {\bibfield  {journal} {\bibinfo  {journal} {Science}\ }\textbf {\bibinfo
  {volume} {348}},\ \bibinfo {pages} {551} (\bibinfo {year}
  {2015})}\BibitemShut {NoStop}%
\bibitem [{\citenamefont {Chevy}\ and\ \citenamefont
  {Salomon}(2016)}]{rev:chevy_salomon}%
  \BibitemOpen
  \bibfield  {author} {\bibinfo {author} {\bibfnamefont {F.}~\bibnamefont
  {Chevy}}\ and\ \bibinfo {author} {\bibfnamefont {C.}~\bibnamefont
  {Salomon}},\ }\href {\doibase 10.1088/0953-4075/49/19/192001} {\bibfield
  {journal} {\bibinfo  {journal} {J. Phys. B}\ }\textbf {\bibinfo {volume}
  {49}},\ \bibinfo {pages} {192001} (\bibinfo {year} {2016})}\BibitemShut
  {NoStop}%
\bibitem [{\citenamefont {Efimov}(1971)}]{efimov1971weakly}%
  \BibitemOpen
  \bibfield  {author} {\bibinfo {author} {\bibfnamefont {V.}~\bibnamefont
  {Efimov}},\ }\href@noop {} {\bibfield  {journal} {\bibinfo  {journal} {Sov.
  J. Nucl. Phys}\ }\textbf {\bibinfo {volume} {12}},\ \bibinfo {pages} {589}
  (\bibinfo {year} {1971})}\BibitemShut {NoStop}%
\bibitem [{\citenamefont {Efimov}(1979)}]{efimov1979low}%
  \BibitemOpen
  \bibfield  {author} {\bibinfo {author} {\bibfnamefont {V.}~\bibnamefont
  {Efimov}},\ }\href@noop {} {\bibfield  {journal} {\bibinfo  {journal} {Sov.
  J. Nucl. Phys.}\ }\textbf {\bibinfo {volume} {29}},\ \bibinfo {pages} {546}
  (\bibinfo {year} {1979})}\BibitemShut {NoStop}%
\bibitem [{\citenamefont {Nishida}(2012)}]{Nishida2012}%
  \BibitemOpen
  \bibfield  {author} {\bibinfo {author} {\bibfnamefont {Y.}~\bibnamefont
  {Nishida}},\ }\href {\doibase 10.1103/PhysRevLett.109.240401} {\bibfield
  {journal} {\bibinfo  {journal} {Phys. Rev. Lett.}\ }\textbf {\bibinfo
  {volume} {109}},\ \bibinfo {pages} {240401} (\bibinfo {year}
  {2012})}\BibitemShut {NoStop}%
\bibitem [{\citenamefont {Tajima}\ and\ \citenamefont
  {Naidon}(2019)}]{Naidon2019}%
  \BibitemOpen
  \bibfield  {author} {\bibinfo {author} {\bibfnamefont {H.}~\bibnamefont
  {Tajima}}\ and\ \bibinfo {author} {\bibfnamefont {P.}~\bibnamefont
  {Naidon}},\ }\href {\doibase 10.1088/1367-2630/ab306b} {\bibfield  {journal}
  {\bibinfo  {journal} {New J. Phys.}\ }\textbf {\bibinfo {volume} {21}},\
  \bibinfo {pages} {073051} (\bibinfo {year} {2019})}\BibitemShut {NoStop}%
\bibitem [{\citenamefont {Nishida}(2015)}]{PhysRevLett.114.115302}%
  \BibitemOpen
  \bibfield  {author} {\bibinfo {author} {\bibfnamefont {Y.}~\bibnamefont
  {Nishida}},\ }\href {\doibase 10.1103/PhysRevLett.114.115302} {\bibfield
  {journal} {\bibinfo  {journal} {Phys. Rev. Lett.}\ }\textbf {\bibinfo
  {volume} {114}},\ \bibinfo {pages} {115302} (\bibinfo {year}
  {2015})}\BibitemShut {NoStop}%
\bibitem [{\citenamefont {Pe\~na Ardila}\ and\ \citenamefont
  {Giorgini}(2015)}]{PhysRevA.92.033612}%
  \BibitemOpen
  \bibfield  {author} {\bibinfo {author} {\bibfnamefont {L.~A.}\ \bibnamefont
  {Pe\~na Ardila}}\ and\ \bibinfo {author} {\bibfnamefont {S.}~\bibnamefont
  {Giorgini}},\ }\href {\doibase 10.1103/PhysRevA.92.033612} {\bibfield
  {journal} {\bibinfo  {journal} {Phys. Rev. A}\ }\textbf {\bibinfo {volume}
  {92}},\ \bibinfo {pages} {033612} (\bibinfo {year} {2015})}\BibitemShut
  {NoStop}%
\bibitem [{\citenamefont {Sun}\ \emph {et~al.}(2017)\citenamefont {Sun},
  \citenamefont {Zhai},\ and\ \citenamefont {Cui}}]{PhysRevLett.119.013401}%
  \BibitemOpen
  \bibfield  {author} {\bibinfo {author} {\bibfnamefont {M.}~\bibnamefont
  {Sun}}, \bibinfo {author} {\bibfnamefont {H.}~\bibnamefont {Zhai}}, \ and\
  \bibinfo {author} {\bibfnamefont {X.}~\bibnamefont {Cui}},\ }\href {\doibase
  10.1103/PhysRevLett.119.013401} {\bibfield  {journal} {\bibinfo  {journal}
  {Phys. Rev. Lett.}\ }\textbf {\bibinfo {volume} {119}},\ \bibinfo {pages}
  {013401} (\bibinfo {year} {2017})}\BibitemShut {NoStop}%
\bibitem [{\citenamefont {Nygaard}\ and\ \citenamefont
  {Zinner}(2014)}]{Nygaard_2014}%
  \BibitemOpen
  \bibfield  {author} {\bibinfo {author} {\bibfnamefont {N.~G.}\ \bibnamefont
  {Nygaard}}\ and\ \bibinfo {author} {\bibfnamefont {N.~T.}\ \bibnamefont
  {Zinner}},\ }\href {\doibase 10.1088/1367-2630/16/2/023026} {\bibfield
  {journal} {\bibinfo  {journal} {New J. Phys.}\ }\textbf {\bibinfo {volume}
  {16}},\ \bibinfo {pages} {023026} (\bibinfo {year} {2014})}\BibitemShut
  {NoStop}%
\bibitem [{\citenamefont {Levinsen}\ \emph {et~al.}(2015)\citenamefont
  {Levinsen}, \citenamefont {Parish},\ and\ \citenamefont
  {Bruun}}]{PhysRevLett.115.125302}%
  \BibitemOpen
  \bibfield  {author} {\bibinfo {author} {\bibfnamefont {J.}~\bibnamefont
  {Levinsen}}, \bibinfo {author} {\bibfnamefont {M.~M.}\ \bibnamefont
  {Parish}}, \ and\ \bibinfo {author} {\bibfnamefont {G.~M.}\ \bibnamefont
  {Bruun}},\ }\href {\doibase 10.1103/PhysRevLett.115.125302} {\bibfield
  {journal} {\bibinfo  {journal} {Phys. Rev. Lett.}\ }\textbf {\bibinfo
  {volume} {115}},\ \bibinfo {pages} {125302} (\bibinfo {year}
  {2015})}\BibitemShut {NoStop}%
\bibitem [{\citenamefont {Christianen}\ \emph {et~al.}()\citenamefont
  {Christianen}, \citenamefont {Cirac},\ and\ \citenamefont
  {Schmidt}}]{christianen2021efimov}%
  \BibitemOpen
  \bibfield  {author} {\bibinfo {author} {\bibfnamefont {A.}~\bibnamefont
  {Christianen}}, \bibinfo {author} {\bibfnamefont {J.~I.}\ \bibnamefont
  {Cirac}}, \ and\ \bibinfo {author} {\bibfnamefont {R.}~\bibnamefont
  {Schmidt}},\ }\href@noop {} {}\Eprint {http://arxiv.org/abs/2108.03175}
  {arXiv:2108.03175 [cond-mat.quant-gas]} \BibitemShut {NoStop}%
\bibitem [{\citenamefont {Naidon}\ and\ \citenamefont
  {Endo}(2017)}]{rev:naidon}%
  \BibitemOpen
  \bibfield  {author} {\bibinfo {author} {\bibfnamefont {P.}~\bibnamefont
  {Naidon}}\ and\ \bibinfo {author} {\bibfnamefont {S.}~\bibnamefont {Endo}},\
  }\href {\doibase 10.1088/1361-6633/aa50e8} {\bibfield  {journal} {\bibinfo
  {journal} {Rep. Prog. Phys.}\ }\textbf {\bibinfo {volume} {80}},\ \bibinfo
  {pages} {056001} (\bibinfo {year} {2017})}\BibitemShut {NoStop}%
\bibitem [{\citenamefont {Taie}\ \emph {et~al.}(2012)\citenamefont {Taie},
  \citenamefont {Yamazaki}, \citenamefont {Sugawa},\ and\ \citenamefont
  {Takahashi}}]{Takahashi2021}%
  \BibitemOpen
  \bibfield  {author} {\bibinfo {author} {\bibfnamefont {S.}~\bibnamefont
  {Taie}}, \bibinfo {author} {\bibfnamefont {R.}~\bibnamefont {Yamazaki}},
  \bibinfo {author} {\bibfnamefont {S.}~\bibnamefont {Sugawa}}, \ and\ \bibinfo
  {author} {\bibfnamefont {Y.}~\bibnamefont {Takahashi}},\ }\href {\doibase
  10.1038/nphys2430} {\bibfield  {journal} {\bibinfo  {journal} {Nat. Phys.}\
  }\textbf {\bibinfo {volume} {8}},\ \bibinfo {pages} {825} (\bibinfo {year}
  {2012})}\BibitemShut {NoStop}%
\bibitem [{\citenamefont {Makotyn}\ \emph {et~al.}(2014)\citenamefont
  {Makotyn}, \citenamefont {Klauss}, \citenamefont {Goldberger}, \citenamefont
  {Cornell},\ and\ \citenamefont {Jin}}]{art:makotyn}%
  \BibitemOpen
  \bibfield  {author} {\bibinfo {author} {\bibfnamefont {P.}~\bibnamefont
  {Makotyn}}, \bibinfo {author} {\bibfnamefont {C.~E.}\ \bibnamefont {Klauss}},
  \bibinfo {author} {\bibfnamefont {D.~L.}\ \bibnamefont {Goldberger}},
  \bibinfo {author} {\bibfnamefont {E.}~\bibnamefont {Cornell}}, \ and\
  \bibinfo {author} {\bibfnamefont {D.~S.}\ \bibnamefont {Jin}},\ }\href
  {\doibase 10.1038/nphys2850} {\bibfield  {journal} {\bibinfo  {journal} {Nat.
  Phys.}\ }\textbf {\bibinfo {volume} {10}},\ \bibinfo {pages} {116} (\bibinfo
  {year} {2014})}\BibitemShut {NoStop}%
\bibitem [{\citenamefont {Klauss}\ \emph {et~al.}(2017)\citenamefont {Klauss},
  \citenamefont {Xie}, \citenamefont {Lopez-Abadia}, \citenamefont {D'Incao},
  \citenamefont {Hadzibabic}, \citenamefont {Jin},\ and\ \citenamefont
  {Cornell}}]{art:klauss}%
  \BibitemOpen
  \bibfield  {author} {\bibinfo {author} {\bibfnamefont {C.~E.}\ \bibnamefont
  {Klauss}}, \bibinfo {author} {\bibfnamefont {X.}~\bibnamefont {Xie}},
  \bibinfo {author} {\bibfnamefont {C.}~\bibnamefont {Lopez-Abadia}}, \bibinfo
  {author} {\bibfnamefont {J.~P.}\ \bibnamefont {D'Incao}}, \bibinfo {author}
  {\bibfnamefont {Z.}~\bibnamefont {Hadzibabic}}, \bibinfo {author}
  {\bibfnamefont {D.~S.}\ \bibnamefont {Jin}}, \ and\ \bibinfo {author}
  {\bibfnamefont {E.~A.}\ \bibnamefont {Cornell}},\ }\href {\doibase
  10.1103/PhysRevLett.119.143401} {\bibfield  {journal} {\bibinfo  {journal}
  {Phys. Rev. Lett.}\ }\textbf {\bibinfo {volume} {119}},\ \bibinfo {pages}
  {143401} (\bibinfo {year} {2017})}\BibitemShut {NoStop}%
\bibitem [{\citenamefont {Eigen}\ \emph {et~al.}(2017)\citenamefont {Eigen},
  \citenamefont {Glidden}, \citenamefont {Lopes}, \citenamefont {Navon},
  \citenamefont {Hadzibabic},\ and\ \citenamefont {Smith}}]{art:eigen17}%
  \BibitemOpen
  \bibfield  {author} {\bibinfo {author} {\bibfnamefont {C.}~\bibnamefont
  {Eigen}}, \bibinfo {author} {\bibfnamefont {J.~A.~P.}\ \bibnamefont
  {Glidden}}, \bibinfo {author} {\bibfnamefont {R.}~\bibnamefont {Lopes}},
  \bibinfo {author} {\bibfnamefont {N.}~\bibnamefont {Navon}}, \bibinfo
  {author} {\bibfnamefont {Z.}~\bibnamefont {Hadzibabic}}, \ and\ \bibinfo
  {author} {\bibfnamefont {R.~P.}\ \bibnamefont {Smith}},\ }\href {\doibase
  10.1103/PhysRevLett.119.250404} {\bibfield  {journal} {\bibinfo  {journal}
  {Phys. Rev. Lett.}\ }\textbf {\bibinfo {volume} {119}},\ \bibinfo {pages}
  {250404} (\bibinfo {year} {2017})}\BibitemShut {NoStop}%
\bibitem [{\citenamefont {Eigen}\ \emph {et~al.}(2018)\citenamefont {Eigen},
  \citenamefont {Glidden}, \citenamefont {Lopes}, \citenamefont {Cornell},
  \citenamefont {Smith},\ and\ \citenamefont {Hadzibabic}}]{art:eigen18}%
  \BibitemOpen
  \bibfield  {author} {\bibinfo {author} {\bibfnamefont {C.}~\bibnamefont
  {Eigen}}, \bibinfo {author} {\bibfnamefont {J.~A.~P.}\ \bibnamefont
  {Glidden}}, \bibinfo {author} {\bibfnamefont {R.}~\bibnamefont {Lopes}},
  \bibinfo {author} {\bibfnamefont {E.~A.}\ \bibnamefont {Cornell}}, \bibinfo
  {author} {\bibfnamefont {R.~P.}\ \bibnamefont {Smith}}, \ and\ \bibinfo
  {author} {\bibfnamefont {Z.}~\bibnamefont {Hadzibabic}},\ }\href {\doibase
  10.1038/s41586-018-0674-1} {\bibfield  {journal} {\bibinfo  {journal} {Nature
  (London)}\ }\textbf {\bibinfo {volume} {563}},\ \bibinfo {pages} {221}
  (\bibinfo {year} {2018})}\BibitemShut {NoStop}%
\bibitem [{\citenamefont {Regal}\ \emph {et~al.}(2004)\citenamefont {Regal},
  \citenamefont {Greiner},\ and\ \citenamefont {Jin}}]{art:regal2004}%
  \BibitemOpen
  \bibfield  {author} {\bibinfo {author} {\bibfnamefont {C.~A.}\ \bibnamefont
  {Regal}}, \bibinfo {author} {\bibfnamefont {M.}~\bibnamefont {Greiner}}, \
  and\ \bibinfo {author} {\bibfnamefont {D.~S.}\ \bibnamefont {Jin}},\ }\href
  {\doibase 10.1103/PhysRevLett.92.040403} {\bibfield  {journal} {\bibinfo
  {journal} {Phys. Rev. Lett.}\ }\textbf {\bibinfo {volume} {92}},\ \bibinfo
  {pages} {040403} (\bibinfo {year} {2004})}\BibitemShut {NoStop}%
\bibitem [{\citenamefont {Zwerger}(2012)}]{book:Zwerger}%
  \BibitemOpen
  \bibfield  {author} {\bibinfo {author} {\bibfnamefont {W.~E.}\ \bibnamefont
  {Zwerger}},\ }\href {\doibase https://doi.org/10.1007/978-3-642-21978-8}
  {\emph {\bibinfo {title} {The BCS-BEC Crossover and the Unitary Fermi Gas
  (Vol. 863 of Lecture Notes in Physics)}}},\ edited by\ \bibinfo {editor}
  {\bibfnamefont {W.}~\bibnamefont {Zwerger}}\ (\bibinfo  {publisher}
  {Springer-Verlag},\ \bibinfo {address} {Berlin},\ \bibinfo {year}
  {2012})\BibitemShut {NoStop}%
\bibitem [{\citenamefont {Ho}(2004)}]{PhysRevLett.92.090402}%
  \BibitemOpen
  \bibfield  {author} {\bibinfo {author} {\bibfnamefont {T.-L.}\ \bibnamefont
  {Ho}},\ }\href {\doibase 10.1103/PhysRevLett.92.090402} {\bibfield  {journal}
  {\bibinfo  {journal} {Phys. Rev. Lett.}\ }\textbf {\bibinfo {volume} {92}},\
  \bibinfo {pages} {090402} (\bibinfo {year} {2004})}\BibitemShut {NoStop}%
\bibitem [{\citenamefont {Van~Regemortel}\ \emph {et~al.}(2018)\citenamefont
  {Van~Regemortel}, \citenamefont {Kurkjian}, \citenamefont {Wouters},\ and\
  \citenamefont {Carusotto}}]{art:mathias}%
  \BibitemOpen
  \bibfield  {author} {\bibinfo {author} {\bibfnamefont {M.}~\bibnamefont
  {Van~Regemortel}}, \bibinfo {author} {\bibfnamefont {H.}~\bibnamefont
  {Kurkjian}}, \bibinfo {author} {\bibfnamefont {M.}~\bibnamefont {Wouters}}, \
  and\ \bibinfo {author} {\bibfnamefont {I.}~\bibnamefont {Carusotto}},\ }\href
  {\doibase 10.1103/PhysRevA.98.053612} {\bibfield  {journal} {\bibinfo
  {journal} {Phys. Rev. A}\ }\textbf {\bibinfo {volume} {98}},\ \bibinfo
  {pages} {053612} (\bibinfo {year} {2018})}\BibitemShut {NoStop}%
\bibitem [{\citenamefont {Sykes}\ \emph {et~al.}(2014)\citenamefont {Sykes},
  \citenamefont {Corson}, \citenamefont {D'Incao}, \citenamefont {Koller},
  \citenamefont {Greene}, \citenamefont {Rey}, \citenamefont {Hazzard},\ and\
  \citenamefont {Bohn}}]{art:sykes}%
  \BibitemOpen
  \bibfield  {author} {\bibinfo {author} {\bibfnamefont {A.~G.}\ \bibnamefont
  {Sykes}}, \bibinfo {author} {\bibfnamefont {J.~P.}\ \bibnamefont {Corson}},
  \bibinfo {author} {\bibfnamefont {J.~P.}\ \bibnamefont {D'Incao}}, \bibinfo
  {author} {\bibfnamefont {A.~P.}\ \bibnamefont {Koller}}, \bibinfo {author}
  {\bibfnamefont {C.~H.}\ \bibnamefont {Greene}}, \bibinfo {author}
  {\bibfnamefont {A.~M.}\ \bibnamefont {Rey}}, \bibinfo {author} {\bibfnamefont
  {K.~R.~A.}\ \bibnamefont {Hazzard}}, \ and\ \bibinfo {author} {\bibfnamefont
  {J.~L.}\ \bibnamefont {Bohn}},\ }\href {\doibase 10.1103/PhysRevA.89.021601}
  {\bibfield  {journal} {\bibinfo  {journal} {Phys. Rev. A}\ }\textbf {\bibinfo
  {volume} {89}},\ \bibinfo {pages} {021601(R)} (\bibinfo {year}
  {2014})}\BibitemShut {NoStop}%
\bibitem [{\citenamefont {Corson}\ and\ \citenamefont
  {Bohn}(2015)}]{art:corson_bohn}%
  \BibitemOpen
  \bibfield  {author} {\bibinfo {author} {\bibfnamefont {J.~P.}\ \bibnamefont
  {Corson}}\ and\ \bibinfo {author} {\bibfnamefont {J.~L.}\ \bibnamefont
  {Bohn}},\ }\href {\doibase 10.1103/PhysRevA.91.013616} {\bibfield  {journal}
  {\bibinfo  {journal} {Phys. Rev. A}\ }\textbf {\bibinfo {volume} {91}},\
  \bibinfo {pages} {013616} (\bibinfo {year} {2015})}\BibitemShut {NoStop}%
\bibitem [{\citenamefont {Gao}\ \emph {et~al.}(2020)\citenamefont {Gao},
  \citenamefont {Sun}, \citenamefont {Zhang},\ and\ \citenamefont
  {Zhai}}]{PhysRevLett.124.040403}%
  \BibitemOpen
  \bibfield  {author} {\bibinfo {author} {\bibfnamefont {C.}~\bibnamefont
  {Gao}}, \bibinfo {author} {\bibfnamefont {M.}~\bibnamefont {Sun}}, \bibinfo
  {author} {\bibfnamefont {P.}~\bibnamefont {Zhang}}, \ and\ \bibinfo {author}
  {\bibfnamefont {H.}~\bibnamefont {Zhai}},\ }\href {\doibase
  10.1103/PhysRevLett.124.040403} {\bibfield  {journal} {\bibinfo  {journal}
  {Phys. Rev. Lett.}\ }\textbf {\bibinfo {volume} {124}},\ \bibinfo {pages}
  {040403} (\bibinfo {year} {2020})}\BibitemShut {NoStop}%
\bibitem [{\citenamefont {Mu\~noz de~las Heras}\ \emph
  {et~al.}(2019)\citenamefont {Mu\~noz de~las Heras}, \citenamefont {Parish},\
  and\ \citenamefont {Marchetti}}]{art:munozdelasheras}%
  \BibitemOpen
  \bibfield  {author} {\bibinfo {author} {\bibfnamefont {A.}~\bibnamefont
  {Mu\~noz de~las Heras}}, \bibinfo {author} {\bibfnamefont {M.~M.}\
  \bibnamefont {Parish}}, \ and\ \bibinfo {author} {\bibfnamefont {F.~M.}\
  \bibnamefont {Marchetti}},\ }\href {\doibase 10.1103/PhysRevA.99.023623}
  {\bibfield  {journal} {\bibinfo  {journal} {Phys. Rev. A}\ }\textbf {\bibinfo
  {volume} {99}},\ \bibinfo {pages} {023623} (\bibinfo {year}
  {2019})}\BibitemShut {NoStop}%
\bibitem [{\citenamefont {Musolino}\ \emph {et~al.}(2019)\citenamefont
  {Musolino}, \citenamefont {Colussi},\ and\ \citenamefont
  {Kokkelmans}}]{art:silviack}%
  \BibitemOpen
  \bibfield  {author} {\bibinfo {author} {\bibfnamefont {S.}~\bibnamefont
  {Musolino}}, \bibinfo {author} {\bibfnamefont {V.~E.}\ \bibnamefont
  {Colussi}}, \ and\ \bibinfo {author} {\bibfnamefont {S.~J. J. M.~F.}\
  \bibnamefont {Kokkelmans}},\ }\href {\doibase 10.1103/PhysRevA.100.013612}
  {\bibfield  {journal} {\bibinfo  {journal} {Phys. Rev. A}\ }\textbf {\bibinfo
  {volume} {100}},\ \bibinfo {pages} {013612} (\bibinfo {year}
  {2019})}\BibitemShut {NoStop}%
\bibitem [{\citenamefont {Colussi}\ \emph
  {et~al.}(2018{\natexlab{a}})\citenamefont {Colussi}, \citenamefont
  {Musolino},\ and\ \citenamefont {Kokkelmans}}]{art:colussimk}%
  \BibitemOpen
  \bibfield  {author} {\bibinfo {author} {\bibfnamefont {V.~E.}\ \bibnamefont
  {Colussi}}, \bibinfo {author} {\bibfnamefont {S.}~\bibnamefont {Musolino}}, \
  and\ \bibinfo {author} {\bibfnamefont {S.~J. J. M.~F.}\ \bibnamefont
  {Kokkelmans}},\ }\href {\doibase 10.1103/PhysRevA.98.051601} {\bibfield
  {journal} {\bibinfo  {journal} {Phys. Rev. A}\ }\textbf {\bibinfo {volume}
  {98}},\ \bibinfo {pages} {051601(R)} (\bibinfo {year}
  {2018}{\natexlab{a}})}\BibitemShut {NoStop}%
\bibitem [{\citenamefont {D'Incao}\ \emph {et~al.}(2018)\citenamefont
  {D'Incao}, \citenamefont {Wang},\ and\ \citenamefont
  {Colussi}}]{art:dincao_efimov}%
  \BibitemOpen
  \bibfield  {author} {\bibinfo {author} {\bibfnamefont {J.~P.}\ \bibnamefont
  {D'Incao}}, \bibinfo {author} {\bibfnamefont {J.}~\bibnamefont {Wang}}, \
  and\ \bibinfo {author} {\bibfnamefont {V.~E.}\ \bibnamefont {Colussi}},\
  }\href {\doibase 10.1103/PhysRevLett.121.023401} {\bibfield  {journal}
  {\bibinfo  {journal} {Phys. Rev. Lett.}\ }\textbf {\bibinfo {volume} {121}},\
  \bibinfo {pages} {023401} (\bibinfo {year} {2018})}\BibitemShut {NoStop}%
\bibitem [{\citenamefont {Colussi}\ \emph
  {et~al.}(2018{\natexlab{b}})\citenamefont {Colussi}, \citenamefont {Corson},\
  and\ \citenamefont {D'Incao}}]{art:3bc_victor}%
  \BibitemOpen
  \bibfield  {author} {\bibinfo {author} {\bibfnamefont {V.~E.}\ \bibnamefont
  {Colussi}}, \bibinfo {author} {\bibfnamefont {J.~P.}\ \bibnamefont {Corson}},
  \ and\ \bibinfo {author} {\bibfnamefont {J.~P.}\ \bibnamefont {D'Incao}},\
  }\href {\doibase 10.1103/PhysRevLett.120.100401} {\bibfield  {journal}
  {\bibinfo  {journal} {Phys. Rev. Lett.}\ }\textbf {\bibinfo {volume} {120}},\
  \bibinfo {pages} {100401} (\bibinfo {year} {2018}{\natexlab{b}})}\BibitemShut
  {NoStop}%
\bibitem [{\citenamefont {Colussi}\ \emph {et~al.}(2019)\citenamefont
  {Colussi}, \citenamefont {van Zwol}, \citenamefont {D'Incao},\ and\
  \citenamefont {Kokkelmans}}]{art:colussi2019}%
  \BibitemOpen
  \bibfield  {author} {\bibinfo {author} {\bibfnamefont {V.~E.}\ \bibnamefont
  {Colussi}}, \bibinfo {author} {\bibfnamefont {B.~E.}\ \bibnamefont {van
  Zwol}}, \bibinfo {author} {\bibfnamefont {J.~P.}\ \bibnamefont {D'Incao}}, \
  and\ \bibinfo {author} {\bibfnamefont {S.~J. J. M.~F.}\ \bibnamefont
  {Kokkelmans}},\ }\href {\doibase 10.1103/PhysRevA.99.043604} {\bibfield
  {journal} {\bibinfo  {journal} {Phys. Rev. A}\ }\textbf {\bibinfo {volume}
  {99}},\ \bibinfo {pages} {043604} (\bibinfo {year} {2019})}\BibitemShut
  {NoStop}%
\bibitem [{\citenamefont {Bodineau}\ \emph {et~al.}(2008)\citenamefont
  {Bodineau}, \citenamefont {Derrida}, \citenamefont {Lecomte},\ and\
  \citenamefont {van Wijland}}]{vanWijland2008}%
  \BibitemOpen
  \bibfield  {author} {\bibinfo {author} {\bibfnamefont {T.}~\bibnamefont
  {Bodineau}}, \bibinfo {author} {\bibfnamefont {B.}~\bibnamefont {Derrida}},
  \bibinfo {author} {\bibfnamefont {V.}~\bibnamefont {Lecomte}}, \ and\
  \bibinfo {author} {\bibfnamefont {F.}~\bibnamefont {van Wijland}},\ }\href
  {\doibase 10.1007/s10955-008-9647-3} {\bibfield  {journal} {\bibinfo
  {journal} {J. Stat. Phys.}\ }\textbf {\bibinfo {volume} {133}},\ \bibinfo
  {pages} {1013} (\bibinfo {year} {2008})}\BibitemShut {NoStop}%
\bibitem [{\citenamefont {Kira}\ and\ \citenamefont
  {Koch}(2011)}]{kira2011semiconductor}%
  \BibitemOpen
  \bibfield  {author} {\bibinfo {author} {\bibfnamefont {M.}~\bibnamefont
  {Kira}}\ and\ \bibinfo {author} {\bibfnamefont {S.~W.}\ \bibnamefont
  {Koch}},\ }\href@noop {} {\emph {\bibinfo {title} {Semiconductor Quantum
  Optics}}}\ (\bibinfo  {publisher} {Cambridge University Press},\ \bibinfo
  {year} {2011})\BibitemShut {NoStop}%
\bibitem [{\citenamefont {Tohyama}\ and\ \citenamefont
  {Schuck}(2015)}]{Schuck2015}%
  \BibitemOpen
  \bibfield  {author} {\bibinfo {author} {\bibfnamefont {M.}~\bibnamefont
  {Tohyama}}\ and\ \bibinfo {author} {\bibfnamefont {P.}~\bibnamefont
  {Schuck}},\ }\href {\doibase 10.1103/PhysRevC.91.034316} {\bibfield
  {journal} {\bibinfo  {journal} {Phys. Rev. C}\ }\textbf {\bibinfo {volume}
  {91}},\ \bibinfo {pages} {034316} (\bibinfo {year} {2015})}\BibitemShut
  {NoStop}%
\bibitem [{\citenamefont {Colussi}\ \emph {et~al.}(2020)\citenamefont
  {Colussi}, \citenamefont {Kurkjian}, \citenamefont {Van~Regemortel},
  \citenamefont {Musolino}, \citenamefont {van~de Kraats}, \citenamefont
  {Wouters},\ and\ \citenamefont {Kokkelmans}}]{art:cumulant2020}%
  \BibitemOpen
  \bibfield  {author} {\bibinfo {author} {\bibfnamefont {V.~E.}\ \bibnamefont
  {Colussi}}, \bibinfo {author} {\bibfnamefont {H.}~\bibnamefont {Kurkjian}},
  \bibinfo {author} {\bibfnamefont {M.}~\bibnamefont {Van~Regemortel}},
  \bibinfo {author} {\bibfnamefont {S.}~\bibnamefont {Musolino}}, \bibinfo
  {author} {\bibfnamefont {J.}~\bibnamefont {van~de Kraats}}, \bibinfo {author}
  {\bibfnamefont {M.}~\bibnamefont {Wouters}}, \ and\ \bibinfo {author}
  {\bibfnamefont {S.~J. J. M.~F.}\ \bibnamefont {Kokkelmans}},\ }\href
  {\doibase 10.1103/PhysRevA.102.063314} {\bibfield  {journal} {\bibinfo
  {journal} {Phys. Rev. A}\ }\textbf {\bibinfo {volume} {102}},\ \bibinfo
  {pages} {063314} (\bibinfo {year} {2020})}\BibitemShut {NoStop}%
\bibitem [{\citenamefont {Kira}(2015{\natexlab{a}})}]{KIRA2015185}%
  \BibitemOpen
  \bibfield  {author} {\bibinfo {author} {\bibfnamefont {M.}~\bibnamefont
  {Kira}},\ }\href {\doibase https://doi.org/10.1016/j.aop.2015.02.030}
  {\bibfield  {journal} {\bibinfo  {journal} {Ann. Phys. (Amsterdam)}\ }\textbf
  {\bibinfo {volume} {356}},\ \bibinfo {pages} {185 } (\bibinfo {year}
  {2015}{\natexlab{a}})}\BibitemShut {NoStop}%
\bibitem [{\citenamefont {Blaizot}\ and\ \citenamefont
  {Ripka}(1986)}]{book:blaizot}%
  \BibitemOpen
  \bibfield  {author} {\bibinfo {author} {\bibfnamefont {J.~P.}\ \bibnamefont
  {Blaizot}}\ and\ \bibinfo {author} {\bibfnamefont {G.}~\bibnamefont
  {Ripka}},\ }\href@noop {} {\emph {\bibinfo {title} {Quantum Theory of Finite
  Systems}}}\ (\bibinfo  {publisher} {The MIT Press},\ \bibinfo {address}
  {Cambridge, Massachusetts},\ \bibinfo {year} {1986})\BibitemShut {NoStop}%
\bibitem [{\citenamefont {Chin}\ \emph {et~al.}(2010)\citenamefont {Chin},
  \citenamefont {Grimm}, \citenamefont {Julienne},\ and\ \citenamefont
  {Tiesinga}}]{art:chin}%
  \BibitemOpen
  \bibfield  {author} {\bibinfo {author} {\bibfnamefont {C.}~\bibnamefont
  {Chin}}, \bibinfo {author} {\bibfnamefont {R.}~\bibnamefont {Grimm}},
  \bibinfo {author} {\bibfnamefont {P.}~\bibnamefont {Julienne}}, \ and\
  \bibinfo {author} {\bibfnamefont {E.}~\bibnamefont {Tiesinga}},\ }\href
  {\doibase 10.1103/RevModPhys.82.1225} {\bibfield  {journal} {\bibinfo
  {journal} {Rev. Mod. Phys.}\ }\textbf {\bibinfo {volume} {82}},\ \bibinfo
  {pages} {1225} (\bibinfo {year} {2010})}\BibitemShut {NoStop}%
\bibitem [{\citenamefont {Flambaum}\ \emph {et~al.}(1999)\citenamefont
  {Flambaum}, \citenamefont {Gribakin},\ and\ \citenamefont
  {Harabati}}]{PhysRevA.59.1998}%
  \BibitemOpen
  \bibfield  {author} {\bibinfo {author} {\bibfnamefont {V.~V.}\ \bibnamefont
  {Flambaum}}, \bibinfo {author} {\bibfnamefont {G.~F.}\ \bibnamefont
  {Gribakin}}, \ and\ \bibinfo {author} {\bibfnamefont {C.}~\bibnamefont
  {Harabati}},\ }\href {\doibase 10.1103/PhysRevA.59.1998} {\bibfield
  {journal} {\bibinfo  {journal} {Phys. Rev. A}\ }\textbf {\bibinfo {volume}
  {59}},\ \bibinfo {pages} {1998} (\bibinfo {year} {1999})}\BibitemShut
  {NoStop}%
\bibitem [{\citenamefont {K\"ohler}\ \emph {et~al.}(2003)\citenamefont
  {K\"ohler}, \citenamefont {Gasenzer},\ and\ \citenamefont
  {Burnett}}]{PhysRevA.67.013601}%
  \BibitemOpen
  \bibfield  {author} {\bibinfo {author} {\bibfnamefont {T.}~\bibnamefont
  {K\"ohler}}, \bibinfo {author} {\bibfnamefont {T.}~\bibnamefont {Gasenzer}},
  \ and\ \bibinfo {author} {\bibfnamefont {K.}~\bibnamefont {Burnett}},\ }\href
  {\doibase 10.1103/PhysRevA.67.013601} {\bibfield  {journal} {\bibinfo
  {journal} {Phys. Rev. A}\ }\textbf {\bibinfo {volume} {67}},\ \bibinfo
  {pages} {013601} (\bibinfo {year} {2003})}\BibitemShut {NoStop}%
\bibitem [{SM()}]{SM}%
  \BibitemOpen
  \href@noop {} {}\bibinfo {note} {See Supplemental Material, which includes
  Refs.~\cite{faddeev2013quantum,glockle1983,taylor2006scattering,PhysRev.47.903,skorniakov1957three,rev:greene,PhysRevLett.108.263001,PhysRevA.90.022106,Salasnich2005,PhysRevA.88.063623,art:werner_fermions,art:leggett,Tanguy2003,Combescot_2001,Fletcher377,PhysRevLett.125.110404},
  for additional details of our calculations.}\BibitemShut {Stop}%
\bibitem [{\citenamefont {Fricke}(1996)}]{FRICKE1996479}%
  \BibitemOpen
  \bibfield  {author} {\bibinfo {author} {\bibfnamefont {J.}~\bibnamefont
  {Fricke}},\ }\href {\doibase https://doi.org/10.1006/aphy.1996.0142}
  {\bibfield  {journal} {\bibinfo  {journal} {Ann. Phys. (N.Y.)}\ }\textbf
  {\bibinfo {volume} {252}},\ \bibinfo {pages} {479} (\bibinfo {year}
  {1996})}\BibitemShut {NoStop}%
\bibitem [{\citenamefont {K\"ohler}\ and\ \citenamefont
  {Burnett}(2002)}]{PhysRevA.65.033601}%
  \BibitemOpen
  \bibfield  {author} {\bibinfo {author} {\bibfnamefont {T.}~\bibnamefont
  {K\"ohler}}\ and\ \bibinfo {author} {\bibfnamefont {K.}~\bibnamefont
  {Burnett}},\ }\href {\doibase 10.1103/PhysRevA.65.033601} {\bibfield
  {journal} {\bibinfo  {journal} {Phys. Rev. A}\ }\textbf {\bibinfo {volume}
  {65}},\ \bibinfo {pages} {033601} (\bibinfo {year} {2002})}\BibitemShut
  {NoStop}%
\bibitem [{\citenamefont {Kira}(2014)}]{kira2014excitation}%
  \BibitemOpen
  \bibfield  {author} {\bibinfo {author} {\bibfnamefont {M.}~\bibnamefont
  {Kira}},\ }\href {https://doi.org/10.1016/j.aop.2014.08.022} {\bibfield
  {journal} {\bibinfo  {journal} {Ann. Phys. (Amsterdam)}\ }\textbf {\bibinfo
  {volume} {351}},\ \bibinfo {pages} {200} (\bibinfo {year}
  {2014})}\BibitemShut {NoStop}%
\bibitem [{\citenamefont {Kira}(2015{\natexlab{b}})}]{kirancomm}%
  \BibitemOpen
  \bibfield  {author} {\bibinfo {author} {\bibfnamefont {M.}~\bibnamefont
  {Kira}},\ }\href {https://doi.org/10.1038/ncomms7624} {\bibfield  {journal}
  {\bibinfo  {journal} {Nat. Commun.}\ }\textbf {\bibinfo {volume} {6}},\
  \bibinfo {pages} {6624} (\bibinfo {year} {2015}{\natexlab{b}})}\BibitemShut
  {NoStop}%
\bibitem [{\citenamefont {K\"ohler}(2002)}]{PhysRevLett.89.210404}%
  \BibitemOpen
  \bibfield  {author} {\bibinfo {author} {\bibfnamefont {T.}~\bibnamefont
  {K\"ohler}},\ }\href {\doibase 10.1103/PhysRevLett.89.210404} {\bibfield
  {journal} {\bibinfo  {journal} {Phys. Rev. Lett.}\ }\textbf {\bibinfo
  {volume} {89}},\ \bibinfo {pages} {210404} (\bibinfo {year}
  {2002})}\BibitemShut {NoStop}%
\bibitem [{\citenamefont {Penrose}(1951)}]{art:penrose_51}%
  \BibitemOpen
  \bibfield  {author} {\bibinfo {author} {\bibfnamefont {O.}~\bibnamefont
  {Penrose}},\ }\href {\doibase 10.1080/14786445108560954} {\bibfield
  {journal} {\bibinfo  {journal} {London, Edinburgh, Dublin Philos. Mag. J.
  Sci.}\ }\textbf {\bibinfo {volume} {42}},\ \bibinfo {pages} {1373} (\bibinfo
  {year} {1951})}\BibitemShut {NoStop}%
\bibitem [{\citenamefont {Penrose}\ and\ \citenamefont
  {Onsager}(1956)}]{art:penrose_onsager}%
  \BibitemOpen
  \bibfield  {author} {\bibinfo {author} {\bibfnamefont {O.}~\bibnamefont
  {Penrose}}\ and\ \bibinfo {author} {\bibfnamefont {L.}~\bibnamefont
  {Onsager}},\ }\href {\doibase 10.1103/PhysRev.104.576} {\bibfield  {journal}
  {\bibinfo  {journal} {Phys. Rev.}\ }\textbf {\bibinfo {volume} {104}},\
  \bibinfo {pages} {576} (\bibinfo {year} {1956})}\BibitemShut {NoStop}%
\bibitem [{\citenamefont {Yang}(1962)}]{art:yang}%
  \BibitemOpen
  \bibfield  {author} {\bibinfo {author} {\bibfnamefont {C.~N.}\ \bibnamefont
  {Yang}},\ }\href {\doibase 10.1103/RevModPhys.34.694} {\bibfield  {journal}
  {\bibinfo  {journal} {Rev. Mod. Phys.}\ }\textbf {\bibinfo {volume} {34}},\
  \bibinfo {pages} {694} (\bibinfo {year} {1962})}\BibitemShut {NoStop}%
\bibitem [{\citenamefont {Pong}\ and\ \citenamefont
  {Law}(2007)}]{PhysRevA.75.043613}%
  \BibitemOpen
  \bibfield  {author} {\bibinfo {author} {\bibfnamefont {Y.~H.}\ \bibnamefont
  {Pong}}\ and\ \bibinfo {author} {\bibfnamefont {C.~K.}\ \bibnamefont {Law}},\
  }\href {\doibase 10.1103/PhysRevA.75.043613} {\bibfield  {journal} {\bibinfo
  {journal} {Phys. Rev. A}\ }\textbf {\bibinfo {volume} {75}},\ \bibinfo
  {pages} {043613} (\bibinfo {year} {2007})}\BibitemShut {NoStop}%
\bibitem [{\citenamefont {Bedaque}\ \emph {et~al.}(1999)\citenamefont
  {Bedaque}, \citenamefont {Hammer},\ and\ \citenamefont {van
  Kolck}}]{PhysRevLett.82.463}%
  \BibitemOpen
  \bibfield  {author} {\bibinfo {author} {\bibfnamefont {P.~F.}\ \bibnamefont
  {Bedaque}}, \bibinfo {author} {\bibfnamefont {H.-W.}\ \bibnamefont {Hammer}},
  \ and\ \bibinfo {author} {\bibfnamefont {U.}~\bibnamefont {van Kolck}},\
  }\href {\doibase 10.1103/PhysRevLett.82.463} {\bibfield  {journal} {\bibinfo
  {journal} {Phys. Rev. Lett.}\ }\textbf {\bibinfo {volume} {82}},\ \bibinfo
  {pages} {463} (\bibinfo {year} {1999})}\BibitemShut {NoStop}%
\bibitem [{\citenamefont {Braaten}\ \emph {et~al.}(2002)\citenamefont
  {Braaten}, \citenamefont {Hammer},\ and\ \citenamefont
  {Mehen}}]{PhysRevLett.88.040401}%
  \BibitemOpen
  \bibfield  {author} {\bibinfo {author} {\bibfnamefont {E.}~\bibnamefont
  {Braaten}}, \bibinfo {author} {\bibfnamefont {H.-W.}\ \bibnamefont {Hammer}},
  \ and\ \bibinfo {author} {\bibfnamefont {T.}~\bibnamefont {Mehen}},\ }\href
  {\doibase 10.1103/PhysRevLett.88.040401} {\bibfield  {journal} {\bibinfo
  {journal} {Phys. Rev. Lett.}\ }\textbf {\bibinfo {volume} {88}},\ \bibinfo
  {pages} {040401} (\bibinfo {year} {2002})}\BibitemShut {NoStop}%
\bibitem [{\citenamefont {Donley}\ \emph {et~al.}(2002)\citenamefont {Donley},
  \citenamefont {Claussen}, \citenamefont {Thompson},\ and\ \citenamefont
  {Wieman}}]{art:donley_ram}%
  \BibitemOpen
  \bibfield  {author} {\bibinfo {author} {\bibfnamefont {E.~A.}\ \bibnamefont
  {Donley}}, \bibinfo {author} {\bibfnamefont {N.~R.}\ \bibnamefont
  {Claussen}}, \bibinfo {author} {\bibfnamefont {S.~T.}\ \bibnamefont
  {Thompson}}, \ and\ \bibinfo {author} {\bibfnamefont {C.~E.}\ \bibnamefont
  {Wieman}},\ }\href {https://doi.org/10.1038/417529a} {\bibfield  {journal}
  {\bibinfo  {journal} {Nature (London)}\ }\textbf {\bibinfo {volume} {417}},\
  \bibinfo {pages} {529} (\bibinfo {year} {2002})}\BibitemShut {NoStop}%
\bibitem [{\citenamefont {Braaten}\ \emph {et~al.}(2003)\citenamefont
  {Braaten}, \citenamefont {Hammer},\ and\ \citenamefont
  {Kusunoki}}]{art:braaten_2003}%
  \BibitemOpen
  \bibfield  {author} {\bibinfo {author} {\bibfnamefont {E.}~\bibnamefont
  {Braaten}}, \bibinfo {author} {\bibfnamefont {H.-W.}\ \bibnamefont {Hammer}},
  \ and\ \bibinfo {author} {\bibfnamefont {M.}~\bibnamefont {Kusunoki}},\
  }\href {\doibase 10.1103/PhysRevLett.90.170402} {\bibfield  {journal}
  {\bibinfo  {journal} {Phys. Rev. Lett.}\ }\textbf {\bibinfo {volume} {90}},\
  \bibinfo {pages} {170402} (\bibinfo {year} {2003})}\BibitemShut {NoStop}%
\bibitem [{\citenamefont {Werner}\ and\ \citenamefont
  {Castin}(2012{\natexlab{a}})}]{art:werner_bosons}%
  \BibitemOpen
  \bibfield  {author} {\bibinfo {author} {\bibfnamefont {F.}~\bibnamefont
  {Werner}}\ and\ \bibinfo {author} {\bibfnamefont {Y.}~\bibnamefont
  {Castin}},\ }\href {\doibase 10.1103/PhysRevA.86.053633} {\bibfield
  {journal} {\bibinfo  {journal} {Phys. Rev. A}\ }\textbf {\bibinfo {volume}
  {86}},\ \bibinfo {pages} {053633} (\bibinfo {year}
  {2012}{\natexlab{a}})}\BibitemShut {NoStop}%
\bibitem [{Note1()}]{Note1}%
  \BibitemOpen
  \bibinfo {note} {The normalization constant is $\protect \langle \phi _s|\phi
  _s\protect \rangle \equiv \DOTSI \intop \ilimits@ d{\protect \bf \Omega
  }|\phi _{s}({\protect \bf \Omega })|^2=-\protect \frac {12\pi }{s}\protect
  \qopname \relax o{sin}\left (\protect \frac {s^\ast \pi }{2}\right )$ $\left
  [\protect \qopname \relax o{cos}\left (\protect \frac {s\pi }{2}\right
  )-\protect \frac {s\pi }{2}\protect \qopname \relax o{sin}\left (\protect
  \frac {s\pi }{2}\right )-\protect \frac {4\pi }{3\protect \sqrt {3}}\protect
  \qopname \relax o{cos}\left (\protect \frac {s\pi }{6}\right )\right
  ]$.}\BibitemShut {Stop}%
\bibitem [{\citenamefont {Braaten}\ \emph {et~al.}(2011)\citenamefont
  {Braaten}, \citenamefont {Kang},\ and\ \citenamefont
  {Platter}}]{art:braaten_c3_first}%
  \BibitemOpen
  \bibfield  {author} {\bibinfo {author} {\bibfnamefont {E.}~\bibnamefont
  {Braaten}}, \bibinfo {author} {\bibfnamefont {D.}~\bibnamefont {Kang}}, \
  and\ \bibinfo {author} {\bibfnamefont {L.}~\bibnamefont {Platter}},\ }\href
  {\doibase 10.1103/PhysRevLett.106.153005} {\bibfield  {journal} {\bibinfo
  {journal} {Phys. Rev. Lett.}\ }\textbf {\bibinfo {volume} {106}},\ \bibinfo
  {pages} {153005} (\bibinfo {year} {2011})}\BibitemShut {NoStop}%
\bibitem [{\citenamefont {Parish}\ \emph {et~al.}(2005)\citenamefont {Parish},
  \citenamefont {Mihaila}, \citenamefont {Timmermans}, \citenamefont
  {Blagoev},\ and\ \citenamefont {Littlewood}}]{art:parish}%
  \BibitemOpen
  \bibfield  {author} {\bibinfo {author} {\bibfnamefont {M.~M.}\ \bibnamefont
  {Parish}}, \bibinfo {author} {\bibfnamefont {B.}~\bibnamefont {Mihaila}},
  \bibinfo {author} {\bibfnamefont {E.~M.}\ \bibnamefont {Timmermans}},
  \bibinfo {author} {\bibfnamefont {K.~B.}\ \bibnamefont {Blagoev}}, \ and\
  \bibinfo {author} {\bibfnamefont {P.~B.}\ \bibnamefont {Littlewood}},\ }\href
  {\doibase 10.1103/PhysRevB.71.064513} {\bibfield  {journal} {\bibinfo
  {journal} {Phys. Rev. B}\ }\textbf {\bibinfo {volume} {71}},\ \bibinfo
  {pages} {064513} (\bibinfo {year} {2005})}\BibitemShut {NoStop}%
\bibitem [{\citenamefont {Werner}(2008)}]{wernerthesis}%
  \BibitemOpen
  \bibfield  {author} {\bibinfo {author} {\bibfnamefont {F.}~\bibnamefont
  {Werner}},\ }\emph {\bibinfo {title} {Atomes froids pi\'eg\'es en interaction
  r\'esonnante: gaz unitaire et probl\`eme \`a trois corps}},\ \href@noop {}
  {Ph.D. thesis},\ \bibinfo  {school} {Universit\'e Pierre et Marie Curie}
  (\bibinfo {year} {2008})\BibitemShut {NoStop}%
\bibitem [{\citenamefont {Altman}\ and\ \citenamefont
  {Vishwanath}(2005)}]{PhysRevLett.95.110404}%
  \BibitemOpen
  \bibfield  {author} {\bibinfo {author} {\bibfnamefont {E.}~\bibnamefont
  {Altman}}\ and\ \bibinfo {author} {\bibfnamefont {A.}~\bibnamefont
  {Vishwanath}},\ }\href {\doibase 10.1103/PhysRevLett.95.110404} {\bibfield
  {journal} {\bibinfo  {journal} {Phys. Rev. Lett.}\ }\textbf {\bibinfo
  {volume} {95}},\ \bibinfo {pages} {110404} (\bibinfo {year}
  {2005})}\BibitemShut {NoStop}%
\bibitem [{\citenamefont {Tenart}\ \emph {et~al.}(2021)\citenamefont {Tenart},
  \citenamefont {Hercé}, \citenamefont {Bureik}, \citenamefont {Dareau},\ and\
  \citenamefont {Clément}}]{Clement2021}%
  \BibitemOpen
  \bibfield  {author} {\bibinfo {author} {\bibfnamefont {A.}~\bibnamefont
  {Tenart}}, \bibinfo {author} {\bibfnamefont {G.}~\bibnamefont {Hercé}},
  \bibinfo {author} {\bibfnamefont {J.-P.}\ \bibnamefont {Bureik}}, \bibinfo
  {author} {\bibfnamefont {A.}~\bibnamefont {Dareau}}, \ and\ \bibinfo {author}
  {\bibfnamefont {D.}~\bibnamefont {Clément}},\ }\href {\doibase
  10.1038/s41567-021-01381-2} {\bibfield  {journal} {\bibinfo  {journal} {Nat.
  Phys.}\ }\textbf {\bibinfo {volume} {17}},\ \bibinfo {pages} {1364–1368}
  (\bibinfo {year} {2021})}\BibitemShut {NoStop}%
\bibitem [{\citenamefont {Schweigler}\ \emph {et~al.}(2017)\citenamefont
  {Schweigler}, \citenamefont {Kasper}, \citenamefont {Erne}, \citenamefont
  {Mazets}, \citenamefont {Rauer}, \citenamefont {Cataldini}, \citenamefont
  {Langen}, \citenamefont {Gasenzer}, \citenamefont {Berges},\ and\
  \citenamefont {Schmiedmayer}}]{Schweigler2017}%
  \BibitemOpen
  \bibfield  {author} {\bibinfo {author} {\bibfnamefont {T.}~\bibnamefont
  {Schweigler}}, \bibinfo {author} {\bibfnamefont {V.}~\bibnamefont {Kasper}},
  \bibinfo {author} {\bibfnamefont {S.}~\bibnamefont {Erne}}, \bibinfo {author}
  {\bibfnamefont {I.}~\bibnamefont {Mazets}}, \bibinfo {author} {\bibfnamefont
  {B.}~\bibnamefont {Rauer}}, \bibinfo {author} {\bibfnamefont
  {F.}~\bibnamefont {Cataldini}}, \bibinfo {author} {\bibfnamefont
  {T.}~\bibnamefont {Langen}}, \bibinfo {author} {\bibfnamefont
  {T.}~\bibnamefont {Gasenzer}}, \bibinfo {author} {\bibfnamefont
  {J.}~\bibnamefont {Berges}}, \ and\ \bibinfo {author} {\bibfnamefont
  {J.}~\bibnamefont {Schmiedmayer}},\ }\href {\doibase 10.1038/nature22310}
  {\bibfield  {journal} {\bibinfo  {journal} {Nature (London)}\ }\textbf
  {\bibinfo {volume} {545}},\ \bibinfo {pages} {323} (\bibinfo {year}
  {2017})}\BibitemShut {NoStop}%
\bibitem [{\citenamefont {Faddeev}\ and\ \citenamefont
  {Merkuriev}(2013)}]{faddeev2013quantum}%
  \BibitemOpen
  \bibfield  {author} {\bibinfo {author} {\bibfnamefont {L.~D.}\ \bibnamefont
  {Faddeev}}\ and\ \bibinfo {author} {\bibfnamefont {S.~P.}\ \bibnamefont
  {Merkuriev}},\ }\href@noop {} {\emph {\bibinfo {title} {Quantum Scattering
  Theory for Several Particle Systems}}},\ Vol.~\bibinfo {volume} {11}\
  (\bibinfo  {publisher} {Springer Science \& Business Media},\ \bibinfo {year}
  {2013})\BibitemShut {NoStop}%
\bibitem [{\citenamefont {Gl\"ockle}(1983)}]{glockle1983}%
  \BibitemOpen
  \bibfield  {author} {\bibinfo {author} {\bibfnamefont {W.}~\bibnamefont
  {Gl\"ockle}},\ }\href@noop {} {\emph {\bibinfo {title} {The Quantum
  Mechanical Few-Body Problem}}},\ edited by\ \bibinfo {editor} {\bibfnamefont
  {W.}~\bibnamefont {Beiglb\"ock}}\ (\bibinfo  {publisher} {Springer Berlin
  Heidelberg},\ \bibinfo {address} {Berlin},\ \bibinfo {year}
  {1983})\BibitemShut {NoStop}%
\bibitem [{\citenamefont {Taylor}(2006)}]{taylor2006scattering}%
  \BibitemOpen
  \bibfield  {author} {\bibinfo {author} {\bibfnamefont {J.~R.}\ \bibnamefont
  {Taylor}},\ }\href@noop {} {\emph {\bibinfo {title} {Scattering Theory: The
  Quantum Theory of Nonrelativistic Collisions}}}\ (\bibinfo  {publisher}
  {Dover Publications},\ \bibinfo {address} {New York},\ \bibinfo {year}
  {2006})\BibitemShut {NoStop}%
\bibitem [{\citenamefont {Thomas}(1935)}]{PhysRev.47.903}%
  \BibitemOpen
  \bibfield  {author} {\bibinfo {author} {\bibfnamefont {L.~H.}\ \bibnamefont
  {Thomas}},\ }\href {\doibase 10.1103/PhysRev.47.903} {\bibfield  {journal}
  {\bibinfo  {journal} {Phys. Rev.}\ }\textbf {\bibinfo {volume} {47}},\
  \bibinfo {pages} {903} (\bibinfo {year} {1935})}\BibitemShut {NoStop}%
\bibitem [{\citenamefont {Skorniakov}\ and\ \citenamefont
  {Ter-Martirosian}(1957)}]{skorniakov1957three}%
  \BibitemOpen
  \bibfield  {author} {\bibinfo {author} {\bibfnamefont {G.}~\bibnamefont
  {Skorniakov}}\ and\ \bibinfo {author} {\bibfnamefont {K.}~\bibnamefont
  {Ter-Martirosian}},\ }\href@noop {} {\bibfield  {journal} {\bibinfo
  {journal} {Soviet Phys. JETP}\ }\textbf {\bibinfo {volume} {4}} (\bibinfo
  {year} {1957})}\BibitemShut {NoStop}%
\bibitem [{\citenamefont {Greene}\ \emph {et~al.}(2017)\citenamefont {Greene},
  \citenamefont {Giannakeas},\ and\ \citenamefont
  {P\'erez-R\'{\i}os}}]{rev:greene}%
  \BibitemOpen
  \bibfield  {author} {\bibinfo {author} {\bibfnamefont {C.~H.}\ \bibnamefont
  {Greene}}, \bibinfo {author} {\bibfnamefont {P.}~\bibnamefont {Giannakeas}},
  \ and\ \bibinfo {author} {\bibfnamefont {J.}~\bibnamefont
  {P\'erez-R\'{\i}os}},\ }\href {\doibase 10.1103/RevModPhys.89.035006}
  {\bibfield  {journal} {\bibinfo  {journal} {Rev. Mod. Phys.}\ }\textbf
  {\bibinfo {volume} {89}},\ \bibinfo {pages} {035006} (\bibinfo {year}
  {2017})}\BibitemShut {NoStop}%
\bibitem [{\citenamefont {Wang}\ \emph {et~al.}(2012)\citenamefont {Wang},
  \citenamefont {D'Incao}, \citenamefont {Esry},\ and\ \citenamefont
  {Greene}}]{PhysRevLett.108.263001}%
  \BibitemOpen
  \bibfield  {author} {\bibinfo {author} {\bibfnamefont {J.}~\bibnamefont
  {Wang}}, \bibinfo {author} {\bibfnamefont {J.~P.}\ \bibnamefont {D'Incao}},
  \bibinfo {author} {\bibfnamefont {B.~D.}\ \bibnamefont {Esry}}, \ and\
  \bibinfo {author} {\bibfnamefont {C.~H.}\ \bibnamefont {Greene}},\ }\href
  {\doibase 10.1103/PhysRevLett.108.263001} {\bibfield  {journal} {\bibinfo
  {journal} {Phys. Rev. Lett.}\ }\textbf {\bibinfo {volume} {108}},\ \bibinfo
  {pages} {263001} (\bibinfo {year} {2012})}\BibitemShut {NoStop}%
\bibitem [{\citenamefont {Naidon}\ \emph {et~al.}(2014)\citenamefont {Naidon},
  \citenamefont {Endo},\ and\ \citenamefont {Ueda}}]{PhysRevA.90.022106}%
  \BibitemOpen
  \bibfield  {author} {\bibinfo {author} {\bibfnamefont {P.}~\bibnamefont
  {Naidon}}, \bibinfo {author} {\bibfnamefont {S.}~\bibnamefont {Endo}}, \ and\
  \bibinfo {author} {\bibfnamefont {M.}~\bibnamefont {Ueda}},\ }\href {\doibase
  10.1103/PhysRevA.90.022106} {\bibfield  {journal} {\bibinfo  {journal} {Phys.
  Rev. A}\ }\textbf {\bibinfo {volume} {90}},\ \bibinfo {pages} {022106}
  (\bibinfo {year} {2014})}\BibitemShut {NoStop}%
\bibitem [{\citenamefont {Salasnich}\ \emph {et~al.}(2005)\citenamefont
  {Salasnich}, \citenamefont {Manini},\ and\ \citenamefont
  {Parola}}]{Salasnich2005}%
  \BibitemOpen
  \bibfield  {author} {\bibinfo {author} {\bibfnamefont {L.}~\bibnamefont
  {Salasnich}}, \bibinfo {author} {\bibfnamefont {N.}~\bibnamefont {Manini}}, \
  and\ \bibinfo {author} {\bibfnamefont {A.}~\bibnamefont {Parola}},\ }\href
  {\doibase 10.1103/PhysRevA.72.023621} {\bibfield  {journal} {\bibinfo
  {journal} {Phys. Rev. A}\ }\textbf {\bibinfo {volume} {72}},\ \bibinfo
  {pages} {023621} (\bibinfo {year} {2005})}\BibitemShut {NoStop}%
\bibitem [{\citenamefont {Kurkjian}\ \emph {et~al.}(2013)\citenamefont
  {Kurkjian}, \citenamefont {Castin},\ and\ \citenamefont
  {Sinatra}}]{PhysRevA.88.063623}%
  \BibitemOpen
  \bibfield  {author} {\bibinfo {author} {\bibfnamefont {H.}~\bibnamefont
  {Kurkjian}}, \bibinfo {author} {\bibfnamefont {Y.}~\bibnamefont {Castin}}, \
  and\ \bibinfo {author} {\bibfnamefont {A.}~\bibnamefont {Sinatra}},\ }\href
  {\doibase 10.1103/PhysRevA.88.063623} {\bibfield  {journal} {\bibinfo
  {journal} {Phys. Rev. A}\ }\textbf {\bibinfo {volume} {88}},\ \bibinfo
  {pages} {063623} (\bibinfo {year} {2013})}\BibitemShut {NoStop}%
\bibitem [{\citenamefont {Werner}\ and\ \citenamefont
  {Castin}(2012{\natexlab{b}})}]{art:werner_fermions}%
  \BibitemOpen
  \bibfield  {author} {\bibinfo {author} {\bibfnamefont {F.}~\bibnamefont
  {Werner}}\ and\ \bibinfo {author} {\bibfnamefont {Y.}~\bibnamefont
  {Castin}},\ }\href {\doibase 10.1103/PhysRevA.86.013626} {\bibfield
  {journal} {\bibinfo  {journal} {Phys. Rev. A}\ }\textbf {\bibinfo {volume}
  {86}},\ \bibinfo {pages} {013626} (\bibinfo {year}
  {2012}{\natexlab{b}})}\BibitemShut {NoStop}%
\bibitem [{\citenamefont {Leggett}(1980)}]{art:leggett}%
  \BibitemOpen
  \bibfield  {author} {\bibinfo {author} {\bibfnamefont {A.}~\bibnamefont
  {Leggett}},\ }\href {https://doi.org/10.1051/jphyscol:1980704} {\bibfield
  {journal} {\bibinfo  {journal} {J. Phys. (Paris) Colloq.}\ }\textbf {\bibinfo
  {volume} {41}},\ \bibinfo {pages} {C7} (\bibinfo {year} {1980})}\BibitemShut
  {NoStop}%
\bibitem [{\citenamefont {Combescot}\ \emph {et~al.}(2003)\citenamefont
  {Combescot}, \citenamefont {Leyronas},\ and\ \citenamefont
  {Tanguy}}]{Tanguy2003}%
  \BibitemOpen
  \bibfield  {author} {\bibinfo {author} {\bibfnamefont {M.}~\bibnamefont
  {Combescot}}, \bibinfo {author} {\bibfnamefont {X.}~\bibnamefont {Leyronas}},
  \ and\ \bibinfo {author} {\bibfnamefont {C.}~\bibnamefont {Tanguy}},\ }\href
  {\doibase 10.1140/epjb/e2003-00003-1} {\bibfield  {journal} {\bibinfo
  {journal} {Eur. Phys. J. B}\ }\textbf {\bibinfo {volume} {31}},\ \bibinfo
  {pages} {17} (\bibinfo {year} {2003})}\BibitemShut {NoStop}%
\bibitem [{\citenamefont {Combescot}\ and\ \citenamefont
  {Tanguy}(2001)}]{Combescot_2001}%
  \BibitemOpen
  \bibfield  {author} {\bibinfo {author} {\bibfnamefont {M.}~\bibnamefont
  {Combescot}}\ and\ \bibinfo {author} {\bibfnamefont {C.}~\bibnamefont
  {Tanguy}},\ }\href {\doibase 10.1209/epl/i2001-00427-7} {\bibfield  {journal}
  {\bibinfo  {journal} {Europhys. Lett.}\ }\textbf {\bibinfo {volume} {55}},\
  \bibinfo {pages} {390} (\bibinfo {year} {2001})}\BibitemShut {NoStop}%
\bibitem [{\citenamefont {Fletcher}\ \emph {et~al.}(2017)\citenamefont
  {Fletcher}, \citenamefont {Lopes}, \citenamefont {Man}, \citenamefont
  {Navon}, \citenamefont {Smith}, \citenamefont {Zwierlein},\ and\
  \citenamefont {Hadzibabic}}]{Fletcher377}%
  \BibitemOpen
  \bibfield  {author} {\bibinfo {author} {\bibfnamefont {R.~J.}\ \bibnamefont
  {Fletcher}}, \bibinfo {author} {\bibfnamefont {R.}~\bibnamefont {Lopes}},
  \bibinfo {author} {\bibfnamefont {J.}~\bibnamefont {Man}}, \bibinfo {author}
  {\bibfnamefont {N.}~\bibnamefont {Navon}}, \bibinfo {author} {\bibfnamefont
  {R.~P.}\ \bibnamefont {Smith}}, \bibinfo {author} {\bibfnamefont {M.~W.}\
  \bibnamefont {Zwierlein}}, \ and\ \bibinfo {author} {\bibfnamefont
  {Z.}~\bibnamefont {Hadzibabic}},\ }\href {\doibase 10.1126/science.aai8195}
  {\bibfield  {journal} {\bibinfo  {journal} {Science}\ }\textbf {\bibinfo
  {volume} {355}},\ \bibinfo {pages} {377} (\bibinfo {year}
  {2017})}\BibitemShut {NoStop}%
\bibitem [{\citenamefont {Sun}\ \emph {et~al.}(2020)\citenamefont {Sun},
  \citenamefont {Zhang},\ and\ \citenamefont {Zhai}}]{PhysRevLett.125.110404}%
  \BibitemOpen
  \bibfield  {author} {\bibinfo {author} {\bibfnamefont {M.}~\bibnamefont
  {Sun}}, \bibinfo {author} {\bibfnamefont {P.}~\bibnamefont {Zhang}}, \ and\
  \bibinfo {author} {\bibfnamefont {H.}~\bibnamefont {Zhai}},\ }\href {\doibase
  10.1103/PhysRevLett.125.110404} {\bibfield  {journal} {\bibinfo  {journal}
  {Phys. Rev. Lett.}\ }\textbf {\bibinfo {volume} {125}},\ \bibinfo {pages}
  {110404} (\bibinfo {year} {2020})}\BibitemShut {NoStop}%
\end{thebibliography}%


\begin{thebibliography}{40}%
\makeatletter
\providecommand \@ifxundefined [1]{%
 \@ifx{#1\undefined}
}%
\providecommand \@ifnum [1]{%
 \ifnum #1\expandafter \@firstoftwo
 \else \expandafter \@secondoftwo
 \fi
}%
\providecommand \@ifx [1]{%
 \ifx #1\expandafter \@firstoftwo
 \else \expandafter \@secondoftwo
 \fi
}%
\providecommand \natexlab [1]{#1}%
\providecommand \enquote  [1]{``#1''}%
\providecommand \bibnamefont  [1]{#1}%
\providecommand \bibfnamefont [1]{#1}%
\providecommand \citenamefont [1]{#1}%
\providecommand \href@noop [0]{\@secondoftwo}%
\providecommand \href [0]{\begingroup \@sanitize@url \@href}%
\providecommand \@href[1]{\@@startlink{#1}\@@href}%
\providecommand \@@href[1]{\endgroup#1\@@endlink}%
\providecommand \@sanitize@url [0]{\catcode `\\12\catcode `\$12\catcode
  `\&12\catcode `\#12\catcode `\^12\catcode `\_12\catcode `\%12\relax}%
\providecommand \@@startlink[1]{}%
\providecommand \@@endlink[0]{}%
\providecommand \url  [0]{\begingroup\@sanitize@url \@url }%
\providecommand \@url [1]{\endgroup\@href {#1}{\urlprefix }}%
\providecommand \urlprefix  [0]{URL }%
\providecommand \Eprint [0]{\href }%
\providecommand \doibase [0]{http://dx.doi.org/}%
\providecommand \selectlanguage [0]{\@gobble}%
\providecommand \bibinfo  [0]{\@secondoftwo}%
\providecommand \bibfield  [0]{\@secondoftwo}%
\providecommand \translation [1]{[#1]}%
\providecommand \BibitemOpen [0]{}%
\providecommand \bibitemStop [0]{}%
\providecommand \bibitemNoStop [0]{.\EOS\space}%
\providecommand \EOS [0]{\spacefactor3000\relax}%
\providecommand \BibitemShut  [1]{\csname bibitem#1\endcsname}%
\let\auto@bib@innerbib\@empty
\bibitem [{\citenamefont {Colussi}\ \emph {et~al.}(2018)\citenamefont
  {Colussi}, \citenamefont {Musolino},\ and\ \citenamefont
  {Kokkelmans}}]{art:colussimk}%
  \BibitemOpen
  \bibfield  {author} {\bibinfo {author} {\bibfnamefont {V.~E.}\ \bibnamefont
  {Colussi}}, \bibinfo {author} {\bibfnamefont {S.}~\bibnamefont {Musolino}}, \
  and\ \bibinfo {author} {\bibfnamefont {S.~J. J. M.~F.}\ \bibnamefont
  {Kokkelmans}},\ }\href {\doibase 10.1103/PhysRevA.98.051601} {\bibfield
  {journal} {\bibinfo  {journal} {Phys. Rev. A}\ }\textbf {\bibinfo {volume}
  {98}},\ \bibinfo {pages} {051601(R)} (\bibinfo {year} {2018})}\BibitemShut
  {NoStop}%
\bibitem [{\citenamefont {Musolino}\ \emph {et~al.}(2019)\citenamefont
  {Musolino}, \citenamefont {Colussi},\ and\ \citenamefont
  {Kokkelmans}}]{art:silviack}%
  \BibitemOpen
  \bibfield  {author} {\bibinfo {author} {\bibfnamefont {S.}~\bibnamefont
  {Musolino}}, \bibinfo {author} {\bibfnamefont {V.~E.}\ \bibnamefont
  {Colussi}}, \ and\ \bibinfo {author} {\bibfnamefont {S.~J. J. M.~F.}\
  \bibnamefont {Kokkelmans}},\ }\href {\doibase 10.1103/PhysRevA.100.013612}
  {\bibfield  {journal} {\bibinfo  {journal} {Phys. Rev. A}\ }\textbf {\bibinfo
  {volume} {100}},\ \bibinfo {pages} {013612} (\bibinfo {year}
  {2019})}\BibitemShut {NoStop}%
\bibitem [{\citenamefont {Colussi}\ \emph {et~al.}(2020)\citenamefont
  {Colussi}, \citenamefont {Kurkjian}, \citenamefont {Van~Regemortel},
  \citenamefont {Musolino}, \citenamefont {van~de Kraats}, \citenamefont
  {Wouters},\ and\ \citenamefont {Kokkelmans}}]{art:cumulant2020}%
  \BibitemOpen
  \bibfield  {author} {\bibinfo {author} {\bibfnamefont {V.~E.}\ \bibnamefont
  {Colussi}}, \bibinfo {author} {\bibfnamefont {H.}~\bibnamefont {Kurkjian}},
  \bibinfo {author} {\bibfnamefont {M.}~\bibnamefont {Van~Regemortel}},
  \bibinfo {author} {\bibfnamefont {S.}~\bibnamefont {Musolino}}, \bibinfo
  {author} {\bibfnamefont {J.}~\bibnamefont {van~de Kraats}}, \bibinfo {author}
  {\bibfnamefont {M.}~\bibnamefont {Wouters}}, \ and\ \bibinfo {author}
  {\bibfnamefont {S.~J. J. M.~F.}\ \bibnamefont {Kokkelmans}},\ }\href
  {\doibase 10.1103/PhysRevA.102.063314} {\bibfield  {journal} {\bibinfo
  {journal} {Phys. Rev. A}\ }\textbf {\bibinfo {volume} {102}},\ \bibinfo
  {pages} {063314} (\bibinfo {year} {2020})}\BibitemShut {NoStop}%
\bibitem [{\citenamefont {Chin}\ \emph {et~al.}(2010)\citenamefont {Chin},
  \citenamefont {Grimm}, \citenamefont {Julienne},\ and\ \citenamefont
  {Tiesinga}}]{art:chin}%
  \BibitemOpen
  \bibfield  {author} {\bibinfo {author} {\bibfnamefont {C.}~\bibnamefont
  {Chin}}, \bibinfo {author} {\bibfnamefont {R.}~\bibnamefont {Grimm}},
  \bibinfo {author} {\bibfnamefont {P.}~\bibnamefont {Julienne}}, \ and\
  \bibinfo {author} {\bibfnamefont {E.}~\bibnamefont {Tiesinga}},\ }\href
  {\doibase 10.1103/RevModPhys.82.1225} {\bibfield  {journal} {\bibinfo
  {journal} {Rev. Mod. Phys.}\ }\textbf {\bibinfo {volume} {82}},\ \bibinfo
  {pages} {1225} (\bibinfo {year} {2010})}\BibitemShut {NoStop}%
\bibitem [{\citenamefont {Makotyn}\ \emph {et~al.}(2014)\citenamefont
  {Makotyn}, \citenamefont {Klauss}, \citenamefont {Goldberger}, \citenamefont
  {Cornell},\ and\ \citenamefont {Jin}}]{art:makotyn}%
  \BibitemOpen
  \bibfield  {author} {\bibinfo {author} {\bibfnamefont {P.}~\bibnamefont
  {Makotyn}}, \bibinfo {author} {\bibfnamefont {C.~E.}\ \bibnamefont {Klauss}},
  \bibinfo {author} {\bibfnamefont {D.~L.}\ \bibnamefont {Goldberger}},
  \bibinfo {author} {\bibfnamefont {E.}~\bibnamefont {Cornell}}, \ and\
  \bibinfo {author} {\bibfnamefont {D.~S.}\ \bibnamefont {Jin}},\ }\href
  {\doibase 10.1038/nphys2850} {\bibfield  {journal} {\bibinfo  {journal} {Nat.
  Phys.}\ }\textbf {\bibinfo {volume} {10}},\ \bibinfo {pages} {116} (\bibinfo
  {year} {2014})}\BibitemShut {NoStop}%
\bibitem [{\citenamefont {Klauss}\ \emph {et~al.}(2017)\citenamefont {Klauss},
  \citenamefont {Xie}, \citenamefont {Lopez-Abadia}, \citenamefont {D'Incao},
  \citenamefont {Hadzibabic}, \citenamefont {Jin},\ and\ \citenamefont
  {Cornell}}]{art:klauss}%
  \BibitemOpen
  \bibfield  {author} {\bibinfo {author} {\bibfnamefont {C.~E.}\ \bibnamefont
  {Klauss}}, \bibinfo {author} {\bibfnamefont {X.}~\bibnamefont {Xie}},
  \bibinfo {author} {\bibfnamefont {C.}~\bibnamefont {Lopez-Abadia}}, \bibinfo
  {author} {\bibfnamefont {J.~P.}\ \bibnamefont {D'Incao}}, \bibinfo {author}
  {\bibfnamefont {Z.}~\bibnamefont {Hadzibabic}}, \bibinfo {author}
  {\bibfnamefont {D.~S.}\ \bibnamefont {Jin}}, \ and\ \bibinfo {author}
  {\bibfnamefont {E.~A.}\ \bibnamefont {Cornell}},\ }\href {\doibase
  10.1103/PhysRevLett.119.143401} {\bibfield  {journal} {\bibinfo  {journal}
  {Phys. Rev. Lett.}\ }\textbf {\bibinfo {volume} {119}},\ \bibinfo {pages}
  {143401} (\bibinfo {year} {2017})}\BibitemShut {NoStop}%
\bibitem [{\citenamefont {Eigen}\ \emph {et~al.}(2018)\citenamefont {Eigen},
  \citenamefont {Glidden}, \citenamefont {Lopes}, \citenamefont {Cornell},
  \citenamefont {Smith},\ and\ \citenamefont {Hadzibabic}}]{art:eigen18}%
  \BibitemOpen
  \bibfield  {author} {\bibinfo {author} {\bibfnamefont {C.}~\bibnamefont
  {Eigen}}, \bibinfo {author} {\bibfnamefont {J.~A.~P.}\ \bibnamefont
  {Glidden}}, \bibinfo {author} {\bibfnamefont {R.}~\bibnamefont {Lopes}},
  \bibinfo {author} {\bibfnamefont {E.~A.}\ \bibnamefont {Cornell}}, \bibinfo
  {author} {\bibfnamefont {R.~P.}\ \bibnamefont {Smith}}, \ and\ \bibinfo
  {author} {\bibfnamefont {Z.}~\bibnamefont {Hadzibabic}},\ }\href {\doibase
  10.1038/s41586-018-0674-1} {\bibfield  {journal} {\bibinfo  {journal} {Nature
  (London)}\ }\textbf {\bibinfo {volume} {563}},\ \bibinfo {pages} {221}
  (\bibinfo {year} {2018})}\BibitemShut {NoStop}%
\bibitem [{\citenamefont {Eigen}\ \emph {et~al.}(2017)\citenamefont {Eigen},
  \citenamefont {Glidden}, \citenamefont {Lopes}, \citenamefont {Navon},
  \citenamefont {Hadzibabic},\ and\ \citenamefont {Smith}}]{art:eigen17}%
  \BibitemOpen
  \bibfield  {author} {\bibinfo {author} {\bibfnamefont {C.}~\bibnamefont
  {Eigen}}, \bibinfo {author} {\bibfnamefont {J.~A.~P.}\ \bibnamefont
  {Glidden}}, \bibinfo {author} {\bibfnamefont {R.}~\bibnamefont {Lopes}},
  \bibinfo {author} {\bibfnamefont {N.}~\bibnamefont {Navon}}, \bibinfo
  {author} {\bibfnamefont {Z.}~\bibnamefont {Hadzibabic}}, \ and\ \bibinfo
  {author} {\bibfnamefont {R.~P.}\ \bibnamefont {Smith}},\ }\href {\doibase
  10.1103/PhysRevLett.119.250404} {\bibfield  {journal} {\bibinfo  {journal}
  {Phys. Rev. Lett.}\ }\textbf {\bibinfo {volume} {119}},\ \bibinfo {pages}
  {250404} (\bibinfo {year} {2017})}\BibitemShut {NoStop}%
\bibitem [{\citenamefont {Faddeev}\ and\ \citenamefont
  {Merkuriev}(2013)}]{faddeev2013quantum}%
  \BibitemOpen
  \bibfield  {author} {\bibinfo {author} {\bibfnamefont {L.~D.}\ \bibnamefont
  {Faddeev}}\ and\ \bibinfo {author} {\bibfnamefont {S.~P.}\ \bibnamefont
  {Merkuriev}},\ }\href@noop {} {\emph {\bibinfo {title} {Quantum Scattering
  Theory for Several Particle Systems}}},\ Vol.~\bibinfo {volume} {11}\
  (\bibinfo  {publisher} {Springer Science \& Business Media},\ \bibinfo {year}
  {2013})\BibitemShut {NoStop}%
\bibitem [{\citenamefont {Gl\"ockle}(1983)}]{glockle1983}%
  \BibitemOpen
  \bibfield  {author} {\bibinfo {author} {\bibfnamefont {W.}~\bibnamefont
  {Gl\"ockle}},\ }\href@noop {} {\emph {\bibinfo {title} {The Quantum
  Mechanical Few-Body Problem}}},\ edited by\ \bibinfo {editor} {\bibfnamefont
  {W.}~\bibnamefont {Beiglb\"ock}}\ (\bibinfo  {publisher} {Springer Berlin
  Heidelberg},\ \bibinfo {address} {Berlin},\ \bibinfo {year}
  {1983})\BibitemShut {NoStop}%
\bibitem [{\citenamefont {Taylor}(2006)}]{taylor2006scattering}%
  \BibitemOpen
  \bibfield  {author} {\bibinfo {author} {\bibfnamefont {J.~R.}\ \bibnamefont
  {Taylor}},\ }\href@noop {} {\emph {\bibinfo {title} {Scattering Theory: The
  Quantum Theory of Nonrelativistic Collisions}}}\ (\bibinfo  {publisher}
  {Dover Publications},\ \bibinfo {address} {New York},\ \bibinfo {year}
  {2006})\BibitemShut {NoStop}%
\bibitem [{\citenamefont {Thomas}(1935)}]{PhysRev.47.903}%
  \BibitemOpen
  \bibfield  {author} {\bibinfo {author} {\bibfnamefont {L.~H.}\ \bibnamefont
  {Thomas}},\ }\href {\doibase 10.1103/PhysRev.47.903} {\bibfield  {journal}
  {\bibinfo  {journal} {Phys. Rev.}\ }\textbf {\bibinfo {volume} {47}},\
  \bibinfo {pages} {903} (\bibinfo {year} {1935})}\BibitemShut {NoStop}%
\bibitem [{\citenamefont {Flambaum}\ \emph {et~al.}(1999)\citenamefont
  {Flambaum}, \citenamefont {Gribakin},\ and\ \citenamefont
  {Harabati}}]{PhysRevA.59.1998}%
  \BibitemOpen
  \bibfield  {author} {\bibinfo {author} {\bibfnamefont {V.~V.}\ \bibnamefont
  {Flambaum}}, \bibinfo {author} {\bibfnamefont {G.~F.}\ \bibnamefont
  {Gribakin}}, \ and\ \bibinfo {author} {\bibfnamefont {C.}~\bibnamefont
  {Harabati}},\ }\href {\doibase 10.1103/PhysRevA.59.1998} {\bibfield
  {journal} {\bibinfo  {journal} {Phys. Rev. A}\ }\textbf {\bibinfo {volume}
  {59}},\ \bibinfo {pages} {1998} (\bibinfo {year} {1999})}\BibitemShut
  {NoStop}%
\bibitem [{\citenamefont {Skorniakov}\ and\ \citenamefont
  {Ter-Martirosian}(1957)}]{skorniakov1957three}%
  \BibitemOpen
  \bibfield  {author} {\bibinfo {author} {\bibfnamefont {G.}~\bibnamefont
  {Skorniakov}}\ and\ \bibinfo {author} {\bibfnamefont {K.}~\bibnamefont
  {Ter-Martirosian}},\ }\href@noop {} {\bibfield  {journal} {\bibinfo
  {journal} {Soviet Phys. JETP}\ }\textbf {\bibinfo {volume} {4}} (\bibinfo
  {year} {1957})}\BibitemShut {NoStop}%
\bibitem [{\citenamefont {Naidon}\ and\ \citenamefont
  {Endo}(2017)}]{rev:naidon}%
  \BibitemOpen
  \bibfield  {author} {\bibinfo {author} {\bibfnamefont {P.}~\bibnamefont
  {Naidon}}\ and\ \bibinfo {author} {\bibfnamefont {S.}~\bibnamefont {Endo}},\
  }\href {\doibase 10.1088/1361-6633/aa50e8} {\bibfield  {journal} {\bibinfo
  {journal} {Rep. Prog. Phys.}\ }\textbf {\bibinfo {volume} {80}},\ \bibinfo
  {pages} {056001} (\bibinfo {year} {2017})}\BibitemShut {NoStop}%
\bibitem [{\citenamefont {Greene}\ \emph {et~al.}(2017)\citenamefont {Greene},
  \citenamefont {Giannakeas},\ and\ \citenamefont
  {P\'erez-R\'{\i}os}}]{rev:greene}%
  \BibitemOpen
  \bibfield  {author} {\bibinfo {author} {\bibfnamefont {C.~H.}\ \bibnamefont
  {Greene}}, \bibinfo {author} {\bibfnamefont {P.}~\bibnamefont {Giannakeas}},
  \ and\ \bibinfo {author} {\bibfnamefont {J.}~\bibnamefont
  {P\'erez-R\'{\i}os}},\ }\href {\doibase 10.1103/RevModPhys.89.035006}
  {\bibfield  {journal} {\bibinfo  {journal} {Rev. Mod. Phys.}\ }\textbf
  {\bibinfo {volume} {89}},\ \bibinfo {pages} {035006} (\bibinfo {year}
  {2017})}\BibitemShut {NoStop}%
\bibitem [{\citenamefont {Wang}\ \emph {et~al.}(2012)\citenamefont {Wang},
  \citenamefont {D'Incao}, \citenamefont {Esry},\ and\ \citenamefont
  {Greene}}]{PhysRevLett.108.263001}%
  \BibitemOpen
  \bibfield  {author} {\bibinfo {author} {\bibfnamefont {J.}~\bibnamefont
  {Wang}}, \bibinfo {author} {\bibfnamefont {J.~P.}\ \bibnamefont {D'Incao}},
  \bibinfo {author} {\bibfnamefont {B.~D.}\ \bibnamefont {Esry}}, \ and\
  \bibinfo {author} {\bibfnamefont {C.~H.}\ \bibnamefont {Greene}},\ }\href
  {\doibase 10.1103/PhysRevLett.108.263001} {\bibfield  {journal} {\bibinfo
  {journal} {Phys. Rev. Lett.}\ }\textbf {\bibinfo {volume} {108}},\ \bibinfo
  {pages} {263001} (\bibinfo {year} {2012})}\BibitemShut {NoStop}%
\bibitem [{\citenamefont {Naidon}\ \emph {et~al.}(2014)\citenamefont {Naidon},
  \citenamefont {Endo},\ and\ \citenamefont {Ueda}}]{PhysRevA.90.022106}%
  \BibitemOpen
  \bibfield  {author} {\bibinfo {author} {\bibfnamefont {P.}~\bibnamefont
  {Naidon}}, \bibinfo {author} {\bibfnamefont {S.}~\bibnamefont {Endo}}, \ and\
  \bibinfo {author} {\bibfnamefont {M.}~\bibnamefont {Ueda}},\ }\href {\doibase
  10.1103/PhysRevA.90.022106} {\bibfield  {journal} {\bibinfo  {journal} {Phys.
  Rev. A}\ }\textbf {\bibinfo {volume} {90}},\ \bibinfo {pages} {022106}
  (\bibinfo {year} {2014})}\BibitemShut {NoStop}%
\bibitem [{\citenamefont {Sykes}\ \emph {et~al.}(2014)\citenamefont {Sykes},
  \citenamefont {Corson}, \citenamefont {D'Incao}, \citenamefont {Koller},
  \citenamefont {Greene}, \citenamefont {Rey}, \citenamefont {Hazzard},\ and\
  \citenamefont {Bohn}}]{art:sykes}%
  \BibitemOpen
  \bibfield  {author} {\bibinfo {author} {\bibfnamefont {A.~G.}\ \bibnamefont
  {Sykes}}, \bibinfo {author} {\bibfnamefont {J.~P.}\ \bibnamefont {Corson}},
  \bibinfo {author} {\bibfnamefont {J.~P.}\ \bibnamefont {D'Incao}}, \bibinfo
  {author} {\bibfnamefont {A.~P.}\ \bibnamefont {Koller}}, \bibinfo {author}
  {\bibfnamefont {C.~H.}\ \bibnamefont {Greene}}, \bibinfo {author}
  {\bibfnamefont {A.~M.}\ \bibnamefont {Rey}}, \bibinfo {author} {\bibfnamefont
  {K.~R.~A.}\ \bibnamefont {Hazzard}}, \ and\ \bibinfo {author} {\bibfnamefont
  {J.~L.}\ \bibnamefont {Bohn}},\ }\href {\doibase 10.1103/PhysRevA.89.021601}
  {\bibfield  {journal} {\bibinfo  {journal} {Phys. Rev. A}\ }\textbf {\bibinfo
  {volume} {89}},\ \bibinfo {pages} {021601(R)} (\bibinfo {year}
  {2014})}\BibitemShut {NoStop}%
\bibitem [{\citenamefont {Gao}\ \emph {et~al.}(2020)\citenamefont {Gao},
  \citenamefont {Sun}, \citenamefont {Zhang},\ and\ \citenamefont
  {Zhai}}]{PhysRevLett.124.040403}%
  \BibitemOpen
  \bibfield  {author} {\bibinfo {author} {\bibfnamefont {C.}~\bibnamefont
  {Gao}}, \bibinfo {author} {\bibfnamefont {M.}~\bibnamefont {Sun}}, \bibinfo
  {author} {\bibfnamefont {P.}~\bibnamefont {Zhang}}, \ and\ \bibinfo {author}
  {\bibfnamefont {H.}~\bibnamefont {Zhai}},\ }\href {\doibase
  10.1103/PhysRevLett.124.040403} {\bibfield  {journal} {\bibinfo  {journal}
  {Phys. Rev. Lett.}\ }\textbf {\bibinfo {volume} {124}},\ \bibinfo {pages}
  {040403} (\bibinfo {year} {2020})}\BibitemShut {NoStop}%
\bibitem [{\citenamefont {Mu\~noz de~las Heras}\ \emph
  {et~al.}(2019)\citenamefont {Mu\~noz de~las Heras}, \citenamefont {Parish},\
  and\ \citenamefont {Marchetti}}]{art:munozdelasheras}%
  \BibitemOpen
  \bibfield  {author} {\bibinfo {author} {\bibfnamefont {A.}~\bibnamefont
  {Mu\~noz de~las Heras}}, \bibinfo {author} {\bibfnamefont {M.~M.}\
  \bibnamefont {Parish}}, \ and\ \bibinfo {author} {\bibfnamefont {F.~M.}\
  \bibnamefont {Marchetti}},\ }\href {\doibase 10.1103/PhysRevA.99.023623}
  {\bibfield  {journal} {\bibinfo  {journal} {Phys. Rev. A}\ }\textbf {\bibinfo
  {volume} {99}},\ \bibinfo {pages} {023623} (\bibinfo {year}
  {2019})}\BibitemShut {NoStop}%
\bibitem [{\citenamefont {Corson}\ and\ \citenamefont
  {Bohn}(2015)}]{art:corson_bohn}%
  \BibitemOpen
  \bibfield  {author} {\bibinfo {author} {\bibfnamefont {J.~P.}\ \bibnamefont
  {Corson}}\ and\ \bibinfo {author} {\bibfnamefont {J.~L.}\ \bibnamefont
  {Bohn}},\ }\href {\doibase 10.1103/PhysRevA.91.013616} {\bibfield  {journal}
  {\bibinfo  {journal} {Phys. Rev. A}\ }\textbf {\bibinfo {volume} {91}},\
  \bibinfo {pages} {013616} (\bibinfo {year} {2015})}\BibitemShut {NoStop}%
\bibitem [{\citenamefont {Blaizot}\ and\ \citenamefont
  {Ripka}(1986)}]{book:blaizot}%
  \BibitemOpen
  \bibfield  {author} {\bibinfo {author} {\bibfnamefont {J.~P.}\ \bibnamefont
  {Blaizot}}\ and\ \bibinfo {author} {\bibfnamefont {G.}~\bibnamefont
  {Ripka}},\ }\href@noop {} {\emph {\bibinfo {title} {Quantum Theory of Finite
  Systems}}}\ (\bibinfo  {publisher} {The MIT Press},\ \bibinfo {address}
  {Cambridge, Massachusetts},\ \bibinfo {year} {1986})\BibitemShut {NoStop}%
\bibitem [{\citenamefont {Leggett}(2006)}]{book:leggett}%
  \BibitemOpen
  \bibfield  {author} {\bibinfo {author} {\bibfnamefont {A.}~\bibnamefont
  {Leggett}},\ }\href {https://books.google.nl/books?id=HnlPAwAAQBAJ} {\emph
  {\bibinfo {title} {Quantum Liquids: Bose condensation and Cooper pairing in
  condensed-matter systems}}},\ Oxford Graduate Texts\ (\bibinfo  {publisher}
  {OUP Oxford},\ \bibinfo {year} {2006})\BibitemShut {NoStop}%
\bibitem [{\citenamefont {K\"ohler}\ and\ \citenamefont
  {Burnett}(2002)}]{PhysRevA.65.033601}%
  \BibitemOpen
  \bibfield  {author} {\bibinfo {author} {\bibfnamefont {T.}~\bibnamefont
  {K\"ohler}}\ and\ \bibinfo {author} {\bibfnamefont {K.}~\bibnamefont
  {Burnett}},\ }\href {\doibase 10.1103/PhysRevA.65.033601} {\bibfield
  {journal} {\bibinfo  {journal} {Phys. Rev. A}\ }\textbf {\bibinfo {volume}
  {65}},\ \bibinfo {pages} {033601} (\bibinfo {year} {2002})}\BibitemShut
  {NoStop}%
\bibitem [{\citenamefont {K\"ohler}(2002)}]{PhysRevLett.89.210404}%
  \BibitemOpen
  \bibfield  {author} {\bibinfo {author} {\bibfnamefont {T.}~\bibnamefont
  {K\"ohler}},\ }\href {\doibase 10.1103/PhysRevLett.89.210404} {\bibfield
  {journal} {\bibinfo  {journal} {Phys. Rev. Lett.}\ }\textbf {\bibinfo
  {volume} {89}},\ \bibinfo {pages} {210404} (\bibinfo {year}
  {2002})}\BibitemShut {NoStop}%
\bibitem [{\citenamefont {Salasnich}\ \emph {et~al.}(2005)\citenamefont
  {Salasnich}, \citenamefont {Manini},\ and\ \citenamefont
  {Parola}}]{Salasnich2005}%
  \BibitemOpen
  \bibfield  {author} {\bibinfo {author} {\bibfnamefont {L.}~\bibnamefont
  {Salasnich}}, \bibinfo {author} {\bibfnamefont {N.}~\bibnamefont {Manini}}, \
  and\ \bibinfo {author} {\bibfnamefont {A.}~\bibnamefont {Parola}},\ }\href
  {\doibase 10.1103/PhysRevA.72.023621} {\bibfield  {journal} {\bibinfo
  {journal} {Phys. Rev. A}\ }\textbf {\bibinfo {volume} {72}},\ \bibinfo
  {pages} {023621} (\bibinfo {year} {2005})}\BibitemShut {NoStop}%
\bibitem [{\citenamefont {Kurkjian}\ \emph {et~al.}(2013)\citenamefont
  {Kurkjian}, \citenamefont {Castin},\ and\ \citenamefont
  {Sinatra}}]{PhysRevA.88.063623}%
  \BibitemOpen
  \bibfield  {author} {\bibinfo {author} {\bibfnamefont {H.}~\bibnamefont
  {Kurkjian}}, \bibinfo {author} {\bibfnamefont {Y.}~\bibnamefont {Castin}}, \
  and\ \bibinfo {author} {\bibfnamefont {A.}~\bibnamefont {Sinatra}},\ }\href
  {\doibase 10.1103/PhysRevA.88.063623} {\bibfield  {journal} {\bibinfo
  {journal} {Phys. Rev. A}\ }\textbf {\bibinfo {volume} {88}},\ \bibinfo
  {pages} {063623} (\bibinfo {year} {2013})}\BibitemShut {NoStop}%
\bibitem [{\citenamefont {Pong}\ and\ \citenamefont
  {Law}(2007)}]{PhysRevA.75.043613}%
  \BibitemOpen
  \bibfield  {author} {\bibinfo {author} {\bibfnamefont {Y.~H.}\ \bibnamefont
  {Pong}}\ and\ \bibinfo {author} {\bibfnamefont {C.~K.}\ \bibnamefont {Law}},\
  }\href {\doibase 10.1103/PhysRevA.75.043613} {\bibfield  {journal} {\bibinfo
  {journal} {Phys. Rev. A}\ }\textbf {\bibinfo {volume} {75}},\ \bibinfo
  {pages} {043613} (\bibinfo {year} {2007})}\BibitemShut {NoStop}%
\bibitem [{\citenamefont {Combescot}\ \emph {et~al.}(2003)\citenamefont
  {Combescot}, \citenamefont {Leyronas},\ and\ \citenamefont
  {Tanguy}}]{Tanguy2003}%
  \BibitemOpen
  \bibfield  {author} {\bibinfo {author} {\bibfnamefont {M.}~\bibnamefont
  {Combescot}}, \bibinfo {author} {\bibfnamefont {X.}~\bibnamefont {Leyronas}},
  \ and\ \bibinfo {author} {\bibfnamefont {C.}~\bibnamefont {Tanguy}},\ }\href
  {\doibase 10.1140/epjb/e2003-00003-1} {\bibfield  {journal} {\bibinfo
  {journal} {Eur. Phys. J. B}\ }\textbf {\bibinfo {volume} {31}},\ \bibinfo
  {pages} {17} (\bibinfo {year} {2003})}\BibitemShut {NoStop}%
\bibitem [{\citenamefont {Combescot}\ and\ \citenamefont
  {Tanguy}(2001)}]{Combescot_2001}%
  \BibitemOpen
  \bibfield  {author} {\bibinfo {author} {\bibfnamefont {M.}~\bibnamefont
  {Combescot}}\ and\ \bibinfo {author} {\bibfnamefont {C.}~\bibnamefont
  {Tanguy}},\ }\href {\doibase 10.1209/epl/i2001-00427-7} {\bibfield  {journal}
  {\bibinfo  {journal} {Europhys. Lett.}\ }\textbf {\bibinfo {volume} {55}},\
  \bibinfo {pages} {390} (\bibinfo {year} {2001})}\BibitemShut {NoStop}%
\bibitem [{\citenamefont {Regal}\ \emph {et~al.}(2004)\citenamefont {Regal},
  \citenamefont {Greiner},\ and\ \citenamefont {Jin}}]{art:regal2004}%
  \BibitemOpen
  \bibfield  {author} {\bibinfo {author} {\bibfnamefont {C.~A.}\ \bibnamefont
  {Regal}}, \bibinfo {author} {\bibfnamefont {M.}~\bibnamefont {Greiner}}, \
  and\ \bibinfo {author} {\bibfnamefont {D.~S.}\ \bibnamefont {Jin}},\ }\href
  {\doibase 10.1103/PhysRevLett.92.040403} {\bibfield  {journal} {\bibinfo
  {journal} {Phys. Rev. Lett.}\ }\textbf {\bibinfo {volume} {92}},\ \bibinfo
  {pages} {040403} (\bibinfo {year} {2004})}\BibitemShut {NoStop}%
\bibitem [{\citenamefont {Braaten}\ \emph {et~al.}(2011)\citenamefont
  {Braaten}, \citenamefont {Kang},\ and\ \citenamefont
  {Platter}}]{art:braaten_c3_first}%
  \BibitemOpen
  \bibfield  {author} {\bibinfo {author} {\bibfnamefont {E.}~\bibnamefont
  {Braaten}}, \bibinfo {author} {\bibfnamefont {D.}~\bibnamefont {Kang}}, \
  and\ \bibinfo {author} {\bibfnamefont {L.}~\bibnamefont {Platter}},\ }\href
  {\doibase 10.1103/PhysRevLett.106.153005} {\bibfield  {journal} {\bibinfo
  {journal} {Phys. Rev. Lett.}\ }\textbf {\bibinfo {volume} {106}},\ \bibinfo
  {pages} {153005} (\bibinfo {year} {2011})}\BibitemShut {NoStop}%
\bibitem [{\citenamefont {Werner}\ and\ \citenamefont
  {Castin}(2012{\natexlab{a}})}]{art:werner_fermions}%
  \BibitemOpen
  \bibfield  {author} {\bibinfo {author} {\bibfnamefont {F.}~\bibnamefont
  {Werner}}\ and\ \bibinfo {author} {\bibfnamefont {Y.}~\bibnamefont
  {Castin}},\ }\href {\doibase 10.1103/PhysRevA.86.013626} {\bibfield
  {journal} {\bibinfo  {journal} {Phys. Rev. A}\ }\textbf {\bibinfo {volume}
  {86}},\ \bibinfo {pages} {013626} (\bibinfo {year}
  {2012}{\natexlab{a}})}\BibitemShut {NoStop}%
\bibitem [{\citenamefont {Strinati}\ \emph {et~al.}(2018)\citenamefont
  {Strinati}, \citenamefont {Pieri}, \citenamefont {R\"opke}, \citenamefont
  {Schuck},\ and\ \citenamefont {Urban}}]{STRINATI20181}%
  \BibitemOpen
  \bibfield  {author} {\bibinfo {author} {\bibfnamefont {G.~C.}\ \bibnamefont
  {Strinati}}, \bibinfo {author} {\bibfnamefont {P.}~\bibnamefont {Pieri}},
  \bibinfo {author} {\bibfnamefont {G.}~\bibnamefont {R\"opke}}, \bibinfo
  {author} {\bibfnamefont {P.}~\bibnamefont {Schuck}}, \ and\ \bibinfo {author}
  {\bibfnamefont {M.}~\bibnamefont {Urban}},\ }\href {\doibase
  https://doi.org/10.1016/j.physrep.2018.02.004} {\bibfield  {journal}
  {\bibinfo  {journal} {Phys. Rep.}\ }\textbf {\bibinfo {volume} {738}},\
  \bibinfo {pages} {1} (\bibinfo {year} {2018})}\BibitemShut {NoStop}%
\bibitem [{\citenamefont {Werner}\ and\ \citenamefont
  {Castin}(2012{\natexlab{b}})}]{art:werner_bosons}%
  \BibitemOpen
  \bibfield  {author} {\bibinfo {author} {\bibfnamefont {F.}~\bibnamefont
  {Werner}}\ and\ \bibinfo {author} {\bibfnamefont {Y.}~\bibnamefont
  {Castin}},\ }\href {\doibase 10.1103/PhysRevA.86.053633} {\bibfield
  {journal} {\bibinfo  {journal} {Phys. Rev. A}\ }\textbf {\bibinfo {volume}
  {86}},\ \bibinfo {pages} {053633} (\bibinfo {year}
  {2012}{\natexlab{b}})}\BibitemShut {NoStop}%
\bibitem [{\citenamefont {Leggett}(1980)}]{art:leggett}%
  \BibitemOpen
  \bibfield  {author} {\bibinfo {author} {\bibfnamefont {A.}~\bibnamefont
  {Leggett}},\ }\href {https://doi.org/10.1051/jphyscol:1980704} {\bibfield
  {journal} {\bibinfo  {journal} {J. Phys. (Paris) Colloq.}\ }\textbf {\bibinfo
  {volume} {41}},\ \bibinfo {pages} {C7} (\bibinfo {year} {1980})}\BibitemShut
  {NoStop}%
\bibitem [{\citenamefont {Pitaevskii}\ and\ \citenamefont
  {Stringari}(2016)}]{book:pitaevskii_stringari}%
  \BibitemOpen
  \bibfield  {author} {\bibinfo {author} {\bibfnamefont {L.}~\bibnamefont
  {Pitaevskii}}\ and\ \bibinfo {author} {\bibfnamefont {S.}~\bibnamefont
  {Stringari}},\ }\href {https://books.google.nl/books?id=yHByCwAAQBAJ} {\emph
  {\bibinfo {title} {Bose-Einstein Condensation and Superfluidity}}},\
  International Series of Monographs on Physics\ (\bibinfo  {publisher} {OUP
  Oxford},\ \bibinfo {year} {2016})\BibitemShut {NoStop}%
\bibitem [{\citenamefont {Fletcher}\ \emph {et~al.}(2017)\citenamefont
  {Fletcher}, \citenamefont {Lopes}, \citenamefont {Man}, \citenamefont
  {Navon}, \citenamefont {Smith}, \citenamefont {Zwierlein},\ and\
  \citenamefont {Hadzibabic}}]{Fletcher377}%
  \BibitemOpen
  \bibfield  {author} {\bibinfo {author} {\bibfnamefont {R.~J.}\ \bibnamefont
  {Fletcher}}, \bibinfo {author} {\bibfnamefont {R.}~\bibnamefont {Lopes}},
  \bibinfo {author} {\bibfnamefont {J.}~\bibnamefont {Man}}, \bibinfo {author}
  {\bibfnamefont {N.}~\bibnamefont {Navon}}, \bibinfo {author} {\bibfnamefont
  {R.~P.}\ \bibnamefont {Smith}}, \bibinfo {author} {\bibfnamefont {M.~W.}\
  \bibnamefont {Zwierlein}}, \ and\ \bibinfo {author} {\bibfnamefont
  {Z.}~\bibnamefont {Hadzibabic}},\ }\href {\doibase 10.1126/science.aai8195}
  {\bibfield  {journal} {\bibinfo  {journal} {Science}\ }\textbf {\bibinfo
  {volume} {355}},\ \bibinfo {pages} {377} (\bibinfo {year}
  {2017})}\BibitemShut {NoStop}%
\bibitem [{\citenamefont {Sun}\ \emph {et~al.}(2020)\citenamefont {Sun},
  \citenamefont {Zhang},\ and\ \citenamefont {Zhai}}]{PhysRevLett.125.110404}%
  \BibitemOpen
  \bibfield  {author} {\bibinfo {author} {\bibfnamefont {M.}~\bibnamefont
  {Sun}}, \bibinfo {author} {\bibfnamefont {P.}~\bibnamefont {Zhang}}, \ and\
  \bibinfo {author} {\bibfnamefont {H.}~\bibnamefont {Zhai}},\ }\href {\doibase
  10.1103/PhysRevLett.125.110404} {\bibfield  {journal} {\bibinfo  {journal}
  {Phys. Rev. Lett.}\ }\textbf {\bibinfo {volume} {125}},\ \bibinfo {pages}
  {110404} (\bibinfo {year} {2020})}\BibitemShut {NoStop}%
\end{thebibliography}%
\end{document}



\title{Supplementary Material: ``Bose-Einstein condensation of Efimovian triples in the unitary Bose gas''}
\author{S. Musolino}
\email{sil.musolino@gmail.com}
\affiliation{Eindhoven University of Technology, PO Box 513, 5600 MB Eindhoven, The Netherlands}
\author{H. Kurkjian}
\affiliation{Laboratoire de Physique Th\'eorique, Universit\'e de Toulouse, CNRS, UPS, 31400 Toulouse, France}
\author{M. Van Regemortel}
\affiliation{Joint Quantum Institute and The Institute for Research in Electronics and Applied Physics, University of Maryland, College Park, 20742 MD, USA}
\author{M. Wouters}
\affiliation{TQC, Universiteit Antwerpen, Universiteitsplein 1, B-2610 Antwerp, Belgium}
\author{S. J. J. M. F. Kokkelmans}
\affiliation{Eindhoven University of Technology, PO Box 513, 5600 MB Eindhoven, The Netherlands}
\author{V. E. Colussi}
\email{colussiv@gmail.com}
\affiliation{INO-CNR BEC Center and Dipartimento di Fisica, Universit\`a di Trento, 38123 Povo, Italy}

\date{\today}

\maketitle

\section{Few-Body Model at unitarity}

The results of the main text are produced using the calibrated two-body model employed in Refs.~\cite{art:colussimk,art:silviack,art:cumulant2020} that describes well the broad, entrance-channel dominated Feshbach resonances used experimentally \cite{art:chin,art:makotyn,art:klauss,art:eigen18,art:eigen17}.  In this section, we briefly outline this model and implications of the finite-range effects on the three-body level.
\newline

\subsection{Calibrated two-body model}
Eq.~(1) of the main text is written in terms of a local potential $V(|\rg-\rg'|)$. In general, a local potential can be always expanded as a sum of nonlocal separable potentials as
\begin{equation}\label{eq:sepexp}
\langle {\bf k}|\hat{V}|{\bf k'}\rangle=\sum_{j=1} g_j\langle{\bf k}|\zeta_j\rangle\langle\zeta_j|{\bf k'}\rangle,
\end{equation}
with form factors $|\zeta_j\rangle$ and interaction strengths $g_j$ \cite{faddeev2013quantum}.
 We make the unitary pole approximation, valid in the unitarity limit~\cite{glockle1983}, replacing the actual potential by a separable one and choosing $s$-wave form factors $\langle {\bf k}|\zeta\rangle=\theta(\Lambda-|{\bf k}|)$ that are functions of the relative momentum where $\theta(x)$ is the unit step function defined such that $\theta(x\geq0)=1$ and $\theta(x<0)=0$. This step function form factor therefore provides a cutoff on the relative two-body momentum. Therefore, the momentum space representation of Eq.(1) of the main text reads as
 \begin{equation}\label{eq:mbhamk}
\hat{H}\!=\!\sum_{\bf k}\epsilon_{\bf k}\hat{a}^\dagger_{\bf k}\hat{a}_{\bf k}
+\frac{g}{2V}\sum_{{\bf p},{\bf p'},{\bf q}}\zeta_{{\bf p}-{\bf p'}+2{\bf q}}\zeta_{{\bf p}-{\bf p'}} \hat{a}^\dagger_{{\bf p}+{\bf q}}\hat{a}^\dagger_{{\bf p'}-{\bf q}}\hat{a}_{\bf p} \hat{a}_{\bf p'}
\end{equation}  
where $\epsilon_{\bf k}=\hbar^2 k^2/2m$ is the kinetic term. Using a separable potential, it is possible to obtain a closed, analytic expression for the Lippmann-Schwinger equation for the two-body $T$ operator $\hat{T}(z)=\hat{V}+\hat{V}\hat{G}_{\mathrm{2B}}^{(0)}(z)\hat{T}(z)$ as
\begin{align}
\hat{T}(z)&=\frac{g|\zeta\rangle\langle \zeta|}{1-g\langle\zeta|\hat{G}_\mathrm{2B}^{(0)}(z)|\zeta\rangle},\label{eq:simplepole}
\end{align}
where $\hat{G}_{\mathrm{2B}}^{(0)}(z)$ is the two-body free Green's function \cite{taylor2006scattering}.  The low energy limit of the on-shell $T$-matrix for $s$-wave scattering is given as always by
\begin{eqnarray}
 \frac{4\pi\hbar^2}{m}a&\underset{|{\bf k}|\to0}{=}&\langle{\bf k},-{\bf k}|\hat{T}(\hbar^2k^2/m+i0)|{\bf k}',-{\bf k}'\rangle,
 \end{eqnarray}
which fixes the interaction strength $g=U_0\Gamma$ where $U_0=4\pi\hbar^2 a/m$ and $\Gamma = (1-2a\Lambda/\pi)^{-1}$, which gives $g=-2\pi^2\hbar^2/m\Lambda$ on resonance.  Taking the limit $\Lambda\to\infty$ would yield $\hat{V}$ equivalent to a renormalized contact potential, which leads also to the unphysical Thomas collapse on the three-body level \cite{PhysRev.47.903}.  Instead, we calibrate as $\Lambda=2/\pi\bar{a}$ in order to reproduce finite-range corrections to the binding energy of the shallow $s$-wave dimer $-\hbar^2/m(a-\bar{a})^2$ away from resonance, where $\bar{a}\approx0.956r_\mathrm{vdW}$ is the mean-scattering length that is set by the van der Waals length $r_\mathrm{vdW}$ for a given atomic species \cite{PhysRevA.59.1998}.    

\subsection{\label{sec:3BWF} Efimov states}
On the three-body level, the spectrum of three-body bound Efimov states is highly sensitive to the finite-range physics of the calibrated two-body model.  The use of a pairwise separable potential renders the vacuum three-body problem solvable following the work of Skorniakov and Ter-Martirosian~\cite{skorniakov1957three}.  Here, we outline this solution as well as the construction and normalization of the three-body wave function in momentum space necessary to evaluate Eq.~(23) of the main text.  To solve the three-body problem in vacuum, we begin with the decomposition into  Faddeev components
\begin{equation}
\begin{split}
\ket{\Psi_\mathrm{3b}}&=\ket{\Psi^{(1)}} +\ket{\Psi^{(2)}} + \ket{\Psi^{(3)}}\\
 &= (1+\hat{P}_+ +\hat{P}_-)\ket{\Psi^{(1)}},
\end{split}
\label{eq:Psi_3b_Faddeev}
\end{equation}
where $\hat{P}_+$ and $\hat{P}_-$ are permutation operators~\cite{rev:naidon, rev:greene}.  Each Faddeev component satisfies a bound state equation in momentum space, given for example by
\begin{equation}
\begin{split}
\Psi^{(1)}(\qg_1, \pg_1) &= G_\mathrm{3B}^{(0)}(q_1, p_1, E) \sum_{\pg'_1, \qg'_1} \braket{\qg_1, \pg_1|\hat{T}_{23}(E)|\pg'_1, \qg'_1}\braket{\pg'_1, \qg'_1|\hat{P}_+ +\hat{P}_-|\Psi^{(1)}},
\end{split}
\label{eq:Fadd_mom_1}
\end{equation}
where $G_\mathrm{3B}^{(0)}(q_1, p_1, E)=1/(E-q_1^2/m-3p_1^2/4m)$ is the vacuum three-body Green`s function and $\hat{T}_{23}= \hat{V}_{23} + \hat{V}_{23}\hat{G}_\mathrm{3B}^{(0)}(E)\hat{T}_{23}$ is the two-body transition matrix obeying the Lippmann-Schwinger equation, and $E$ is the binding energy. Here, we have parametrized the Faddeev component by the Jacobi vectors ${\bf q}_1=({\bf k}_2-{\bf k}_3)/2$ and ${\bf p}_1=(2{\bf k}_1-{\bf k}_2-{\bf k}_3)/3$.  Following the original formulation of Ref.~\cite{skorniakov1957three}, we make the ansatz 
\begin{equation}
  \Psi^{(1)}(\qg_1, \pg_1)= G_\mathrm{3B}^{(0)}(q_1, p_1, E) \zeta (\qg_1) \mathcal{F}(\pg_1), 
 \label{eq:Fadd_ans}
\end{equation}
and inserting this ansatz in Eq.~\eqref{eq:Fadd_mom_1} yields the  integral equation
\begin{equation}
\mathcal{F}(\pg_1)= 2 g\,\tau\left(E- \frac{3 p_1^2}{4m}\right)\sum_{\pg'_1}\dfrac{\zeta(2\pg'_1+\pg_1)\zeta(2\pg_1+\pg'_1)}{E- \frac{\pg_1^2}{m}- \frac{\pg_1^{'2}}{m}-\frac{\pg_1 \cdot\pg'_1 }{m}}\mathcal{F}(\pg'_1).
\label{eq:F}
\end{equation}
where $\tau (z) = 1/(1-g \braket{\zeta|\hat{G}_\mathrm{3B}^{(0)}|\zeta})$. The Efimov trimer binding energies correspond to the nontrivial solutions of Eq.~\eqref{eq:F}.  Here, we quote the previous results of Refs.~\cite{art:colussimk,art:cumulant2020} for the first few binding energies at unitarity using the separable step function form factor:  $\kappa_*/\Lambda\approx 0.317$ for the ground Efimov trimer and $\kappa^{(1)}/\Lambda\approx0.0131$ for the first excited Efimov trimer.  We note that our result for $\kappa_*$ is comparable to existing results from broad Feshbach resonances using more realistic interaction potentials~\cite{PhysRevLett.108.263001,PhysRevA.90.022106}

Using the binding energy of a specific Efimov trimer, the corresponding trimer wave function can be constructed via Eqs.~\eqref{eq:Psi_3b_Faddeev} and \eqref{eq:Fadd_ans}. 
In order to normalize the trimer wave function, we calculate the corresponding normalization constant $\mathcal{N}$ by solving the integral
\begin{equation}
\begin{split}
\mathcal{N}&= \braket{\Psi_\mathrm{3b}|\Psi_\mathrm{3b}}=3 \braket{\Psi^{(1)}|\Psi^{(1)}}+6 \braket{\Psi^{(1)}|P_+|\Psi^{(1)}}\\
&=\sum_{\qg_1, \pg_1} \Psi^{(1)\ast}(q_1, p_1) \left[\Psi^{(1)}(q_1, p_1) + \Psi^{(1)}\left(|\frac{\qg_1}{2}+\frac{3\pg_1}{4}|, |\qg_1-\frac{\pg_1}{2}|\right) \right]
\end{split}
\label{eq:norm_psi3b}
\end{equation}
where we have used that $\braket{\Psi^{(1)}| \Psi^{(1)}}=\braket{\Psi^{(2)}| \Psi^{(2)}}=\braket{\Psi^{(3)}| \Psi^{(3)}}$ and
the fact that the mixed terms  $\braket{\Psi^{(i)}| \Psi^{(j)}}$ are identical $\forall i\neq j$ and $i, j=\{1,2,3\}$ due to the particle exchange symmetry.
Therefore, the three-body wave function is normalized according to
\begin{equation}
 \begin{split}
  1&=\frac{1}{\mathcal{N}} \sum_{\qg_1, \pg_1} |\Psi_\mathrm{3b}(\qg_1, \pg_1)|^2\\
   &=\frac{1}{\mathcal{N}} \sum_{\kg_1, \kg_2} |-1|^3 |\Psi_\mathrm{3b}(\kg_1, \kg_2)|^2,
 \end{split}
\label{eq:norm}
\end{equation}
where $|-1|^3$ is the Jacobian for the change of variable from Jacobi  to single-particle momenta coordinates $(\qg_1, \pg_1, \mathbf{Q}) \to (\kg_1, \kg_2, \kg_3)$.  Finally, we note that because both $\langle \Psi_\mathrm{3b}^{(0)}|\Psi_\mathrm{3b}^{(0)}\rangle=1$ and $\langle\varphi_0^{(3)}|\varphi_0^{(3)}\rangle=1$, it is guaranteed by construction that $P_\mathrm{3b}^{(0)}= |\braket{ \varphi_0^{(3)}| \Psi_\mathrm{3b}^{(0)}}|^2\leq 1$.

\section{\label{sec:eoms} Cumulant Equations of Motion}

To describe the coupled-correlation dynamics, we first introduce the cumulant of a $p$-body operator as 
\be
\meanvlr{\prod_{i=0}^l\hat{a}_{\kk_i}^\dagger {\prod_{j=0}^m \hat{a}_{\kk_j'}}}_c=(-1)^m \prod_{i=0}^l \frac{\partial}{\partial x_i} \prod_{j=0}^m \frac{\partial}{\partial y_j^*} \ln\left. \meanvlr{\eee^{\sum_{i=0}^l x_i \hat{a}_{\kk_i}^\dagger} \eee^{\sum_{j=0}^m y_j^*\hat{a}_{\kk_j'}}}\right\vert_{{\bf x},{\bf y}=0}. \label{formalcumulant}
\ee
which we refer to as a ``$p$-uplet''.  Next, we reproduce the equations of motion used to produce the results of the main text.  We note that this system of equations was analyzed in depth in previous work \cite{art:cumulant2020}, and the explicit expressions are reproduced here for reasons of completeness.  As demonstrated in that work, the many equivalent models of the quenched unitary Bose gas found in the literature \cite{art:sykes,PhysRevLett.124.040403,art:munozdelasheras,art:colussimk,art:silviack,art:corson_bohn} can be grouped under the umbrella of the doublet model, which describes only the dynamics of the singlets and doublets.  Here, instead we consider the dynamics of singlets, doublets, and triplets, and we neglect quadruplets and higher order cumulants.

The dynamics of the singlet ($\psi_0$) is given by the Gross-Pitaevskii equation for the Hamiltonian in Eq.(1) of the main text
\begin{align}\label{eq:psi_sep} 
i \hbar\partial_t \psi_0=&g\left(\zeta_0^2n_0+\frac{2}{V}\sum_{\bf l}\zeta_{\bf l}^2n_{\bf l}\right)\psi_0+\frac{g\psi_0^*}{V}\sum_{\bf l}\zeta_0\zeta_{2{\bf l}}c_{\bf l}+\frac{g}{V^{3/2}}\sum_{{\bf l},{\bf s}}\zeta_{\bf l}\zeta_{2{\bf s}-{\bf l}}M^*_{{\bf l},{\bf s}},
\end{align}

and the condensate phase derivative is given by 
\begin{align}
\hbar\frac{\dd\theta_0}{\dd t}=&-\frac{1}{2n_0} \bb{\psi_0^*\ii\hbar \frac{\dd\psi_0}{\dd t}-\ii\hbar \frac{\dd\psi_0^*}{\dd t}\psi_0},\\
=&-\Bigg[g\zeta_0^2n_0+\frac{2g}{V}\sum_{\bf l} \zeta_{\bf l}^2 n_{\bf l}+\frac{g}{V}\sum_{\bf l}\zeta_0\zeta_{2{\bf l}}\Re c_{\bf l}+\frac{g}{\sqrt{n_0V^3}}\sum_{{\bf l},{\bf s}}\zeta_{\bf l}\zeta_{2{\bf s}-{\bf l}}M^*_{{\bf l},{\bf s}}\Bigg].\label{eq:thetadot_sep}
\end{align} 

The doublet equations of motion are 
\begin{widetext}
\begin{eqnarray}\label{eq:n_num}
i\hbar\partial_t n_{\bf k}&=&2i\ \text{Im}\left[\Delta_{\bf k}c_{\bf k}^*+2g\sqrt{\frac{n_0}{V}}\sum_{\bf l}\zeta_{2{\bf k}-{\bf l}}\zeta_{\bf l} M_{{\bf l},{\bf k}}+g\sqrt{\frac{n_0}{V}}\sum_{\bf l}\zeta_{\bf k}\zeta_{2{\bf l}-{\bf k}}M^*_{{\bf k},{\bf l}}\right],\\
\label{eq:n_sep}
i\hbar\partial_t c_{\bf k}&=&2E_{\bf k}c_{\bf k}+(1+2n_{\bf k})\Delta_{\bf k}+4g\sqrt{\frac{n_0}{V}}\sum_{\bf l}\zeta_{\bf l+k}\zeta_{\bf l-k} M^*_{{\bf l},{\bf k}}+2g\sqrt{\frac{n_0}{V}}\sum_{\bf l}\zeta_{\bf k}\zeta_{2{\bf l}-{\bf k}}R_{{\bf k},{\bf l}},\label{eq:c_sep}
\end{eqnarray}
\end{widetext}
 where the Hartree-Fock hamiltonian and pairing field \cite{book:blaizot} are defined as
\begin{align}\label{eq:bog_param}
&E_{\bf k}=\epsilon_{\bf k}+2g\left[\zeta_{\bf k}^2n_0+\frac{1}{V}\sum_{{\bf l}}\zeta_{{\bf k}-{\bf l}}^2n_{\bf l}\right],\\
&\Delta_{\bf k}=g\zeta_{2{\bf k}}\left[\zeta_0n_0+\frac{1}{V}\sum_{\bf l}\zeta_{2{\bf l}}c_{\bf l}\right].
\end{align}
At momenta large compared to the many-body scales, one has
\begin{equation}
i\hbar\partial_t c_{\bf k}\approx2E_{\bf k}c_{\bf k}+\frac{g\zeta_{2{\bf k}}}{V}\sum_{\bf l}\zeta_{2{\bf l}}c_{\bf l},\label{eq:c_large}
\end{equation}
which is identical to the  two-body Schr\"odinger equation \cite{art:cumulant2020,art:colussimk,book:leggett} and justifies the expansion Eq.~(12) of the main text.  Additionally, the inhomogeneous drive terms involving $n_0$ in Eq.~\eqref{eq:c_sep} and $c$ in Eq.~\eqref{eq:psi_sep} describe the direct coherent exchange between atomic and pair condensates via low energy two-body scattering as described in Refs.~\cite{PhysRevA.65.033601,art:cumulant2020}.

The triplet equations of motion are given by
\begin{widetext}
\begin{eqnarray}
\label{eq:M_sep}
i\hbar\partial_t M_{\vec{k},\vec{q}} &=& \Big( E_\vec{k} - E_\vec{q} - E_{\vec{k}-\vec{q}}  \Big) M_{\vec{k},\vec{q}} 
-  \Delta^*_{\bf k-q}M^\ast_{\vec{q},\vec{k}} -\Delta^*_{\bf q} M^\ast_{\vec{k}-\vec{q},\vec{k}} + \Delta_{\bf k} R_{\vec{k},\vec{q}}^\ast + \mathcal{M}^{H_3}_{\bf k,q} +\mathcal{M}^{H_4}_{\bf k,q},\\
\label{eq:R_sep}
i\hbar\partial_t R_{\vec{k},\vec{q}} &=& \Big( E_\vec{k} + E_\vec{q} + E_{\vec{k}-\vec{q}} \Big) R_{\vec{k},\vec{q}} + \Delta_{\bf k} M^\ast_{\vec{k},\vec{q}} + \Delta_{\bf q}M^\ast_{\vec{q},\vec{k}} +\Delta_{\bf k-q} M^\ast_{\vec{k}-\vec{q},\vec{k}} + \mathcal{R}^{H_3}_{\bf k,q} +\mathcal{R}^{H_4}_{\bf k,q},
\end{eqnarray}
\end{widetext}
where $\mathcal{M}^{H_3}_{\bf k,q}$ and  $\mathcal{R}^{H_3}_{\bf k,q}$ contains doublet products and therefore represent the doublet sources, and  are given by
\begin{widetext}
\begin{eqnarray}
\label{eq:MH3_sep}
 \frac{\mathcal{M}^{H_3}_{\bf k,q}}{\sqrt{n_0/V} }&=&2g\left(\zeta_{\bf 2k-q}\zeta_{\bf q}c^*_{\vec{k}-\vec{q}}n_\vec{q} + \zeta_{\bf k+q}\zeta_{\bf k-q}n_{\vec{k}-\vec{q}}c^*_\vec{q} - n_\vec{k}(\zeta_{\bf k+q}\zeta_{\bf k-q}c^*_\vec{q}+\zeta_{\bf q}\zeta_{\bf 2k-q}c^*_{\vec{k}-\vec{q}})\right)\nonumber\\
&&+ 2g\Big(\zeta_{\bf 2q-k}\zeta_{\bf k}n_{\vec{k}-\vec{q}}n_\vec{q}-\zeta_{\bf 2q-k}\zeta_{\bf k}n_\vec{k}(1+n_\vec{q} +n_{\vec{k}-\vec{q}})  - c_\vec{k}(\zeta_{\bf 2k-q}\zeta_{\bf q}c^*_\vec{q}+\zeta_{\bf k+q}\zeta_{\bf k-q}c^*_{\vec{k}-\vec{q}})\Big), \\
\label{eq:RH3_sep}
 \frac{\mathcal{R}^{H_3}_{\bf k,q}}{\sqrt{n_0/V} } &=&2g\Big(\zeta_{\bf 2q-k}\zeta_{\bf k}c_\vec{k}(1+n_\vec{q}+n_{\vec{k}-\vec{q}}) + \zeta_{\bf 2k-q}\zeta_{\bf q}c_\vec{q}(1+n_\vec{k}+n_{\vec{k}-\vec{q}}) +\zeta_{\bf k+q}\zeta_{\bf k-q} c_{\vec{k}-\vec{q}}(1+n_\vec{k}+n_{\vec{q}})\Big)\nonumber\\
&&+2g\Big(\zeta_{\bf k-q}\zeta_{\bf  k+q}c_\vec{k} c_\vec{q} + \zeta_{\bf q}\zeta_{\bf 2k-q}c_\vec{k} c_{\vec{k}-\vec{q}} + \zeta_{\bf k}\zeta_{\bf 2q-k}c_\vec{q} c_{\vec{k}-\vec{q}} \Big).
\end{eqnarray}
\end{widetext}
Instead, $\mathcal{M}^{H_4}_{\bf k,q}$ and $\mathcal{R}^{H_4}_{\bf k,q}$ contain products of doublets and triplets and the full expression can be found in Ref.~\cite{art:cumulant2020}.  Following that work, we do not numerically simulate the full expressions, taking only the dominant $1 + n + n$ terms required to produce the correct form of the interacting few-body Hamiltonian and many-body T-matrix.  As discussed in Ref.~\cite{art:cumulant2020}, this approximation of the full expressions can be used at times $t\lesssim t_\mathrm{n}$ before the quantum depletion becomes significant. Therefore, one has 
\begin{widetext}
\begin{eqnarray}
\label{eq:MH4_sep}
\mathcal{M}^{H_4}_{\bf k,q}&\approx& 
 -\frac{g}{V} \sum_\vec{l} \zeta_{\bf 2q-k}\zeta_{\bf 2l-k}M_{\vec{k},\vec{l}} \big( n_{\vec{k}-\vec{q}} +n_\vec{q} +1 \big),\\
\label{eq:RH4_sep}
\mathcal{R}^{H_4}_{\bf k,q}&\approx& \frac{g}{V}\sum_\vec{l}  \Bigg(\zeta_{\bf 2q-k}\zeta_{\bf 2l-k}R_{\vec{l},\vec{k}} \big( n_\vec{q}+ n_{\vec{k}-\vec{q}} + 1 \big) + \zeta_{\bf 2k-q}\zeta_{\bf 2l-q} R_{\vec{l},\vec{q}} \big( n_\vec{k} + n_{\vec{k}-\vec{q}}+ 1\big) \nonumber\\
&&+ \zeta_{\bf k+q}\zeta_{\bf k-q+2l}R_{\vec{l},\vec{k}-\vec{q}} \big(n_\vec{k} + n_\vec{q} +1 \big) \Bigg),
\end{eqnarray}
\end{widetext}
which significantly reduces the computational burden.  
We direct the interested reader to Appendix B of Ref.~\cite{art:cumulant2020} for an in-depth discussion of the numerical simulation used to produce the results of the main text and of the violation of energy conservation muddies the long-time dynamics in the triplet model.  

At momenta large compared to the many-body scales, one has in analogy with Eq.~\eqref{eq:c_large} the simplification
\begin{equation}
i\hbar\partial_t R_{\vec{k},\vec{q}} \approx \Big( E_\vec{k} + E_\vec{q} + E_{\vec{k}-\vec{q}} \Big) R_{\vec{k},\vec{q}} + \frac{g}{V}\sum_\vec{l}  \Bigg(\zeta_{\bf 2q-k}\zeta_{\bf 2l-k}R_{\vec{l},\vec{k}}  + \zeta_{\bf 2k-q}\zeta_{\bf 2l-q} R_{\vec{l},\vec{q}} + \zeta_{\bf k+q}\zeta_{\bf k-q+2l}R_{\vec{l},\vec{k}-\vec{q}} \Bigg),\label{eq:R_large}
\end{equation}
which is identical to the three-body Schr\"odinger equation \cite{art:cumulant2020,art:colussimk} and justifies the expansion Eq.~(13) of the main text.  Furthermore, unlike Eq.~\eqref{eq:c_sep}, there are no inhomogeneous drive terms in Eq.~\eqref{eq:RH3_sep} depending solely on $n_0$ or solely on $c$.  As discussed in Ref.~\cite{PhysRevLett.89.210404}, these terms are responsible for a direct coupling between the atomic condensate and triples via low energy three-body scattering.

\section{\label{sec:sum_cond} Counting composite bosons}
In this section, we explain why a renormalisation is needed to interpret the 
macroscopic eigenvalues of the density matrices as condensate populations.
We first outline the difficulties encountered when trying to count composite
bosons in our cumulant model, before then illustrating the general problem by considering a simple Fock states of pairs.
\subsection{Renormalization of the pair and triple condensate fractions in the cumulant model}
\begin{figure}[t!]
  \centering
\includegraphics[scale=0.45]{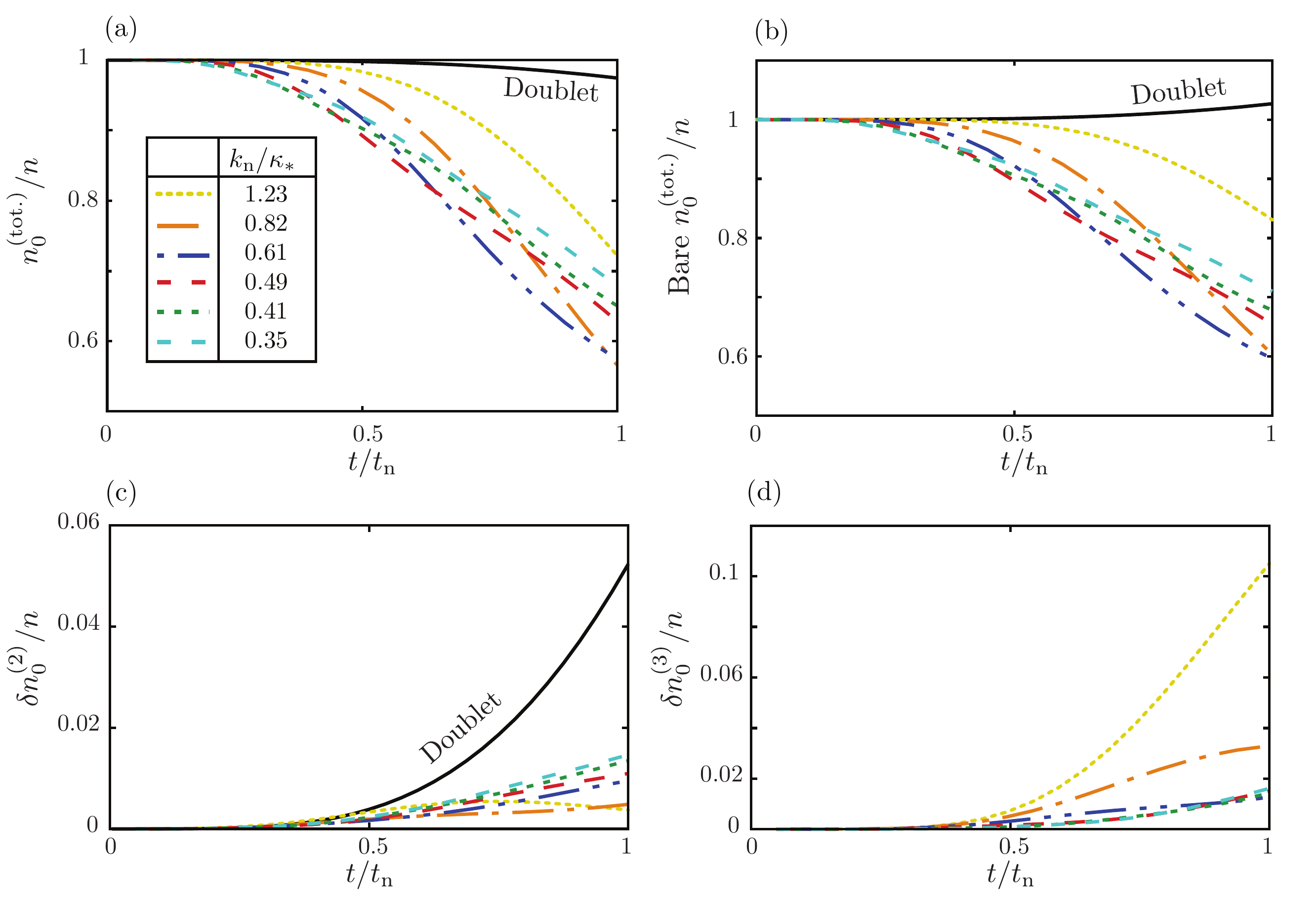}
\caption{Total condensed fraction using the (a) renormalized and (b) bare condensate operators.  The difference between the bare and renormalized (c) pair and (d) triple condensate fractions as a function of the density and time spent in the unitary regime.  The black lines indicate results in the universal doublet model.}
\label{fig:SM1}
\end{figure}
We start by giving explicitly the factorized expression of the 2-body density matrix (valid both in the doublet and triplet models):
\begin{multline}
\rho^{(2)}(\rr_1,\rr_2;\rr_1',\rr_2')\equiv
\meanv{\delta\hat\psi^{\dagger}(\rr_1')\delta\hat\psi^{\dagger}(\rr_2')\delta\hat\psi(\rr_2)\delta\hat\psi(\rr_1)}\\
=N_0^{(2)}\bbcro{\varphi_0^{(2)}(\rr_1'-\rr_2')}^* \varphi_0^{(2)}(\rr_1-\rr_2)+\meanv{\delta\hat\psi^{\dagger}(\rr_1')\delta\hat\psi(\rr_1)}\meanv{\delta\hat\psi^{\dagger}(\rr_2')\delta\hat\psi(\rr_2)}+\meanv{\delta\hat\psi^{\dagger}(\rr_1')\delta\hat\psi(\rr_2)}\meanv{\delta\hat\psi^{\dagger}(\rr_2')\delta\hat\psi(\rr_1)}.
\label{factorizedrho2}
\end{multline}
Here, the Hartree and Fock terms fall off to zero when $|\rr_1-\rr_1'|$ or $|\rr_1-\rr_2'|$ exceed a few Fermi lengths. 
On the contrary the anomalous term, written here directly in terms of the pair condensate wave function Eq.~(7), 
is responsible for long-range ordering. The associated macroscopic eigenvalue is explicitly
\be
N_0^{(2)}=\int \dd^3 r_1 \dd^3 r_2 |\meanv{\delta\hat\psi(\rr_1)\delta\hat\psi(\rr_2)}|^2=\sum_\kk |c_\kk|^2.
\ee

For fermionic pair condensates, this macroscopic eigenvalue is usually
interpreted as the number of fermions in the pair condensate \cite{book:leggett,Salasnich2005,PhysRevA.88.063623}.
Here, this interpretation appears problematic since in the doublet model
(the bosonic equivalent of BCS theory) $N_0^{(2)}$ always exceeds the number
of bosons available for pairing\footnote{We note here in passing the relation
between $N_0^{(2)}$ and the variance of the number of bosons $\hat N_{\rm ex}=\hat N-N_0$ outside the
one-body condensate: $N_0^{(2)}=\text{Var} \hat N_{\rm ex}-\sum_\kk n_\kk(1+n_\kk)\underset{\rm doublet}{=}\text{Var}\hat N_{\rm ex}/2$.
This follows from expressing the trace of $\rho^{(2)}$ in two ways: $\text{Tr}\hat\rho^{(2)}=\text{Var} \hat N_{\rm ex} + \meanv{\hat N_{\rm ex}}(\meanv{\hat N_{\rm ex}}-1)$ using the definition of $\rho^{(2)}$ (first line of Eq.~\eqref{factorizedrho2}) and the fact $\hat N_{\rm ex}$ fluctuates in our broken-symmetry state, and 
$\text{Tr}\hat\rho^{(2)}=N_0^{(2)}+\meanv{\hat N_{\rm ex}}^2+\sum_\kk n_\kk^2$ from the decomposed form on the second line of Eq.~\eqref{factorizedrho2}.}
\be
N_0^{(2)}\underset{\rm doublet}{=} \sum_\kk n_\kk(1+n_\kk) \geq N-N_0,
\label{N02ineq}
\ee
as numerically confirmed in Fig.~\ref{fig:SM1}(b).  This result follows from the relation $|c_\kk|^2=n_\kk(1+n_\kk)$ valid for the doublet
model, based on a Gaussian Ansatz. We note that the inequality would be reversed in the case of Fermi statistics:
\be
N_0^{(2),{\rm fermions}}\underset{\rm BCS}{=} \sum_\kk n_\kk(1-n_\kk) \leq N_{\rm fermions}
\label{N02ineq_fermi}
\ee
which explains why the renormalisation procedure was not previously proposed for fermionic pair condensates.

The origin of the overcounting in  Eq.~\eqref{N02ineq} becomes 
clear when one expresses $N_0^{(2)}$ in terms of the (bare) 
pair condensate operator (defined in Eq.~(8)):
\be
\frac{N_0^{(2)}}{2}=\meanv{\hat b_0^{(2)\dagger} \hat b_0^{(2)}}.
\ee
Here the ``superbosonic'' commutation relation obeyed by $\hat b_0^{(2)}$, ($\meanv{[\hat b_0^{(2)},\hat b_0^{(2)^\dagger}]}\geq 1$, see Eq.~(9)) forbids interpreting $\hat b_0^{(2)\dagger} \hat b_0^{(2)}$
as a number operator and hence $N_0^{(2)}$ as a number of bosons. With this in mind,
the renormalization procedure Eq.~(11) appears as a natural way to introduce
a bosonic number operator $\hat B_0^{(2)\dagger} \hat B_0^{(2)}$, whose average value
\be
\meanv{\hat B_0^{(2)\dagger} \hat B_0^{(2)}}=\frac{(N_0^{(2)})^2}{2\sum_\kk |c_\kk|^2(1+2n_\kk)}
\ee
we interpret as the number of pairs in the condensate.
To support this interpretation, we remark that in the doublet model $\meanv{\hat B_0^{(2)\dagger} \hat B_0^{(2)}}$ (contrary to
$N_0^{(2)}/2$) is always lower\footnote{To demonstrate the inequality in \eqref{B02ineq}, we compute
\be
\meanv{\hat B_0^{(2)\dagger} \hat B_0^{(2)}}-\frac{N_{\rm ex}}{2}=\frac{\sum_{\kk,\kk'}n_\kk n_{\kk'}  (1+n_{\kk'})(n_\kk -2 n_{\kk'})}{2\sum_\kk n_\kk(1+n_\kk)(1+2n_\kk)}.
\ee 
Exchanging indices in the sum of the numerator, one sees this quantity is negative for all distributions of $\{n_\kk\}$:
\bea
\sum_{\kk,\kk'} n_\kk n_{\kk'}(n_\kk -2 n_{\kk'}) &=& -\sum_{\kk,\kk'} n_\kk^2 n_{\kk'}, \\
\sum_{\kk,\kk'} n_\kk n_{\kk'}^2(n_\kk -2 n_{\kk'}) &=& \sum_{\kk,\kk'} n_\kk n_{\kk'}(n_\kk n_{\kk'}-n_\kk^2-n_{\kk'}^2)= -\sum_{\kk,\kk'} n_\kk n_{\kk'}\bbcro{(n_\kk -n_{\kk'})^2+n_\kk n_{\kk'}}.
\eea} than the maximal number of pairs
\be
\meanv{\hat B_0^{(2)\dagger} \hat B_0^{(2)}}\underset{\rm doublet}{=} \frac{(N_0^{(2)})^2}{2\sum_\kk n_\kk(1+n_\kk)(1+2n_\kk)}\leq\frac{N-N_0}{2},
\label{B02ineq}
\ee
as shown in Fig.~\ref{fig:SM1}(a). 
We note that the inequality is saturated in the weakly-excitated regime ($n_\kk\ll 1$) where $\meanv{\hat B_0^{(2)\dagger} \hat B_0^{(2)}} \simeq N_0^{(2)}/2  \simeq (N-N_0)/2$.

In the triplet model, the macroscopic eigenvalue of the three-body density matrix is explicitly
\be
N_0^{(3)}=\int \dd^3 r_1 \dd^3 r_2 \dd^3 r_3 |\delta\hat\psi(\rr_1)\delta\hat\psi(\rr_2) \delta\hat\psi(\rr_3)|^2=\sum_{\kk,\qq} |R_{\kk,\qq}|^2.
\ee
As shown in Fig.~\ref{fig:SM1} (b), the sum of the weighted eigenvalues $2N_0^{(2)}/2!+3N_0^{(3)}/3!$ never exceeds the number of bosons outside the one-body condensate. Still, we consider the renormalized numbers $\meanv{\hat B_0^{(2)\dagger} \hat B_0^{(2)}}$ and
\be
\meanv{\hat B_0^{(3)\dagger} \hat B_0^{(3)}}=\frac{\bbcro{N_0^{(3)}}^2}{6\sum_{\kk,\qq} |R_{\kk,\qq}|^2 (1+3n_\kk(1+ n_\qq))}
\ee
as better estimates of the pair and trimer condensate occupation numbers. In the time window we explore, the corrections $\delta N_0^{(p)}=N_0^{(p)}/p!-\meanv{\hat B_0^{(p)\dagger} \hat B_0^{(p)}}$ remain however small, as illustrated by Fig.~\ref{fig:SM1} (c) and (d). 

\subsection{Counting pairs in a Fock state of pairs}
We now argue that the difficulty in counting the number of composite bosons is not specific to cumulants models.  Rather, it occurs generally whenever the composite creation operator 
violates the bosonic commutation relations as was discussed for fermionic pairs 
previously, c.f. Ref.~\cite{PhysRevA.75.043613,Tanguy2003,Combescot_2001}.
We begin by considering the Fock state of pairs composed of $N$ composite bosons 
\be
\ket{N}=\frac{\bb{b^\dagger}^N\ket{0}}{\sqrt{\mathcal{N}(N)}},
\label{Fockpair}
\ee
with the normalization $\mathcal{N}(N)=\bra{0}b^N \bb{b^\dagger}^N \ket{0}$. As in Eq.~(8) (with $p=2$), 
the pair operator $\hat b$ is a linear superposition of two-body operators $\hat a_\alpha \hat a_\beta$.
Note that the discussion here is general and remains valid when $\hat b$ describes pairs of fermions \cite{Combescot_2001} or
distinguishable bosons.
To quantify the deviation of $\hat b$ from bosonicity, we define
\bea
\delta\hat C &=& \left[ \hat b,\hat b^\dagger\right]-1 \\
\delta^{2}\hat C &=& \left[\left[ \hat b,\hat b^\dagger\right],\hat b^\dagger\right]= \left[\delta\hat C ,\hat b^\dagger\right]
\eea
We note that $\delta^{2}\hat C$ is a linear superposition of 
bilinear creation operators $\hat a_\alpha^\dagger \hat a_\beta^\dagger$ and therefore commutes with $\hat b^\dagger$.
Remarking that the result (25) of Ref.~\cite{Combescot_2001}
is not restricted to pairs of fermions, we have
\be
\bra{N}\hat b^\dagger \hat b\ket{N}=N\bb{1+\frac{\meanv{\delta \hat C}}{2}}+O(1/N) \label{bdagb}
\ee
with $\meanv{\delta \hat C}\equiv \bra{N}\delta \hat C \ket{N} $, neglecting terms small in the thermodynamic limit\footnote{
By iteratively displacing the annihilation operator $\hat b$ to the right, at the cost of introducing the residual commutator
$\delta \hat C$ in successive locations, we obtain
\be
\bra{N}\hat b^\dagger \hat b\ket{N}=\frac{\bra{0}\hat b^N \hat b^\dagger \hat b (\hat b^\dagger)^N\ket{0}}{\mathcal{N}(N)}
=N+\frac{\sum_{p=1}^{N}\bra{0}\hat b^N (\hat b^\dagger)^p \delta \hat C (\hat b^\dagger)^{N-p}\ket{0}}{\mathcal{N}(N)}
\ee
To simplify this expression, we consider the sequence  $u_p\equiv\bra{0}\hat b^N (\hat b^\dagger)^{N-p} \delta \hat C (\hat b^\dagger)^{p}\ket{0}$.
The fact that $\delta^{2}\hat C$ commutes with $\hat b^\dagger$ ensure that $u_p$ is an arithmetic sequence:
\be
u_p-u_{p-1}=\bra{0}\hat b^N (\hat b^\dagger)^{N-p} \delta^2 \hat C (\hat b^\dagger)^{p-1}\ket{0}=\bra{0}\hat b^N (\hat b^\dagger)^{N-p+1} \delta^2 \hat C (\hat b^\dagger)^{p-2}\ket{0}=u_{p-1}-u_{p-2}
\ee
Together with $u_0=0$ (consequence of $\delta \hat C\ket{0}=0$), this shows that the general term of the sequence is
\be
u_p=\bra{0}\hat b^N (\hat b^\dagger)^{N-p} \delta \hat C (\hat b^\dagger)^{p}\ket{0}=\frac{p}{N}\bra{0}\hat b^N  \delta \hat C (\hat b^\dagger)^{N}\ket{0}=\frac{p}{N}u_N
\ee
With this property, we obtain
\be
\bra{N}\hat b^\dagger \hat b\ket{N}=N+\bra{N}\delta \hat C \ket{N} \bb{\sum_{p=1}^N \frac{N-p}{N}}=N\bb{1+\frac{N-1}{2N}\meanv{\delta \hat C}}
\ee
and eventually Eq.~\eqref{bdagb} above.
}. 
Equation \eqref{bdagb} shows that the operator $\hat b^\dagger \hat b$ overcounts the number of pairs, when
$\hat b$ is not a bosonic operator.

We note that the intuition that the state \eqref{Fockpair} contains $N$ pairs of bosons matches however an observable reality.  Consider the scenario where the pair wave function $\phi$ is
adiabatically tuned by an external parameter, such as the interaction strength, until it describes tightly bound dimers 
such that $\delta \hat C\approx 0$ and the bosonic commutation relations are well-satisfied. Then $N$ initially delocalised bosonic pairs convert
into a condensate of $N$ localized dimers whose population can be measured.

The conclusion we draw from this paradox is that the number of pairs 
(or, more generally, of composite bosons) cannot be measured 
directly by four-particle correlators such as the two-body
density matrix or $\hat b^\dagger \hat b$, at least not
in the regime where the size of the pairs is comparable to the interparticle spacing. 
In the simple example considered in this subsection, we knew the exact number of pairs $N$ because 
we had access to the exact many-body state \eqref{Fockpair}. This is not the
case of most many-body theories (including the cumulant model), 
which describe only low-order correlations. Experimentally, knowing the many-body state exactly requires
a prohibitive quantum state tomography.

This discussion also shows that care must be taken when comparing the number
of molecules measured after interaction sweeps as in Refs.~\cite{art:regal2004,art:klauss}
to the eigenvalues of the two-body density matrix, as is typically done for Cooper pairs of fermions \cite{Salasnich2005}.
A detailed analysis of the adiabatic sweep is required,
especially with theories whose access to high-order many-body correlations is limited.

\section{\label{sec:RK} Local relations}

In this section, we derive first the particular form of the contact relations Eqs.~(15) and (16), and then use this to obtain the proportionality constants in Eqs.~(12) and (13) in order to obtain well-behaved macroscopic order parameters satisfying $\mathcal{C}_p=|\Psi_0^{(p)}|^2$ in the triplet model.  

First, we derive Eq.~(15) of the main text from the cumulant expansion of the local contact relation \cite{art:braaten_c3_first} 
\begin{align}
\mathcal{C}_2=\frac{m^2 g^2}{\hbar^4}\langle(\hat{\psi}^\dagger)^2 \hat{\psi}^2\rangle=&\frac{m^2 g^2}{\hbar^4}\left[|\psi_0|^4+4n({\bf 0})|\psi_0|^2+2n({\bf 0})^2+|c({\bf 0})|^2+(\psi_0^*)^2c({\bf 0})+\psi_0^2c^*({\bf 0})\right.\nonumber\\
&\left.+2(M({\bf 0})\psi_0+M^*({\bf 0})\psi_0^*)+Q({\bf 0})\right]\label{eq:c2expsm},
\end{align}
where we have suppressed the internal degrees into the notation `$({\bf 0})$' to indicate local evaluation of the cumulants.  Here, $Q$ represents the quadruplet cumulant with $Q({\bf 0})=\langle(\hat{\psi}^\dagger)^2 \hat{\psi}^2\rangle_c$.  From the correspondence between the cumulant equations of motion and few-body Schr\"odinger equations at large momenta (see Ref.~\cite{art:cumulant2020} for the lengthy equation of motion for $Q$) and the known local lattice expression for the zero-energy two-body scattering wave function \cite{art:werner_fermions}
\begin{equation}
|\phi({\bf 0})|^2 = \frac{16\pi^2\hbar^4}{m^2 g^2},\label{eq:cshortlat}
\end{equation}
we infer the scaling of each cumulant in the expansion of Eq.~\eqref{eq:c2expsm} with the cutoff in the limit ($\Lambda/ k_\mathrm{n}\to\infty)$ at unitarity ($|a|\to\infty$)
\begin{equation}\label{eq:unitarityscaling}
\psi_0\propto\Lambda^0,\ n({\bf 0})\propto \Lambda^0,\ c({\bf 0})\propto \Lambda^1,\ M({\bf 0})\propto \Lambda^1,\ Q({\bf 0})\propto \Lambda^2.  
\end{equation}
Therefore, we find that the local contact relation reduces in this limit at unitarity to
\begin{align}
&\mathcal{C}_2=\frac{m^2 g^2}{\hbar^4}\left[|c({\bf 0})|^2+Q({\bf 0})\right]\label{eq:c2exp0},
\end{align}
where the first term describes the contribution of the pair order parameter, and the second term is $\delta\mathcal{C}_2$ of the main text.  In the triplet cumulant model, the quadruplets are set to zero by construction, and so one expects, analogous to BCS theory \cite{STRINATI20181}, that $\delta\mathcal{C}_2\neq 0$ requires the inclusion of pairing fluctuations. 

Next, we derive Eq.~(16) of the main text by considering first the cumulant expansion of the local contact relation  \cite{art:braaten_c3_first} 
\begin{equation}
\mathcal{C}_3=-\frac{m^2g^2}{2\hbar^4\Lambda^2}\left(H'+\frac{J'}{a\Lambda}\right)\langle(\hat{\psi}^\dagger)^3\hat{\psi}^3\rangle.\label{eq:c3local}
\end{equation}
The functions $H$ and $J$ are log-periodic in $\Lambda$ as
\begin{align}
&H(\ln(\Lambda/\Lambda_*))=h_0\frac{C-s_0S}{C+s_0S},\\
&J(\ln(\Lambda/\Lambda_*))=\frac{j_0+j_1(2SC)+j_2(C^2-S^2)}{(C+s_0S)^2},
\end{align}
where $C=\cos(s_0\ln(\Lambda/\Lambda_*))$ and $S=\sin(s_0\ln(\Lambda/\Lambda_*))$ and with universal constants $A=89.262$, $\phi=-0.669$, $h_0=0.879$, $j_0=-0.148$, $j_1=-0.892$, $j_2=-0.087$, and renormalization scale $s_0\ln(\Lambda_*/\kappa_*)=0.971\mod \pi$.  The $'$ notation indicates a partial derivative with respect to $\ln(\Lambda/\Lambda_*)$.  In the limit ($\Lambda/ k_\mathrm{n}\to\infty)$ at unitarity, one finds that $m^2g^2/2\hbar^4\Lambda^2$ scales as $1/\Lambda^4$, such that only terms scaling at least as $\Lambda^4$ in the cumulant expansion of $\langle(\hat{\psi}^\dagger)^3\hat{\psi}^3\rangle$ remain.  From Eq.~\eqref{eq:unitarityscaling} we can see that any combinations of these cumulants will not contribute in this limit.  Instead, we infer the local lattice expression for the zero-energy three-body scattering wave function from Refs.~\cite{art:werner_bosons,art:braaten_c3_first}
\begin{equation}
|\Phi(0,{\bf 0})|^2 = \frac{s_0^2 \sqrt{3}\Lambda^2\hbar^4}{4m^2 g^2}\left[-H'-\frac{J'}{a\Lambda}\right]^{-1}.\label{eq:Rshortlat}
\end{equation}
which displays the desired $\Lambda^4$ scaling at unitarity.  Therefore, the cumulants whose equations of motion correspond to three-body Schr\"odinger equations at large momenta can contribute to the cumulant expansion of Eq.~\eqref{eq:c3local} in the ($\Lambda/ k_\mathrm{n}\to\infty)$ limit at unitarity
\begin{equation}
\mathcal{C}_3=-\frac{m^2g^2}{2\hbar^4\Lambda^2}\left(H'+\frac{J'}{a\Lambda}\right)\left[|R({\bf 0})|^2+ S({\bf 0})\right],\label{eq:contactconnect3sm}
\end{equation}
where the first term describes the contribution of the triple order parameter, and the second term is $\delta\mathcal{C}_3$ of the main text.  In the triplet cumulant model, the sextuplet $S({\bf 0})=\langle(\hat{\psi}^\dagger)^3\hat{\psi}^3\rangle_c$ is set to zero by construction, and so one expects analogously that $\delta\mathcal{C}_3\neq 0$ requires the inclusion of tripling fluctuations.

The next step is to take the local limit of Eqs.~(12) and (13)
\begin{align}
&c({\bf 0},t)\underset{r\to0}{=} \alpha_{(2)}\Psi^{(2)}_0(t)\phi({\bf 0}),\label{eq:cshortsm} \\
&R({\bf 0},t)\underset{R\to0}{=} \alpha_{(3)} \Psi^{(3)}_0(t)\Phi({\bf 0}),\label{eq:Rshortsm}
\end{align}
where the $\alpha_{(p)}$'s are the undetermined proportionality constants.  We note that the local cumulants can be evaluated on a numerical grid as 
\begin{equation}
c({\bf 0},t)=\frac{1}{V}\sum_{\bf k} c_{\bf k},\quad R({\bf 0},t)=\frac{1}{V^{3/2}}\sum_{{\bf k},{\bf q}}R_{{\bf k},{\bf q}}.
\end{equation}
Next, we plug Eqs.~\eqref{eq:cshortsm} and \eqref{eq:Rshortsm} into the cumulant expanded contact relations Eqs.~\eqref{eq:c2exp0} and \eqref{eq:contactconnect3sm} and equate the contributions of $|c({\bf 0})|^2$ and $|R({\bf 0})|^2$, respectively, to obtain
\begin{equation}
\alpha_{(2)}=\frac{1}{4\pi},\quad \alpha_{(3)}=\frac{2^{3/2}}{3^{1/4}s_0}
\end{equation}
Here, several comments about our ``contact' convention for the macroscopic order parameters are in order.  We note that multiplying the $\alpha_{(p)}$'s by $m/\hbar^2$ produces a pair order parameter with units of energy.  This is typically done in theories of the two-component Fermi gas due to the connection between the order parameter and the gap in the weakly-attractive BCS regime \cite{art:leggett}.  We note that in this context the pair wave function is typically referred to as ``$F$'' (c.f. Refs.~\cite{book:pitaevskii_stringari}).  We have chosen to omit these factors to make the interpretation of $|\Psi_0^{(p)}|^2$  as a probability density more apparent.  Explicitly, one finds that integrating the macroscopic order parameters over the entire system yields then the condensed contribution to the extensive contacts $C_2$ and $C_3$.  Explicitly, in the zero-range limit of the triplet model one has the extensive relation $\int dV |\psi_0|^2=N_0$ and $\int dV |\Psi_0^{(p)}|^2=C_p$, which demonstrates the connection with the extensive contacts ($\int dV \mathcal{C}_{p}=C_p$) and analogy with the order parameter of the atomic condensate.  Finally, we qualify that the converse relationship that pair or triple condensation is implied by a nonzero contact is not necessarily true as evidenced by measurements and predictions in the nondegenerate unitary regime \cite{Fletcher377,PhysRevLett.125.110404} in which case $\mathcal{C}_p=\delta\mathcal{C}_p$, and the triplet model becomes insufficient.

\section{\label{sec:RK} triple wave function averaging}
\begin{figure}[t!]
  \centering
\includegraphics[scale=0.5]{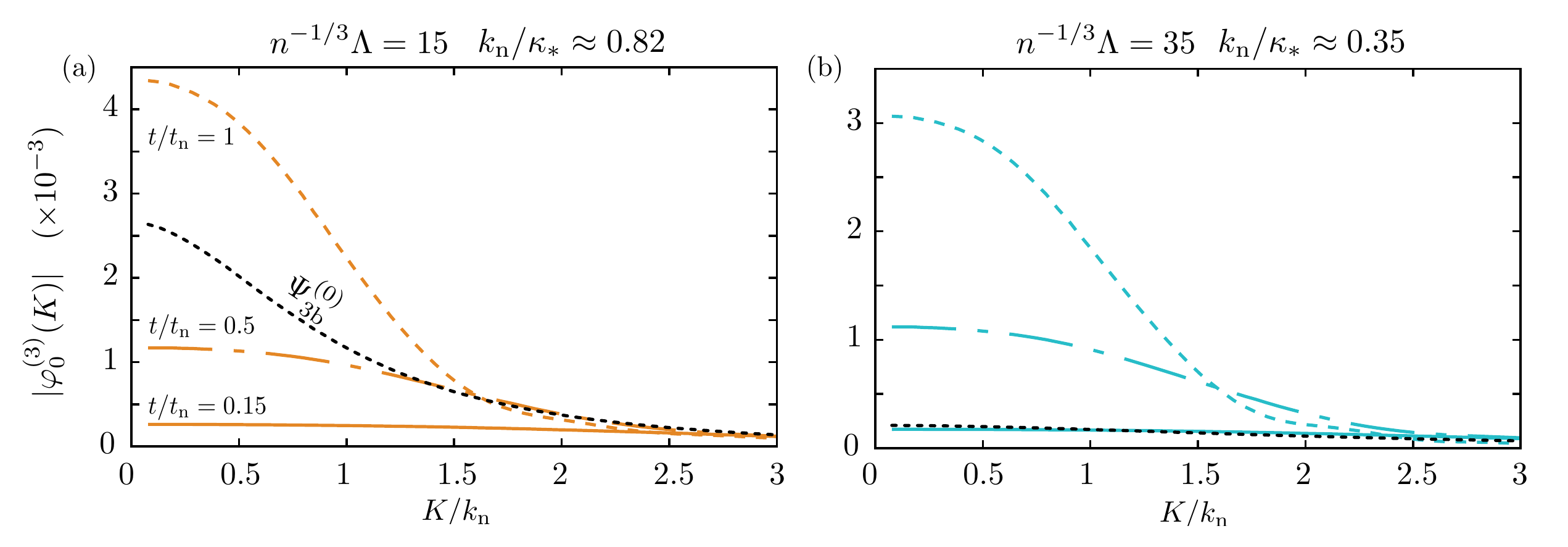}
\caption{Comparison of the normalized triple condensate wave functions ($|\varphi_0^{(3)}\rangle$) over various times and densities corresponding to Fig.~3(b) with the vacuum wave function of the ground-state Efimov trimer ($|\Psi^{(0)}_\mathrm{3b}\rangle$).  The wave functions have been averaged using the procedure outlined in Sec.~\ref{sec:RK}.}
\label{fig:SM2}
\end{figure}

In order to study the internal structure of the triple condensate wave function in Fig.~3(b), we have averaged over the internal configurations to reduce the overall dimensionality.  The relevant cumulant $R_{{\bf k},{\bf q}}=\langle \hat{a}_{\bf q-k}\hat{a}_{\bf k}\hat{a}_{\bf -q}\rangle$ describes a three-body configuration with individual momenta ${\bf k}$, $-{\bf q}$ and ${\bf q-k}$, which can be reduced to a dependence on the norms of the two single-particle momenta $k$ and $q$ and the angle between them ${\bf \hat k}\cdot{\bf \hat q}=\cos\theta$. Here, we introduce the hypermomentum $K$, characterizing the overall three-body momentum scale, which is generally defined in terms of the two Jacobi wave vectors as $K^2= q_1^2+(3/4)p_1^2$ and can be written in terms of the specific parametrization of the $R$ cumulant as  $K^2=k^2+q^2 + kq\cos\theta$. Starting from the 3-dimensional array $R_{{\bf k},{\bf q}}$, we average all the components with the same hypermomentum.
Numerically, the averaging of $R(K)$ is accomplished by the sum
\begin{equation}
R(K) =\frac{1}{N_\mathrm{count}(K)} \sum_{ijl} R(k_i, q_j, \cos\theta_l) P_{ijl}(K),\label{eq:RK}
\end{equation}
where the summation is taken over all indices of the 3D grid array $(k_i,q_j,\cos\theta_l)$.  Here, $P_{ijl}(K)$ is a conditional array, which reads 1 for indices $(i,j,l)$ corresponding to a configuration with hypermomentum $K-\Delta K \leq \sqrt{k_i^2 + q_j^2 + k_i q_j \cos\theta_l} < K+ \Delta K$ falling within a bin of fixed width $2\Delta K$ and reads 0 otherwise.  In order to take the average, we divide by the number $N_\mathrm{count}(K)=\sum_{ijl}P_{ijl}(K)$, which records the total number of suitable configurations counted for a fixed hypermomentum.  We note that each configuration $(i,j,l)$ corresponds to a distinct hyperangle ${\bf \Omega}_{ijl}$, and therefore Eq.~\eqref{eq:RK} is equivalent to preforming the hyperangular average at fixed hypermomentum.

We apply this averaging procedure also the to the ground-state Efimov trimer ($|\Psi^{(0)}_\mathrm{3b}\rangle$) and compare against the normalized triple condensate wave functions ($|\varphi_0^{(3)}\rangle$) in Fig.~\ref{fig:SM2}, which includes additional time and densities to supplement the comparison made in the inset of Fig.~3(b).  Here, we see in Fig.~\ref{fig:SM2}(a) that there is a strong resemblance between the wave functions between times $t/t_\mathrm{n}=0.5$ and $1$.  In Fig.~\ref{fig:SM2}(b), there is a close resemblance instead at early times near $t/t_\mathrm{n}=0.15$.  For both densities, these time windows coincide with the corresponding peaks of $P^{(0)}_\mathrm{3b}$ found in Fig.~4, which reinforces the conclusions of the main text and illustrates explicitly resemblance between condensed triples and Efimov trimers at various times and densities.

\bibliographystyle{apsrev4-1}
\bibliography{biblio}